\documentclass[11pt]{article}

\usepackage[a4paper, margin=1in]{geometry}
\usepackage{setspace}

\usepackage[T1]{fontenc}
\usepackage{lmodern}

\usepackage{graphicx}
\usepackage{float}
\usepackage[section]{placeins}  
\graphicspath{{figures/}}

\usepackage{booktabs}
\usepackage{array}
\usepackage{longtable}
\usepackage{tabularx}
\usepackage{multirow}
\usepackage{threeparttable}
\usepackage{adjustbox}

\usepackage{amsmath}
\usepackage{amssymb}
\usepackage{amsthm}

\usepackage{enumitem}

\newtheorem{definition}{Definition}[section]
\newtheorem{axiom}{Axiom}[section]

\usepackage[round,authoryear]{natbib}
\usepackage{xcolor}
\definecolor{brandburgundy}{RGB}{128,0,32}

\usepackage{tcolorbox}

\usepackage{tikz}
\usetikzlibrary{shapes,arrows,positioning,calc}

\usepackage{url}
\usepackage[colorlinks=true,
            linkcolor=brandburgundy,
            citecolor=brandburgundy,
            urlcolor=brandburgundy,
            breaklinks=true,
            pdftitle={ASRI: An Aggregated Systemic Risk Index for Cryptocurrency Markets},
            pdfauthor={Murad Farzulla, Andrew Maksakov},
            pdfkeywords={systemic risk, cryptocurrency, DeFi, stablecoin, contagion, risk index}]{hyperref}

\def\UrlBreaks{\do\/\do-\do_}
\expandafter\def\expandafter\UrlBreaks\expandafter{\UrlBreaks\do\a\do\b\do\c\do\d\do\e\do\f\do\g\do\h\do\i\do\j\do\k\do\l\do\m\do\n\do\o\do\p\do\q\do\r\do\s\do\t\do\u\do\v\do\w\do\x\do\y\do\z}

\usepackage{titlesec}
\titlespacing*{\section}{0pt}{1.5ex plus 0.5ex minus 0.2ex}{1ex plus 0.2ex}
\titlespacing*{\subsection}{0pt}{1.2ex plus 0.4ex minus 0.2ex}{0.8ex plus 0.2ex}

\titleformat{\section}{\normalfont\large\bfseries\color{brandburgundy}}{\thesection}{0.5em}{}
\titleformat{\subsection}{\normalfont\normalsize\bfseries\color{brandburgundy}}{\thesubsection}{0.5em}{}
\titleformat{\subsubsection}{\normalfont\small\bfseries\color{brandburgundy}}{\thesubsubsection}{0.5em}{}

\usepackage{fancyhdr}

\newcommand{\papernum}{DAI-2509}

\newcommand{\paperdate}{January 2026}

\newcommand{\paperurl}{https://systems.ac/3/DAI-2509}

\fancypagestyle{firstpage}{
  \fancyhf{}
  \fancyfoot[C]{\small\thepage}

}

\begin{document}

\setstretch{1.15}
\thispagestyle{firstpage}

\begin{center}
{\small\textsc{\href{https://dissensus.ai}{Dissensus} Working Paper Series}}\\[0.2em]
{\small \href{\paperurl}{\papernum}}
\end{center}

\vspace{1.5em}

\begin{center}
{\LARGE\bfseries ASRI: An Aggregated Systemic Risk Index for Cryptocurrency Markets}\\[0.5em]
{\large\itshape An Interpretable Crypto-Native Stress Composite for Retrospective Systemic-Risk Discrimination}\\[1.5em]

{\large Murad Farzulla}\textsuperscript{1,2,*}, \quad {\large Andrew Maksakov}\textsuperscript{1}\\[0.8em]

{\small
  \textsuperscript{1}\href{https://dissensus.ai}{Dissensus}, London, UK \quad
  \textsuperscript{2}King's College London, London, UK%
}\\[0.5em]

{\footnotesize
  \textsuperscript{*}Correspondence: \href{mailto:murad@dissensus.ai}{murad@dissensus.ai}
  \quad
  ORCID: \href{https://orcid.org/0009-0002-7164-8704}{0009-0002-7164-8704}%
}\\[0.3em]
{\footnotesize \paperdate}
\end{center}

\begin{abstract}
\noindent Cryptocurrency markets have grown to represent over \$3 trillion in capitalisation, yet practitioners lack an interpretable, channel-decomposed composite for retrospectively characterising crypto-native systemic stress. This paper introduces the Aggregated Systemic Risk Index (ASRI), a composite stress measure built around crypto-native channels---stablecoin total-value-locked (TVL) drawdown, stablecoin-issuer concentration, and TVL volatility---augmented by a cross-market \emph{contagion} channel that we implement as a traditional-finance (TradFi) stress \emph{proxy} (Treasury yields, VIX, and the yield-curve spread) rather than as a directly measured DeFi-TradFi exposure. Methodologically the index belongs to the financial-network tradition of connectedness and contagion measurement: the contagion channel is a cross-market connectedness construct, the four weighted sub-indices form an explicit risk-transmission decomposition, and a Diebold--Yilmaz (2012) connectedness series computed on those sub-indices serves as the network benchmark. We derive an axiomatic foundation and quantitative formulas for four weighted sub-indices---Stablecoin Concentration Risk (30\%), DeFi Liquidity Risk (25\%), Contagion Risk (25\%), and Regulatory Opacity Risk (20\%)---and evaluate the index \emph{retrospectively}, as an interpretable discriminative monitoring framework, against four historical crisis episodes: the Terra/Luna collapse (May 2022), the Celsius/3AC contagion (June 2022), the FTX bankruptcy (November 2022), and the SVB banking crisis (March 2023). We do not claim a validated real-time forecaster; the contribution is a transparent, reproducible construction together with a methodological account of how autocorrelation- and block-structure-robust inference reshapes the apparent strength of crisis-detection claims. \textbf{Empirical Results:} An event-study test for an elevated cumulative abnormal signal around the four onsets is inconclusive: the signal sums a 41-day window of 30-day-rolling components and is heavily serially correlated (AR(1) $\approx 0.8$--$0.9$), and a placebo test on non-crisis dates shows the finite-sample (fixed-$b$) reference has essentially no crisis-specificity on this smooth trending series---roughly sixty percent of arbitrary dates clear the nominal 5\% threshold, and the four crisis onsets sit at or below the placebo median. We therefore do not read the raw event-study significance values as crisis detection, and rest the empirical case on the fair-baseline day-level analysis below. Threshold-based operational detection (fixed ASRI $\geq 50$) identifies the same three events (Celsius/3AC, FTX, SVB) with an average first-crossing lead of $\approx$19 days (partly inflated by a running-maximum drawdown offset; $\approx$5 days under a responsive 30-day percentage-change specification). Benchmarking against a Diebold--Yilmaz (2012) connectedness series computed on the four sub-indices, ASRI shows higher \emph{retrospective} day-level discrimination on this sample (AUROC 0.866 vs.\ 0.670; AUPRC 0.298 vs.\ 0.121); but a fair-baseline comparison on identical labels demotes this headline. ASRI's discrimination (0.866) is \emph{not} statistically distinguishable from its single strongest sub-index (Contagion Risk, 0.851) or from the first principal component of its sub-indices (PC1, 0.858)---both fall inside ASRI's bootstrap confidence interval, and aggregation buys only $\approx +0.008$--$0.015$ AUROC over PC1 and the best single channel. A standalone equity-volatility series (VIX) also matches ASRI's headline discrimination (AUROC 0.875 vs.\ 0.866, indistinguishable at $p = 0.58$), so to first order the crisis labelling is a macro-volatility detection problem and ASRI's discriminative edge over an off-the-shelf series is not established. ASRI's measured advantage is confined to the Diebold--Yilmaz benchmark (0.670), which is itself \emph{constructed from} ASRI's four sub-indices (a circular comparator) and is so weak that an off-the-shelf Crypto Fear \& Greed sentiment index outperforms it (0.789 vs.\ 0.670). Under a moving-block bootstrap (block length $L=25$ days) the marginal confidence intervals are roughly four times wider than naive i.i.d.\ resampling implies; ASRI's day-level edge over D-Y is positive in every resample (AUROC $+0.194$, 95\% CI $[+0.093, +0.298]$; AUPRC $+0.182$, 95\% CI $[+0.070, +0.337]$), but this only establishes superiority over that circular, unusually weak comparator, and with only four independent crisis episodes the $\approx$0.008--0.015 AUROC gaps separating ASRI from its best single channel and PC1 are indistinguishable on the marginal intervals---the small paired-bootstrap edge that does separate them ($\leq 0.016$ AUROC, absent in AUPRC) is a structural artefact of ASRI nesting those baselines, not a discriminative gain. We therefore read the value of aggregation as interpretive---channel attribution, lead-time, and regime structure in a single auditable composite---rather than as a gain in discriminative power. Walk-forward thresholds calibrated only on pre-crisis data flag all four episodes, but at a high false-positive cost for the early crises (Terra/Luna and Celsius/3AC alarm on 47--59\% of the index history at $\approx$11--13\% precision), so we read the 4/4 walk-forward rate as evidence against look-ahead bias rather than as a clean prediction record. Stationarity tests reject the unit root for ASRI and the DeFi Liquidity sub-index ($p < 0.01$); Stablecoin and Opacity Risk are trend-stationary, while Contagion Risk is non-stationary on the released sample. A three-regime Hidden Markov Model labels Low Risk, Moderate, and Crisis states with persistence exceeding 97\%---interpretive structure over an autocorrelated trend rather than statistically distinct regimes---and the reproducible full-sample structural-stability (Chow) test at the sample midpoint is a clean non-rejection ($p = 0.78$). All detection, precision, and lead-time results rest on only four crisis events---the binding statistical-power limitation of this study---and roughly 43\% of the backtested series' average level is carried by static default values (which, being constant, contribute little of its time-variation), with the contagion channel's time-variation $\approx$75\% attributable to the macro (Treasury/VIX/yield-curve) proxy rather than to a measured interconnection. Out-of-sample 2024--2025 readings show no sustained false alarms and correctly classify the February 2025 Bybit hack (\$1.5B, the largest exchange theft in history) as non-systemic due to the absence of contagion channels. We therefore present ASRI as an interpretable, retrospective monitoring framework and methodological contribution---designed to target crypto-native vulnerabilities (stablecoin concentration, liquidity drawdown, opacity) that traditional measures such as SRISK and CoVaR are not built to capture---rather than as a statistically validated early-warning system.

\vspace{0.5em}

\vspace{0.3em}
\noindent\textbf{JEL Codes:} G01 (Financial Crises), G15 (International Financial Markets), G23 (Non-bank Financial Institutions)
\end{abstract}

\vspace{1em}

\clearpage
\tableofcontents
\thispagestyle{firstpage}


\clearpage
\setstretch{1.2}

\section{Introduction}

The cryptocurrency market has evolved from a niche technological experiment into a multi-trillion dollar asset class with growing interconnections to traditional finance. As of December 2025, the total cryptocurrency market capitalisation exceeds \$3 trillion, with stablecoins alone representing over \$140 billion in circulation \citep{defillama2025}. This growth has been accompanied by a series of cascading failures that revealed systemic vulnerabilities previously unrecognised: the Terra/Luna collapse eliminated \$40 billion in value within 72 hours; the subsequent Celsius and Three Arrows Capital failures triggered margin calls across centralised exchanges; and the FTX bankruptcy demonstrated how opaque counterparty relationships could propagate losses across the entire ecosystem.

Despite this systemic importance, no unified, channel-decomposed composite exists for retrospectively characterising crypto-native systemic stress. Existing measures either focus exclusively on cryptocurrency price volatility \citep{liu2021risks}, apply traditional banking metrics that miss DeFi-specific dynamics \citep{brownlees2017srisk}, or provide sentiment-based indicators without quantitative grounding \citep{alternative2023fear}.

This paper introduces the Aggregated Systemic Risk Index (ASRI), a composite stress measure constructed around crypto-native channels---stablecoin TVL drawdown, stablecoin-issuer concentration, and TVL volatility---with a cross-market \emph{contagion} channel implemented as a traditional-finance (TradFi) stress \emph{proxy} (Treasury yields, VIX, and the yield-curve spread) rather than as a directly measured DeFi-TradFi exposure. We are explicit about this throughout: ASRI does not measure realised DeFi-TradFi interconnection: it composes interpretable crypto-stress channels and proxies the macro/contagion dimension. We are equally explicit that ``crypto-native'' here characterises the index's \emph{design intent}, not what drives the realised series. A component-level accounting of the backtested index (Appendix~\ref{subsubsec:series_composition}, Table~\ref{tab:series_composition}) attributes only $\approx$37\% of its \emph{average level} (mean composition) to dynamic crypto-native inputs, against $\approx$43\% carried by static default values---which, being constant, contribute little or none of its time-variation---and $\approx$20\% by the macro (Treasury/VIX/yield-curve) proxy. ASRI is therefore better described as crypto-native in \emph{construction} than as predominantly crypto-native in realised \emph{variation}, a distinction we keep in view when interpreting the empirical results. The index comprises four weighted sub-indices:

\begin{enumerate}
    \item \textbf{Stablecoin Concentration Risk (30\%)}: Measures reserve composition, peg stability, and Treasury exposure across major stablecoins
    \item \textbf{DeFi Liquidity Risk (25\%)}: Captures protocol concentration, leverage dynamics, and smart contract vulnerability
    \item \textbf{Contagion Risk (25\%)}: Quantifies TradFi linkage intensity, tokenised RWA growth, and cross-market correlation shifts
    \item \textbf{Regulatory Opacity Risk (20\%)}: Assesses transparency scores, regulatory arbitrage metrics, and compliance infrastructure
\end{enumerate}

The ASRI framework addresses three critical gaps in existing risk monitoring:

First, \textit{composability risk}. DeFi protocols interact through smart contract calls that create dependency chains invisible to external observers. When one protocol fails, composable integrations can transmit losses instantaneously across the ecosystem---a dynamic that traditional contagion models, designed for bilateral counterparty relationships, cannot capture.

Second, \textit{stablecoin-Treasury linkages}. Major stablecoins now hold significant Treasury bill positions, creating a direct transmission channel between US monetary policy and DeFi liquidity conditions. Rate hikes that increase Treasury yields simultaneously reduce stablecoin reserve valuations and incentivise capital rotation out of yield-bearing DeFi positions.

Third, \textit{regulatory arbitrage dynamics}. The fragmented regulatory landscape creates opacity about counterparty risk exposures, custody arrangements, and reserve attestation reliability. Traditional banking metrics assume regulatory disclosure requirements that do not exist for offshore or unregulated platforms.

The urgency of dedicated crypto risk monitoring is underscored by recent TVP-VAR evidence from \citet{malik2025contagion}, who identify a structural break in Q3 2020 after which the cryptocurrency market shifted from net receiver to net transmitter of financial risk spillovers to traditional markets. Their finding that crypto is no longer a ``digital diversifier'' but an active source of systemic contagion motivates ASRI's focus on crypto-native stress channels and a macro-stress proxy for the cross-market boundary, while underscoring that the boundary itself is better measured directly than proxied---a limitation we keep in view throughout.

Conceptually and methodologically, ASRI sits within the financial-network-science tradition that treats systemic risk as a property of connectedness between markets rather than of individual institutions in isolation. That literature runs from Granger-causality and connectedness networks among financial firms \citep{billio2012econometric,diebold2012connectedness,diebold2014network}, through network-propagation measures of distress such as DebtRank \citep{battiston2012debtrank}, to recent mappings of the risk-transmission channels that link traditional and decentralised finance \citep{aufiero2025mapping}. ASRI's contagion sub-index is a market-connectedness construct in this lineage---built on cross-market correlation and co-movement---and its four-channel decomposition is a transmission-channel structure in which each sub-index names a pathway along which stress can propagate. We benchmark the composite against a Diebold--Yilmaz (2012) connectedness series computed on those sub-indices, the canonical variance-decomposition network measure of total spillovers, while remaining explicit (Appendix~\ref{subsec:dy_comparison}) about the limits of that particular comparator.

We are equally explicit about what the empirical evaluation does \emph{not} establish. A fair-baseline comparison on identical crisis labels (Section~\ref{subsec:fair_baselines}) shows that ASRI's day-level discrimination is, on the marginal block-bootstrap intervals, indistinguishable from that of its single strongest sub-index (Contagion Risk) and from the first principal component of its four sub-indices: aggregation adds only $\approx +0.008$--$0.015$ AUROC over the best single channel and PC1, a gap well within the four-event sampling uncertainty (a paired bootstrap separates them only by this small, structural margin---ASRI nests both baselines---not a meaningful discriminative gain). A standalone equity-volatility series (VIX) likewise matches ASRI's day-level discrimination on these labels (0.875 vs.\ 0.866, statistically indistinguishable), so to first order labelling these crises is a macro-volatility detection problem. ASRI's empirical contribution is therefore not that the four-channel composite out-discriminates simpler constructions---it does not, measurably---but that it packages crypto-native stress into a single, auditable, channel-decomposed series whose value is interpretive: it attributes stress to nameable transmission channels, carries operational lead-time for the detected crises, and exhibits coherent regime structure, none of which an opaque single feature or principal component delivers.

This paper proceeds as follows. Section~2 reviews the literature on systemic risk measurement, cryptocurrency market dynamics, and existing crypto risk indices. Section~3 develops the ASRI framework, specifying the axiomatic foundation and quantitative formulas for each sub-index and making explicit which components are crypto-native, which are macro proxies, and which are static defaults. Section~4 describes data sources and implementation architecture. Section~5 presents a \emph{retrospective} evaluation against historical crisis events---event study analysis under autocorrelation-robust inference, a block-bootstrap benchmark comparison, regime detection, robustness tests, and out-of-sample specificity testing. Section~6 discusses theoretical and practical implications. Section~7 concludes.

\section{Literature Review}

\subsection{Traditional Systemic Risk Measures}

The 2008 financial crisis catalysed extensive research on systemic risk measurement. \citet{adrian2016covar} introduced Conditional Value-at-Risk (CoVaR), measuring the VaR of the financial system conditional on an institution being in distress. \citet{acharya2017measuring} developed SRISK, estimating the expected capital shortfall of a financial institution during a systemic crisis. \citet{brownlees2017srisk} extended this framework with LRMES (Long-Run Marginal Expected Shortfall), capturing an institution's contribution to aggregate capital shortfall.

These measures share a common architecture: they model systemic risk as arising from bilateral exposures between regulated financial institutions with observable balance sheets and regulatory capital requirements. This architecture is fundamentally unsuited to DeFi, where:

\begin{itemize}
    \item Protocols are not institutions with capital requirements
    \item Exposures are embedded in smart contract code rather than disclosed counterparty relationships
    \item ``Failure'' may manifest as liquidity drainage rather than insolvency
    \item Contagion propagates through token price collapses and oracle manipulation rather than credit defaults
\end{itemize}

\subsection{Cryptocurrency Market Risk}

Research on cryptocurrency-specific risk has focused primarily on volatility dynamics and market microstructure. \citet{aste2019crypto} establishes a foundational connection between emotional dynamics and economic structure in cryptocurrency markets, demonstrating that market structure emerges from the interplay of sentiment-driven trading and network topology---a perspective that informs ASRI's integration of sentiment-adjacent indicators with structural risk channels. \citet{liu2021risks} identify three common factors driving cryptocurrency returns: market (aggregate crypto exposure), size, and momentum. \citet{makarov2020trading} document arbitrage frictions across exchanges that allow price dislocations to persist, creating opportunities for informed traders and risks for liquidity providers. \citet{farzulla2025market} demonstrate that infrastructure disruptions generate 5.7$\times$ larger volatility shocks than regulatory events, suggesting that technical vulnerabilities represent a more severe systemic risk channel than policy uncertainty.

\citet{griffin2020tether} provide evidence that Tether issuance patterns correlate with Bitcoin price movements, raising questions about stablecoin reserve integrity. This finding has implications for systemic risk: if stablecoin issuance is not fully collateralised, DeFi liquidity pools dependent on stablecoin inflows may be vulnerable to sudden redemption pressures.

\subsection{Crisis Episode Analysis}

The cryptocurrency crises of 2022--2023 generated substantial empirical scholarship validating theoretical concerns about systemic fragility. \citet{liu2023anatomy} provide the definitive analysis of the Terra/Luna collapse, documenting how algorithmic stablecoin mechanics created a reflexive depegging spiral that eliminated \$40 billion in value within 72 hours. \citet{ma2025stablecoin} extend this analysis to demonstrate that stablecoin run dynamics exhibit centralisation in arbitrage activity: during stress, redemption becomes concentrated among sophisticated actors, creating adverse selection that accelerates depegging.

The FTX collapse of November 2022 has received detailed independent analysis. \citet{vidaltomas2023ftx} trace FTX's downfall to the prior Terra/Luna ecosystem collapse, which triggered a liquidity crisis that exposed FTX's reliance on leveraged positions in its native FTT token and opaque intercompany transfers with Alameda Research. Their on-chain analysis demonstrates how centralised exchange fragility---masked by token self-valuation and inadequate reserve transparency---can trigger cascading failures across the broader ecosystem. This independent account of the FTX contagion pathway is qualitatively consistent with ASRI's event study for this one episode---in which the FTX collapse produced the highest recorded index value (84.7)\footnote{The released series' aggregate column differs slightly from a strict Equation~\ref{eq:asri} recomposition of its released sub-index columns, a consequence of the documented post-generation sub-index repair (mean gap $+0.84$ points, maximum $6.34$; the recomposed maximum is $81.06$ against the released $84.70$). All statistics reported in this paper use the released series, the dataset of record.} and the largest absolute increase from baseline of any validation event---though, as a single-event narrative correspondence, it does not constitute statistical out-of-sample validation.

The March 2023 Silicon Valley Bank crisis demonstrated bidirectional contagion between traditional and decentralised finance. \citet{diop2024svb} document the USDC depeg following SVB's collapse---the first major case of TradFi stress propagating into DeFi through stablecoin reserve exposure. \citet{imf2026firesales} model fire sale scenarios under systemic stablecoin stress, while \citet{eichengreen2025stablecoin} develop a framework for quantifying devaluation risk across stablecoin designs. \citet{vidaltomas2025crypto} examine the evolving dynamic relationship between cryptocurrency and traditional financial markets, finding evidence of increasing integration over time---a trend that strengthens the case for monitoring DeFi-TradFi interconnection as a systemic risk channel. More broadly, \citet{aufiero2025mapping} provide a systematic mapping of microscopic and systemic risk transmission channels between TradFi and DeFi, formalising the ``crosstagion'' pathways through which instability propagates across the conventional-decentralised boundary.

\subsection{DeFi-Specific Risks}

The academic literature on DeFi risk is nascent but growing rapidly. \citet{gudgeon2020defi} provide a systematisation of DeFi protocol architectures, identifying liquidation cascades as a primary risk channel. \citet{perez2021liquidations} analyse liquidation events across lending protocols, finding that liquidator competition can exacerbate price volatility during stress periods.

\citet{werner2022sok} present a systematisation of knowledge on DeFi covering lending, decentralised exchanges, and derivatives. They identify composability---the ability of protocols to permissionlessly interact through smart contract calls---as both a source of innovation and systemic vulnerability. When protocols share liquidity pools or collateral mechanisms, failures can propagate faster than human intervention can respond. Recent work on whitepaper-market alignment \citep{farzulla2025whitepaper} suggests that stated protocol objectives may diverge from realised market behaviour, creating additional opacity in risk assessment.

Recent work has deepened understanding of DeFi-specific systemic channels. \citet{bartoletti2022composability} formalise composability risk through MEV (Maximal Extractable Value), demonstrating that protocol interactions create value extraction opportunities that destabilise liquidity during stress. \citet{darlin2022debt} analyse debt-financed collateral structures, identifying leverage amplification mechanisms. Cross-chain bridge infrastructure represents acute vulnerability; \citet{notland2024bridges} systematise knowledge on bridge exploits, analysing 60 bridges and 34 attacks (2021--2023) to identify 13 architectural components linked to 8 vulnerability types. \citet{zhang2026correlation} propose a correlation-based fragility indicator for detecting dangerous protocol synchronisation.

\subsection{Stablecoin Risk}

Stablecoins occupy a critical position in DeFi infrastructure, serving as the primary medium of exchange and store of value across protocols. \citet{lyons2023stablecoin} analyse stablecoin run dynamics, finding that algorithmic stablecoins are particularly vulnerable to reflexive depegging spirals. The Terra/Luna collapse of May 2022 validated this theoretical concern empirically.

\citet{gorton2023taming} frame stablecoins through the lens of 19th-century ``wildcat banking,'' where private currency issuance without adequate regulatory oversight led to periodic banking panics. They argue that stablecoin reserves require the same transparency and examination that bank reserves receive---a standard that most major stablecoins currently fail to meet.

\citet{deblasis2023stablecoin} conduct comparative performance analysis across the Terra, Celsius, and FTX episodes, finding that fiat-collateralised stablecoins exhibit more resilience than algorithmic designs. \citet{antonakakis2020tvpvar} introduce the TVP-VAR connectedness framework that enables measurement of time-varying spillover dynamics without arbitrary rolling-window selection---a methodological advance particularly relevant for cryptocurrency markets where structural breaks occur frequently.

\subsection{Methodological Considerations}

A recent finding warrants consideration: \citet{rapos2025contagion} challenge the assumption of strong Bitcoin-equity market contagion, finding spillovers remain limited even during stress episodes. This finding does not invalidate contagion measurement but suggests correlation-based measures should be interpreted cautiously: elevated correlation may reflect common shocks rather than causal transmission.

\subsection{Existing Crypto Risk Indices}

Several commercial indices attempt to quantify cryptocurrency market risk:

\textbf{Fear \& Greed Index} \citep{alternative2023fear}: Aggregates sentiment indicators (social media, volatility, dominance, trends) into a 0--100 scale. Limitation: purely sentiment-based with no fundamental risk component.

\textbf{Crypto Climate Index (CCI)}: Combines on-chain metrics with market data. Limitation: focuses on individual asset health rather than systemic interconnection.

\textbf{Glassnode Risk Assessment}: Provides sophisticated on-chain analytics for Bitcoin and Ethereum. Limitation: asset-specific rather than ecosystem-wide; minimal DeFi coverage.

In the academic literature, \citet{guo2024cgri} propose the Cryptocurrency General Risk Index (CGRI), which tracks and sources cryptocurrency risk through a composite of market, sentiment, and macro indicators. While CGRI provides a useful aggregate risk signal for the cryptocurrency asset class as a whole, it operates at the market-sentiment level without decomposing risk into the structural transmission channels---stablecoin reserve linkages, DeFi composability dependencies, tokenised RWA exposures, and regulatory opacity---that characterise DeFi-TradFi interconnection. ASRI differs by targeting these specific contagion pathways, enabling identification of \textit{which} structural channel is generating stress rather than only \textit{whether} aggregate risk is elevated.

Most recently, \citet{shah2025defi} constructs a Unified DeFi Risk Index (DeFi-RI) integrating credit, liquidity, and governance risk into a composite scoring model. Shah's decomposition captures governance fragility and protocol-level credit exposure that ASRI currently underweights, but does not address the cross-boundary transmission channels---stablecoin-Treasury linkages, TradFi contagion pathways, and regulatory opacity---that constitute ASRI's primary focus. The two indices are thus complementary rather than competing: DeFi-RI monitors intra-DeFi protocol health while ASRI monitors the DeFi-TradFi boundary where systemic risk materialises.

None of these indices systematically captures the DeFi-TradFi interconnection dynamics that ASRI is designed to monitor: stablecoin reserve composition, composable protocol dependencies, tokenised RWA linkages, or regulatory arbitrage exposure.

\section{ASRI Framework}

\subsection{Conceptual Foundation}

The ASRI framework rests on three theoretical principles:

\textbf{Principle 1: Interconnection Creates Systemic Risk}. Following \citet{battiston2012debtrank}, systemic risk arises not from the failure of individual nodes but from the propagation of distress through network connections. \citet{aste2025ifn} provides the theoretical and methodological foundations for information filtering networks---sparse graph representations that extract statistically significant dependency structures from high-dimensional data---offering a principled approach to identifying which connections carry genuine risk information versus noise. In DeFi, these connections manifest through shared collateral pools, composable protocol integrations, and token price correlations.

\textbf{Principle 2: Risk Transmission Requires Channels}. We identify four primary channels through which DeFi stress can propagate to traditional finance (and vice versa): stablecoin reserve linkages, tokenised RWA exposures, crypto-equity correlations, and regulatory enforcement actions. Each channel corresponds to an ASRI sub-index.

\textbf{Principle 3: Opacity Amplifies Risk}. Systemic risk is exacerbated when counterparties cannot accurately assess exposures. The prevalence of unaudited reserves, undisclosed custody arrangements, and regulatory arbitrage structures in crypto markets justifies a dedicated opacity sub-index.

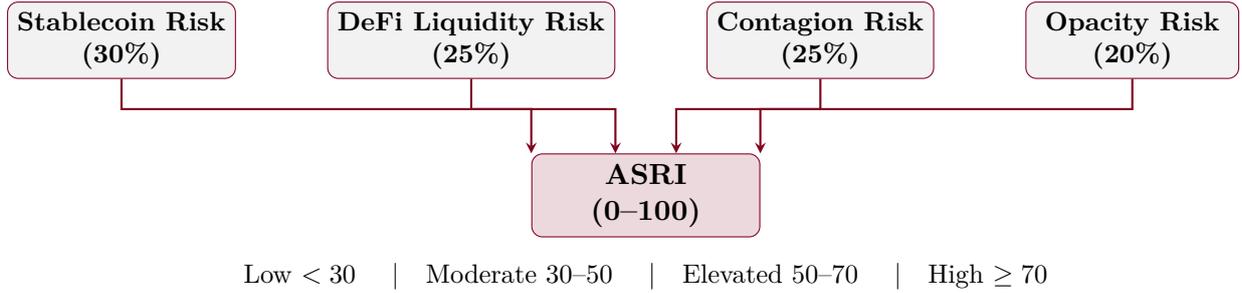
\begin{figure*}[t]
\centering
\begin{tikzpicture}[
    node distance=1cm,
    box/.style={rectangle, draw=brandburgundy, fill=gray!10, rounded corners, minimum width=2.8cm, minimum height=0.8cm, align=center, font=\small\bfseries},
    mainbox/.style={rectangle, draw=brandburgundy, fill=brandburgundy!15, rounded corners, minimum width=3cm, minimum height=1cm, align=center, font=\normalsize\bfseries},
    arrow/.style={->, >=stealth, thick, brandburgundy}
]
\node[box] (scr) {Stablecoin Risk\\(30\%)};
\node[box, right=1.2cm of scr] (dlr) {DeFi Liquidity Risk\\(25\%)};
\node[box, right=1.2cm of dlr] (cr) {Contagion Risk\\(25\%)};
\node[box, right=1.2cm of cr] (or) {Opacity Risk\\(20\%)};

\node[mainbox, below=1.5cm of $(dlr)!0.5!(cr)$] (asri) {ASRI\\(0--100)};

\draw[arrow] (scr.south) -- ++(0,-0.4) -| (asri.north west);
\draw[arrow] (dlr.south) -- ++(0,-0.4) -| ([xshift=-0.4cm]asri.north);
\draw[arrow] (cr.south) -- ++(0,-0.4) -| ([xshift=0.4cm]asri.north);
\draw[arrow] (or.south) -- ++(0,-0.4) -| (asri.north east);

\node[font=\small, below=0.2cm of asri, align=center] {Low $<30$ \quad|\quad Moderate $30$--$50$ \quad|\quad Elevated $50$--$70$ \quad|\quad High $\geq70$};
\end{tikzpicture}
\caption{ASRI Framework Architecture: Four weighted sub-indices aggregate into a normalised composite risk measure with defined alert thresholds.}
\label{fig:framework}
\end{figure*}

Figure~\ref{fig:framework} illustrates the ASRI architecture, showing how the four sub-indices aggregate into the composite risk measure.

\subsection{Axiomatic Foundation}\label{subsec:axiomatic}

We establish the formal properties that any coherent systemic risk index must satisfy, demonstrating that ASRI adheres to these axioms. Our axiomatic framework draws on the coherent risk measure literature \citep{artzner1999coherent} while extending it to the specific requirements of cryptocurrency market monitoring, where the absence of central counterparties and the prevalence of cross-venue arbitrage necessitate distinct aggregation properties.

\begin{definition}[Systemic Risk Index]
Let $\mathcal{S} = \{S_1, \ldots, S_n\}$ denote a set of sub-indices measuring distinct risk dimensions. A \emph{systemic risk index} is a mapping $\rho: \mathcal{S} \to [0, 100]$ that aggregates component risks into a scalar measure of system-wide vulnerability.
\end{definition}

For ASRI, we have $\mathcal{S} = \{\text{SCR}, \text{DLR}, \text{CR}, \text{OR}\}$ with the aggregation function:
\begin{equation}
    \text{ASRI}(\mathcal{S}) = \sum_{i=1}^{4} w_i S_i, \quad \text{where } \sum_{i=1}^{4} w_i = 1 \text{ and } w_i > 0 \ \forall i
\end{equation}

We now state and verify five axioms that characterise well-behaved systemic risk indices.

\begin{axiom}[Monotonicity]\label{ax:mono}
For any sub-index $S_j \in \mathcal{S}$, if $S_j' > S_j$ while $S_i' = S_i$ for all $i \neq j$, then $\rho(\mathcal{S}') > \rho(\mathcal{S})$.
\end{axiom}

\begin{axiom}[Boundedness]\label{ax:bound}
The index satisfies $\rho(\mathcal{S}) \in [0, 100]$ for all feasible states $\mathcal{S}$.
\end{axiom}

\begin{axiom}[Decomposability]\label{ax:decomp}
For any realisation of ASRI, there exists a unique attribution $\{c_1, \ldots, c_n\}$ such that $\sum_{i=1}^{n} c_i = \text{ASRI}(\mathcal{S})$ and $c_i$ represents the contribution of sub-index $S_i$.
\end{axiom}

\begin{axiom}[Aggregation Neutrality]\label{ax:neutral}
Linear aggregation preserves ordinal rankings: if $\mathcal{S}^A$ and $\mathcal{S}^B$ represent two market states with $S_i^A \geq S_i^B$ for all $i$ and strict inequality for at least one $j$, then $\rho(\mathcal{S}^A) > \rho(\mathcal{S}^B)$.
\end{axiom}

\begin{axiom}[Concentration Sensitivity]\label{ax:conc}
Let $H_t$ denote any market-concentration measure entering ASRI---specifically the stablecoin-issuer Herfindahl--Hirschman index $\text{HHI}_t$ in SCR (Eq.~\ref{eq:stablecoin}) or the top-10 protocol-TVL concentration $\text{Conc}_t$ in DLR (Eq.~\ref{eq:defi}). The index satisfies $\partial \rho / \partial H_t > 0$ for each such measure when concentration risk is elevated.
\end{axiom}

Proofs of Axioms~\ref{ax:mono}--\ref{ax:conc}, together with the relationship to the coherent-risk-measure framework of \citet{artzner1999coherent}, are given in Appendix~\ref{app:axiom_proofs}.
\subsection{Weight Selection Justification}
\label{subsec:weight_justification}

Sub-index weights were selected based on theoretical importance and precedent from traditional systemic risk literature:

\begin{itemize}
    \item \textbf{Stablecoin Concentration Risk (30\%)}: Stablecoins are the foundational liquidity layer for DeFi. Their failure would immediately impact all protocols dependent on stablecoin-denominated liquidity pools. The highest weight reflects this critical infrastructure role.

    \item \textbf{DeFi Liquidity Risk (25\%)}: Protocol concentration and leverage dynamics directly determine the ecosystem's resilience to market stress. We assign this channel no leading or ``canary'' role: with only four co-located crisis episodes the data do not identify which channel moves earliest (Section~\ref{subsec:weights}), and the earlier draft's hard-coded DLR weight that motivated such a claim does not reproduce from the weighting code.

    \item \textbf{Contagion Risk (25\%)}: DeFi-TradFi linkages represent the primary channel through which crypto stress could affect traditional finance (and vice versa). Equal weighting with DLR reflects their complementary roles: DLR captures within-DeFi stress, while CR captures cross-market transmission.

    \item \textbf{Opacity Risk (20\%)}: While important, opacity is an amplifying factor rather than a primary risk driver. Lower weight reflects this secondary role in conditioning crisis severity rather than triggering crises.
\end{itemize}

\begin{figure}[htbp]
\centering
\includegraphics[width=\textwidth]{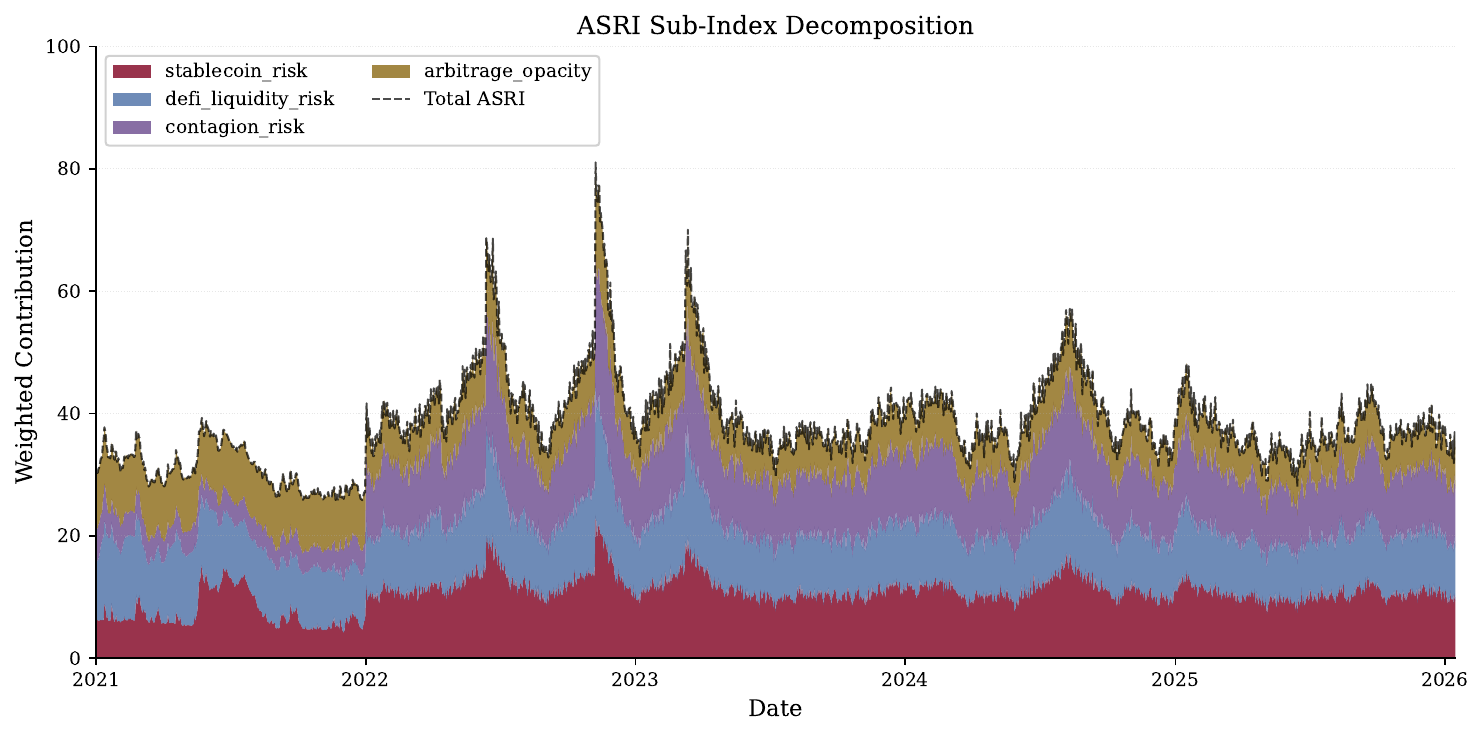}
\caption{ASRI Decomposition by Sub-Index Contribution Over Time. Stacked area chart showing how Stablecoin Risk (SCR, 30\%), DeFi Liquidity Risk (DLR, 25\%), Contagion Risk (CR, 25\%), and Regulatory Opacity (OR, 20\%) contribute to the aggregate ASRI. The decomposition property guarantees $c_i = w_i S_i$ such that total area equals ASRI at each time point. During crisis periods, shifting relative contributions reveal which transmission channels dominate aggregate stress.}
\label{fig:decomposition}
\end{figure}

Sensitivity analysis (Appendix~\ref{subsec:sensitivity}) tests robustness to alternative weight specifications. Section~\ref{subsec:weights} compares theoretical weights against four data-driven alternatives (PCA, Elastic Net, CRITIC, entropy) and finds that they disagree substantially among themselves; ablation analysis further shows detection is invariant at 3/4 to removing any single channel, so the decomposition is retained for interpretability rather than as an individually load-bearing causal structure. Figure~\ref{fig:decomposition} visualises this decomposition empirically, showing how each sub-index contribution evolves over the sample period.

\subsection{Stablecoin Concentration Risk (30\%)}

The Stablecoin Risk sub-index captures reserve composition vulnerabilities, peg stability, and concentration across issuers:

\begin{equation}
\text{SCR}_t = 0.4 \cdot \text{TVL}_t + 0.3 \cdot \text{Treasury}_t + 0.2 \cdot \text{HHI}_t + 0.1 \cdot \text{Vol}_t
\label{eq:stablecoin}
\end{equation}

where:

\begin{itemize}
    \item $\text{TVL}_t = 1 - \frac{\text{Stablecoin TVL}_t}{\max_{\tau \leq t}(\text{Stablecoin TVL}_\tau)}$ measures stablecoin TVL \textit{drawdown} from historical maximum---declining TVL increases risk\footnote{The inversion ensures countercyclical behaviour: when TVL collapses during crises, TVL$_t$ rises towards 1 (high risk); at historical peak, TVL$_t = 0$ (low risk). See Appendix~\ref{app:components} for implementation details.}\footnote{\emph{Construction limitation (running-maximum saturation).} The normalisation is against a running (expanding) maximum, so once total DeFi TVL fell below half of its late-2021 peak in mid-2022 and did not reclaim it, this term is clipped at its ceiling for $\approx$85\% of 2022--2024 and contributes a near-constant $\approx$12-point offset to ASRI ($w_{\text{SCR}}\cdot w_{\text{TVL}}\cdot 100 = 0.30\times0.40\times100$) across both crisis and calm post-2022 days, rather than tracking episode-specific stress. Because operational detection uses a \emph{fixed} threshold, this offset mechanically lifts post-2022 threshold crossings. Section~\ref{subsec:limitations} reports the effect of two alternative normalisations on detection.}

    \item $\text{Treasury}_t = \frac{\text{T-Bill Reserves}_t}{\text{Total Stablecoin Reserves}_t}$ captures Treasury exposure concentration

    \item $\text{HHI}_t = \sum_{i=1}^{n} s_i^2$ is the Herfindahl-Hirschman Index of stablecoin market share concentration

    \item $\text{Vol}_t$ is the 30-day realised volatility of weighted-average stablecoin peg deviation
\end{itemize}

\textbf{Data Sources}: DeFi Llama (stablecoin TVL), attestation reports (reserve composition), CoinGecko (price feeds for volatility calculation).

\subsection{DeFi Liquidity Risk (25\%)}

The DeFi Liquidity sub-index captures protocol concentration, leverage dynamics, and smart contract vulnerability:

\begin{equation}
\begin{split}
\text{DLR}_t = {} & 0.35 \cdot \text{Conc}_t + 0.25 \cdot \text{TVLVol}_t \\
& + 0.20 \cdot \text{SC}_t + 0.10 \cdot \text{Flash}_t + 0.10 \cdot \text{Lev}_t
\end{split}
\label{eq:defi}
\end{equation}

where:

\begin{itemize}
    \item $\text{Conc}_t$ is the HHI of TVL across top-10 DeFi protocols

    \item $\text{TVLVol}_t$ is the 30-day volatility of total DeFi TVL

    \item $\text{SC}_t$ is a composite smart contract risk score based on audit status, time since deployment, and exploit history

    \item $\text{Flash}_t$ measures flash loan volume spikes relative to 90-day average

    \item $\text{Lev}_t$ captures 30-day change in aggregate leverage ratios across lending protocols
\end{itemize}

\emph{Implementation overlap.} In the historical backtest $\text{Flash}_t$ is not sourced from a direct on-chain flash-loan feed but is derived from TVL-volatility dynamics, so it partially overlaps with $\text{TVLVol}_t$; the two terms are therefore not fully independent and their combined weight should be read as partially double-counting TVL-volatility stress rather than as two distinct liquidity signals. A direct on-chain flash-loan-volume metric is the preferable specification and is flagged for future revision.

\textbf{Data Sources}: DeFi Llama (TVL, protocol data), Token Terminal (flash loan data), DefiSafety (audit scores).

\subsection{Contagion Risk (25\%)}

The Contagion Risk sub-index quantifies DeFi-TradFi linkage intensity and cross-market transmission channels:

\begin{equation}
\begin{split}
\text{CR}_t = {} & 0.30 \cdot \text{RWA}_t + 0.25 \cdot \text{Bank}_t \\
& + 0.20 \cdot \text{Link}_t + 0.15 \cdot \text{Corr}_t + 0.10 \cdot \text{Bridge}_t\,
\end{split}
\label{eq:contagion}
\end{equation}

where:

\begin{itemize}
    \item $\text{RWA}_t$ is the 30-day growth rate of tokenised real-world asset TVL

    \item $\text{Bank}_t$ is a normalised score of bank crypto exposure from regulatory filings (OCC, ECB)

    \item $\text{Link}_t$ measures stablecoin flows to TradFi-connected entities

    \item $\text{Corr}_t$ is the 30-day rolling correlation between BTC/ETH and S\&P 500

    \item $\text{Bridge}_t$ is a composite of cross-chain bridge volume and recent exploit frequency
\end{itemize}

\textbf{Data Sources}: RWA.xyz (tokenised assets), DeFi Llama (bridge data), FRED (equity indices), regulatory filings.

\textbf{The Role of $\text{Bank}_t$.} The banking stress proxy $\text{Bank}_t$ occupies a theoretically central position in the Contagion Risk sub-index: it is the primary channel through which traditional financial sector distress propagates into cryptocurrency markets. Conceptually, $\text{Bank}_t$ captures the health of banks with crypto exposure---institutions whose balance sheet stress directly affects crypto-TradFi linkages through stablecoin reserve exposure, custodial relationships, and lending facilities. The March 2023 SVB crisis demonstrated this channel concretely: banking sector stress transmitted into DeFi through USDC's exposure to SVB deposits, causing a stablecoin depeg that briefly destabilised the broader ecosystem. In the empirical implementation, $\text{Bank}_t$ is operationalised as a Treasury-VIX composite (see Appendix~\ref{app:components}), reflecting the two primary mechanisms through which banking stress manifests: Treasury yield movements that affect bank capital ratios via mark-to-market losses, and equity volatility that signals broader risk-off conditions constraining bank lending and risk appetite. This proxy achieves daily frequency, overcoming the 45--90 day publication lag of quarterly regulatory filings from which the theoretical specification derives. We flag a theory-versus-evidence tension that the fair-baseline analysis (Section~\ref{subsec:fair_baselines}, Table~\ref{tab:fair_baselines}) makes explicit. The mark-to-market rationale treats the Treasury-yield \emph{level} as the primary channel, yet on our crisis labels that level discriminates at chance (AUROC 0.501) and, entering $\text{Bank}_t$ at $60\%$ weight, it \emph{lowers} the composite's discrimination from 0.875 (VIX alone) to 0.773---a $0.102$ reduction. The reduction is plausibly a trend artefact rather than a refutation of the channel: the level specification imports the 2021--2024 monetary-tightening trend ($\approx$0.9 percentage points per year) into $\text{Bank}_t$ rather than isolating crisis-specific banking stress. That removing such a trend recovers discrimination is shown by the reproducible detrended-Contagion channel, which rises to 0.912 (Table~\ref{tab:fair_baselines}); a change- or detrended-based Treasury input is therefore the preferable specification, which we flag as a refinement for future revisions.

\textbf{Implementation Note}: Bank$_t$ and Link$_t$ are implemented as high-frequency proxies because quarterly regulatory filings have 45--90 day publication lags. Link$_t$ uses yield curve spread. These proxies capture the same underlying stress dynamics at daily frequency (see Appendix~\ref{app:components} for full specification and Table~\ref{tab:proxies} for proxy validation).

\subsection{Regulatory Opacity Risk (20\%)}

The Opacity Risk sub-index assesses transparency deficits and regulatory arbitrage exposure:

\begin{equation}
\begin{split}
\text{OR}_t = {} & 0.25 \cdot \text{Unreg}_t + 0.25 \cdot \text{Multi}_t \\
& + 0.20 \cdot \text{Cust}_t + 0.15 \cdot \text{Sent}_t + 0.15 \cdot (100 - \text{Trans}_t)\,
\end{split}
\label{eq:opacity}
\end{equation}

where:

\begin{itemize}
    \item $\text{Unreg}_t$ is the ratio of unregulated to regulated platform volume

    \item $\text{Multi}_t$ captures multi-issuer stablecoin scheme exposure

    \item $\text{Cust}_t$ is custody concentration in non-audited jurisdictions

    \item $\text{Sent}_t$ is regulatory sentiment score from NLP analysis of SEC/ESRB/FSB announcements

    \item $\text{Trans}_t$ is a protocol-transparency score. The theoretical specification is reserve-attestation frequency and coverage; the implemented backtest proxies this by \emph{audit coverage}---the share of non-zero-TVL protocols with at least one audit (Appendix~\ref{app:components}, Table~\ref{tab:proxies})---because consistent historical attestation calendars are unavailable
\end{itemize}

\textbf{Data Sources}: DefiSafety audit coverage (implemented proxy); regulatory filings, news APIs (GDELT), and attestation calendars (theoretical specification).

\subsection{Aggregate ASRI Calculation}

The final ASRI is computed as a weighted sum of normalised sub-indices:

\begin{equation}
\text{ASRI}_t = 0.30 \cdot \text{SCR}_t + 0.25 \cdot \text{DLR}_t + 0.25 \cdot \text{CR}_t + 0.20 \cdot \text{OR}_t
\label{eq:asri}
\end{equation}

Normalisation uses min-max scaling over the historical sample to produce a 0--100 index:

\begin{equation}
\text{ASRI}^{\text{norm}}_t = 100 \times \frac{\text{ASRI}_t - \min_{\tau}(\text{ASRI}_\tau)}{\max_{\tau}(\text{ASRI}_\tau) - \min_{\tau}(\text{ASRI}_\tau)}
\label{eq:normalisation}
\end{equation}

Note that the component mappings detailed in Appendix~\ref{app:components}---the piecewise HHI risk score, the bounded \texttt{normalize}$(\cdot)$ transforms, and the $\times 100$ scalings---constrain each sub-index to $[0, 100]$; the displayed sub-index expressions (Equations~\ref{eq:stablecoin}--\ref{eq:opacity}) give the underlying raw inputs (for example, the TVL-drawdown ratio in $[0,1]$ and the raw HHI) prior to those mappings, and are not themselves bounded to $[0,100]$. Because the Appendix~\ref{app:components} mappings already bound the sub-indices, post-hoc min-max normalisation is unnecessary in practice. Equation~\ref{eq:normalisation} documents the theoretical relationship between raw and normalised values; empirical analyses in Section~5 use raw weighted aggregates directly. Collinearity among sub-indices is assessed via Variance Inflation Factors and condition number analysis (Section~\ref{subsubsec:collinearity}); all diagnostics confirm linear aggregation is well-conditioned.

\textbf{Alert Thresholds}:
\begin{itemize}
    \item $\text{ASRI} < 30$: Low systemic risk
    \item $30 \leq \text{ASRI} < 50$: Moderate systemic risk
    \item $50 \leq \text{ASRI} < 70$: Elevated systemic risk
    \item $\text{ASRI} \geq 70$: High systemic risk
\end{itemize}

\textbf{Operational Alert Rule}: An alert is triggered when ASRI $\geq 50$ (Elevated threshold) for at least one trading day. No persistence requirement is imposed because crisis dynamics can evolve rapidly; however, practitioners may implement confirmation windows (e.g., 3-day persistence) to reduce noise at the cost of lead time. Threshold selection follows a precision-recall trade-off documented in Table~\ref{tab:precision_recall}: the 50 threshold maximises crisis-level recall (75\%, i.e.\ 3/4 events; Terra/Luna is missed) while accepting moderate precision (30.1\%); raising the threshold to 70 improves precision to 41.7\% but lowers crisis-level recall to 50\% (2/4 events). These thresholds were chosen based on interpretability (round numbers mapping to verbal risk categories) rather than statistical optimisation; ROC-based calibration is deferred to future work with larger crisis samples. A structural caveat applies to any \emph{fixed} cut-off: because the Contagion sub-index is non-stationary on the released sample (Table~\ref{tab:stationarity}), the composite's baseline level can drift, so a constant threshold such as 50 does not correspond to a stable exceedance probability across the full history. The walk-forward rule of Section~\ref{subsec:walkforward}, which sets the alert at a trailing pre-crisis percentile (the 90th percentile of the index's own training-period history) rather than at a fixed level, is the more robust operational specification under this drift; we recommend it for deployment and retain the fixed-50 rule here for interpretability and comparability with the verbal risk bands. A fuller remedy---recalibrating the alert on a differenced or rolling-baseline series so that a fixed nominal cut-off maps to a stable exceedance probability, rather than reading a raw level against a static line on a drifting index---is a natural extension we flag for future work; the four-event sample here is too small to calibrate such a drift-adjusted threshold with power, which is why we report the fixed-50 detection alongside the drift-robust walk-forward percentile rather than substituting one for the other.

\section{Data and Implementation}

\subsection{Data Sources}

Table~\ref{tab:datasources} summarises the data sources for each ASRI component.

\begin{table*}[t]
\centering
\caption{ASRI Data Sources by Sub-Index}
\label{tab:datasources}
\small
\begin{tabularx}{\textwidth}{lXlll}
\toprule
\textbf{Sub-Index} & \textbf{Component} & \textbf{Source} & \textbf{Frequency} & \textbf{Tier} \\
\midrule
\multirow{4}{*}{Stablecoin Risk}
& TVL & DeFi Llama & Daily & 1 \\
& Treasury Reserves & Attestation Reports & Monthly & 2 \\
& Market Share & CoinGecko & Daily & 1 \\
& Peg Volatility & CoinGecko & Daily & 1 \\
\midrule
\multirow{4}{*}{DeFi Liquidity}
& Protocol TVL & DeFi Llama & Daily & 1 \\
& Flash Loan Volume & Token Terminal & Daily & 1 \\
& Smart Contract Scores & DefiSafety & Weekly & 2 \\
& Leverage Ratios & DeFi Llama & Daily & 1 \\
\midrule
\multirow{6}{*}{Contagion Risk}
& RWA TVL & RWA.xyz / DeFi Llama & Daily & 1 \\
& Bank Exposure & OCC/ECB Filings & Quarterly & 2 \\
& TradFi Linkages & On-chain Analysis & Weekly & 2 \\
& Equity Correlation & FRED/Yahoo Finance & Daily & 1 \\
& Bridge Volume & DeFi Llama & Daily & 1 \\
& Bridge Exploits & DeFi Llama & Daily & 1 \\
\midrule
\multirow{5}{*}{Opacity Risk}
& Platform Regulation & Manual Tracking & Weekly & 2 \\
& Custody Concentration & Public Disclosures & Monthly & 2 \\
& Regulatory Sentiment & GDELT/SEC Filings & Daily & 2 \\
& Attestation Frequency & Calendar Tracking & Daily & 2 \\
& Transparency Scores & DefiSafety & Weekly & 2 \\
\bottomrule
\end{tabularx}
\vspace{0.5em}
\raggedright\footnotesize
\textit{Abbreviations:} TVL = Total Value Locked; RWA = Real-World Assets; OCC = Office of the Comptroller of the Currency; ECB = European Central Bank; GDELT = Global Database of Events, Language, and Tone; SEC = Securities and Exchange Commission.
\end{table*}

\textbf{Tier 1} sources provide daily automated API access; \textbf{Tier 2} sources require manual collection, web scraping, or have lower update frequency.

\subsection{Data Quality Framework}

Data lag assumptions follow the $t-1$ convention: values observed at midnight UTC on date $t$ are attributed to ASRI$_{t-1}$ to avoid look-ahead bias.\footnote{To ensure backtests use only real-time information, min-max bounds in Equation~\ref{eq:normalisation} are theoretical; empirical analyses use raw ASRI values computed from bounded sub-indices without full-sample normalisation.} Several ASRI components draw on lower-than-daily sources (quarterly OCC/ECB bank-exposure filings, monthly stablecoin attestations, weekly on-chain linkage metrics): where a high-frequency proxy exists---most importantly the Treasury--VIX composite that stands in for $\text{Bank}_t$---we use it, and where none exists we carry the last observation forward with explicit confidence degradation. The full missing-data and mixed-frequency protocol is given in Appendix~\ref{app:data_quality_protocol}, and the pseudo-real-time evaluation (Appendix~\ref{subsec:realtime}) confirms that this protocol preserves detection performance under realistic publication lags. The system's four-layer data pipeline---ingestion, normalisation, computation, and publication---and its technology stack are described in Appendix~\ref{app:tech_arch}.
\section{Empirical Validation}

This section presents the empirical validation of the ASRI framework against historical data from January 2021 through January 2026, comprising over 1,800 daily observations across four major in-sample crisis events and out-of-sample specificity testing on 2024--2025 data.

\subsection{Crisis Taxonomy and Operational Definitions}\label{subsec:crisis-taxonomy}

Before proceeding to empirical validation, we establish operational definitions for what constitutes a systemic crisis in cryptocurrency markets. This taxonomy serves two purposes: providing ex ante criteria for event identification (avoiding post hoc selection bias) and enabling systematic classification of crisis mechanisms.

\subsubsection{Operational Crisis Definition}

Following the crisis identification methodology of \citet{laeven2013systemic} and adapted for high-frequency digital asset markets, we define a \textit{systemic stress event} as satisfying three jointly necessary conditions:

\begin{definition}[Systemic Stress Event]\label{def:crisis}
A period $[t_0, t_1]$ constitutes a systemic stress event if and only if:
\begin{enumerate}[label=(\roman*)]
    \item \textbf{Magnitude}: Aggregate market capitalisation decline $\geq 15\%$ within a 7-day window, or single-asset collapse $\geq 50\%$ for assets with market cap $\geq \$10$B;
    \item \textbf{Contagion}: Cross-asset correlation surge, measured as $\bar{\rho}_{t} - \bar{\rho}_{t-30} \geq 0.20$ where $\bar{\rho}$ denotes the mean pairwise correlation across major assets;
    \item \textbf{Duration}: Elevated stress conditions persist for $\geq 5$ trading days, distinguishing systemic events from flash crashes.
\end{enumerate}
\end{definition}

This definition deliberately excludes \textit{stress episodes}---periods of elevated volatility without systemic propagation. For instance, single-asset drawdowns (e.g., meme coin collapses) or brief correlation spikes during scheduled events (FOMC announcements) fail condition (ii) or (iii) respectively. The thresholds are calibrated to cryptocurrency market dynamics; traditional finance definitions \citep[e.g.,][]{reinhart2009time} typically require banking sector involvement, which maps imperfectly to decentralised systems.

\subsubsection{Crisis Typology}

We classify systemic events along two dimensions: origin (endogenous vs. exogenous) and primary transmission mechanism (liquidity vs. solvency). This yields a typology summarised in Table~\ref{tab:crisis_typology}.

\begin{table}[htbp]
\centering
\small
\caption{Crisis Typology for Cryptocurrency Markets}
\label{tab:crisis_typology}
\begin{tabular}{@{}lp{5cm}p{5cm}@{}}
\toprule
\textbf{Type} & \textbf{Characteristics} & \textbf{Historical Examples} \\
\midrule
\textbf{Type I: Endogenous} & Originates within DeFi/crypto ecosystem; propagates through on-chain liquidity channels, collateral cascades, or protocol failures & Terra/Luna (May 2022): algorithmic stablecoin death spiral triggering \$40B TVL collapse \\
\addlinespace
\textbf{Type II: Exogenous} & External shock (TradFi, regulatory, macroeconomic) propagates into crypto markets through stablecoin pegs, institutional exposure, or sentiment channels & SVB Crisis (March 2023): banking contagion $\rightarrow$ USDC depeg $\rightarrow$ DeFi stress \\
\addlinespace
\textbf{Type III: Hybrid} & Combined dynamics---crypto-native entity failures with significant TradFi counterparty exposure amplifying propagation & Celsius/3AC (June 2022); FTX (November 2022): CeFi insolvencies with cross-market contagion \\
\bottomrule
\end{tabular}
\end{table}

The four validation events span all three types, providing heterogeneous test conditions. Notably, Type III events (Celsius/3AC, FTX) exhibit the longest stress durations in our sample, consistent with \citet{brunnermeier2009deciphering}'s observation that hybrid crises produce more persistent dislocation due to opacity in cross-market exposures.

\subsubsection{Detection versus Prediction}

A critical methodological distinction: ASRI is \emph{designed} as a \textit{detection} instrument with intended \textit{leading properties}, not a pure prediction model. This is a statement of design intent, not a validated capability---consistent with our framing of ASRI as an interpretable, retrospective monitor rather than a validated early-warning system (Section~\ref{subsec:limitations}). Following \citet{borio2009assessing}, we distinguish:

\begin{itemize}
    \item \textbf{Detection}: Contemporaneous identification that systemic stress is occurring or imminent (hours to days horizon)---the \emph{design} criterion, not a validated forecast. Design criterion: Does ASRI breach threshold $\tau$ before or coincident with observable market stress?
    \item \textbf{Prediction}: Forecasting crisis probability over medium horizons (weeks to months). Would require different validation methodology (e.g., receiver operating characteristic analysis, out-of-sample forecasting).
\end{itemize}

Our retrospective evaluation focuses on detection performance: lead time relative to price cascade initiation, signal persistence during stress periods, and false-positive rates during non-crisis windows. The lead times documented below characterise ASRI as a threshold-based monitor evaluated in hindsight, not as a validated long-horizon forecasting tool; we read positive lead times as descriptive of the historical episodes rather than as a forward-looking guarantee.

\paragraph{Measurement Definitions.}
For consistency across all empirical analyses, we define:
\begin{itemize}
    \item \textbf{Detection threshold}: ASRI $\geq 50$ (``Elevated'' risk category). A crisis is detected if ASRI exceeds this threshold on any day in the 30-day pre-crisis window.
    \item \textbf{Lead time (threshold-based)}: Days between the \textit{first} threshold crossing (ASRI $\geq 50$) and crisis onset ($t = 0$, defined as price cascade initiation).
    \item \textbf{Lead time (event study)}: Days between first observation where ASRI exceeds 1.5 standard deviations above the estimation-window mean and crisis onset. This captures early stress signals relative to the pre-event baseline.
    \item \textbf{Crisis onset ($t = 0$)}: The first trading day exhibiting observable price cascade---typically a 10\%+ single-day decline in major assets or stablecoin depeg initiation.
    \item \textbf{False positive}: Any day where ASRI $\geq 50$ outside of the $[-30, +30]$ window surrounding a validated crisis event.
\end{itemize}
All hypothesis tests are two-tailed at $\alpha = 0.05$ unless otherwise specified. Confidence intervals are reported as 95\% intervals using bootstrap resampling (1,000 iterations) for detection rates and analytical standard errors for regression coefficients.

\subsection{Data and Sample}

Table~\ref{tab:descriptive} presents descriptive statistics for ASRI and its component sub-indices.

\begin{table}[H]
\begin{threeparttable}
\centering
\caption{Descriptive Statistics of ASRI Components}
\label{tab:descriptive}
\small
\begin{tabular}{@{}l*{6}{r}@{}}
\toprule
Variable & $N$ & Mean & Std & Min & Max & Skew \\
\midrule
ASRI            & 1,841 & 39.2 & 7.8 & 25.8 & 84.7 & 1.46 \\
Stablecoin Risk & 1,841 & 35.5 & 9.1 & 14.0 & 78.2 & 0.27 \\
DeFi Liquidity  & 1,841 & 42.3 & 7.5 & 27.9 & 90.0 & 1.79 \\
Contagion Risk  & 1,841 & 39.1 & 13.8 & 12.1 & 87.9 & $-$0.34 \\
Opacity Risk    & 1,841 & 36.9 & 6.9 & 22.6 & 70.2 & 0.77 \\
\bottomrule
\end{tabular}
\begin{tablenotes}
\small
\item Sample: January 2021 -- January 2026 (daily).
\end{tablenotes}
\end{threeparttable}
\end{table}

The ASRI ranges from 25.8 (low risk) to 84.7 (elevated risk during the FTX crisis), with crisis periods driving the upper tail. Positive skewness (1.46) reflects the asymmetric distribution: most observations cluster in the moderate band (30--50) while systemic stress events generate right-tail outliers, consistent with the design objective of early warning rather than tail risk measurement.

\begin{figure}[htbp]
\centering
\includegraphics[width=\textwidth]{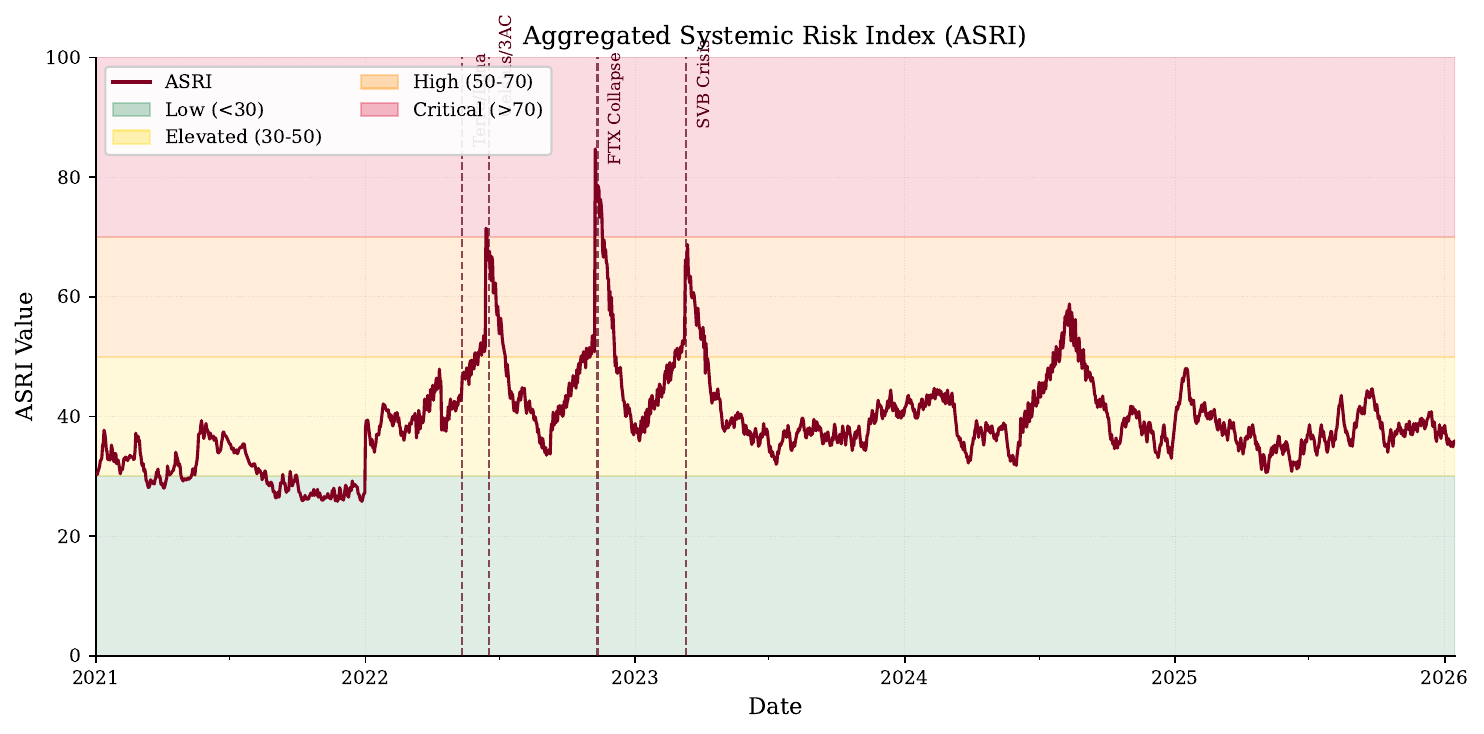}
\caption{ASRI Index Timeseries (January 2021 -- January 2026). Shaded bands denote operational risk regimes: Low ($<$30), Moderate (30--50), Elevated (50--70), and High ($\geq$70). Vertical dashed lines mark crisis event onsets: Terra/Luna collapse (May 2022), Celsius/3AC contagion (June 2022), FTX collapse (November 2022), and SVB banking crisis (March 2023). The index exhibits characteristic spikes during stress periods with rapid mean-reversion during recovery phases.}
\label{fig:asri_timeseries}
\end{figure}

Figure~\ref{fig:asri_timeseries} presents the complete ASRI timeseries from January 2021 through January 2026, with the four validated crisis events annotated and operational risk regime bands indicated.

The visual pattern confirms the statistical properties reported in Table~\ref{tab:descriptive}: the index spends most of its time in the Low-to-Moderate bands (25--50), with transient spikes into Elevated and High zones during the four major crises.

\subsection{Stationarity Tests}

Valid time series analysis requires stationary sub-indices. Table~\ref{tab:stationarity} reports Augmented Dickey-Fuller (ADF) \citep{dickey1979distribution} and KPSS \citep{kwiatkowski1992testing} test results.

\begin{table}[H]
\begin{threeparttable}
\centering
\caption{Stationarity Test Results}
\label{tab:stationarity}
\small
\begin{tabular}{@{}l*{3}{r}l@{}}
\toprule
Variable & ADF Stat & ADF $p$ & KPSS & Conclusion \\
\midrule
ASRI            & $-$4.05 & 0.001 & 0.65 & Stationary \\
Stablecoin Risk & $-$4.09 & 0.001 & 0.81 & Trend-stat. \\
DeFi Liquidity  & $-$4.32 & $<$0.001 & 0.69 & Stationary \\
Contagion Risk  & $-$2.73 & 0.069 & 2.06 & Non-stat. \\
Opacity Risk    & $-$4.30 & $<$0.001 & 2.94 & Trend-stat. \\
\bottomrule
\end{tabular}
\begin{tablenotes}
\small
\item ADF: Augmented Dickey-Fuller (lag selection via AIC, intercept included); KPSS: Kwiatkowski-Phillips-Schmidt-Shin (Bartlett kernel, automatic bandwidth). Computed on the released full-sample series (2021--2026).
\item KPSS critical values: 0.463 (5\%), 0.739 (1\%). Values exceeding 0.739 indicate, at best, trend-stationarity (the level is not covariance-stationary).
\item ASRI and DeFi Liquidity reject the unit root with KPSS consistent with (level/trend) stationarity. Stablecoin and Opacity Risk reject the unit root but exhibit elevated KPSS, indicating deterministic trends. Contagion Risk fails to reject the unit root (ADF $p = 0.069$; KPSS $= 2.06$) and is treated as non-stationary.
\end{tablenotes}
\end{threeparttable}
\end{table}

ASRI and the DeFi Liquidity sub-index are stationary; Stablecoin and Opacity Risk reject the unit root but display deterministic trends (elevated KPSS), and are best characterised as trend-stationary. Contagion Risk is the exception: it fails to reject the unit root on the released sample (ADF $p = 0.069$ full-sample, $p = 0.18$ on the 2021--2024 window) and carries a large KPSS statistic, so its \emph{level} is not covariance-stationary. We retain level-based analysis for the composite and the three stationary/trend-stationary channels, and flag that inferences resting specifically on the Contagion level (e.g.\ its variance-decomposition contribution) should be read with this caveat. Differencing Contagion Risk yields a stationary series, but we report levels for interpretability and consistency with the operational threshold definition.

\subsection{Event Study Analysis}\label{sec:event_study}

We apply formal event study methodology to assess ASRI behaviour around four major crisis events. Following \citet{mackinlay1997event}, we estimate ``normal'' ASRI levels from a 60-observation pre-event window (days $-90$ to $-31$ relative to event onset) and compute Cumulative Abnormal Signal (CAS) over the event window.\footnote{Unlike asset returns, ASRI is bounded on $[0,100]$. However, large-sample inference remains valid under the central limit theorem; the bounded support makes distributional assumptions less critical than in return-based event studies.}

\subsubsection{Event Study Specification}

\textbf{Normal Model.} We employ a constant mean model for expected ASRI during the estimation window:
\begin{equation}
\mathbb{E}[\text{ASRI}_t] = \hat{\mu} = \frac{1}{T_{\text{est}}} \sum_{\tau = -90}^{-31} \text{ASRI}_\tau
\label{eq:normal_model}
\end{equation}
where the estimation window spans $t = -90$ to $t = -31$ relative to the event date (60 trading days), providing a pre-event baseline uncontaminated by crisis dynamics.

\textbf{Abnormal Signal.} The abnormal signal on day $t$ is defined as the deviation from the expected level:
\begin{equation}
\text{AS}_t = \text{ASRI}_t - \mathbb{E}[\text{ASRI}_t] = \text{ASRI}_t - \hat{\mu}
\label{eq:abnormal_signal}
\end{equation}

The Cumulative Abnormal Signal (CAS) aggregates abnormal signals over the event window $[t_1, t_2]$:
\begin{equation}
\text{CAS}_{[t_1, t_2]} = \sum_{t = t_1}^{t_2} \text{AS}_t
\label{eq:cas}
\end{equation}
We use an event window of $[t_1, t_2] = [-30, +10]$, capturing the pre-crisis buildup period and immediate aftermath.

\textbf{Variance Estimation.} The variance of the abnormal signal is estimated from the estimation window:
\begin{equation}
\hat{\sigma}^2_{\text{AS}} = \frac{1}{T_{\text{est}} - 1} \sum_{\tau = -90}^{-31} \left( \text{ASRI}_\tau - \hat{\mu} \right)^2
\label{eq:variance}
\end{equation}

A naive standard error that assumes independent abnormal signals would be $\hat{\sigma}_{\text{AS}}\sqrt{T_{\text{event}}}$, with $T_{\text{event}} = 41$. This independence assumption is, however, untenable for ASRI: the cumulative abnormal signal sums a 41-day window of a series whose components are 30-day rolling quantities, so the abnormal-signal series is strongly positively autocorrelated by construction (event-window AR(1) $\approx 0.78$--$0.90$ across events). A Ljung--Box test decisively rejects white noise---$p$-values from $10^{-15}$ to $10^{-28}$ on the event-window abnormal series, and as low as $2\times10^{-58}$ on the estimation-window residuals---so the naive variance understates $\mathrm{Var}(\text{CAS})$ by roughly an order of magnitude. We therefore report Newey--West \citep{newey1987} heteroskedasticity- and autocorrelation-consistent (HAC) standard errors, obtained by regressing the event-window abnormal series on a constant with a HAC covariance estimator at lag $L = 20$ (chosen to span the 30-day rolling-window construction; the intercept HAC $t$ equals the CAS HAC $t$ because $\text{CAS} = n\bar{\text{AS}}$):
\begin{equation}
\text{SE}_{\text{HAC}}(\text{CAS}) = n \times \text{SE}_{\text{HAC}}(\bar{\text{AS}}),
\label{eq:se_cas}
\end{equation}
yielding $\text{SE}_{\text{HAC}}(\text{CAS}) = 58.3$ (Terra/Luna), $149.5$ (Celsius/3AC), $231.3$ (FTX), and $132.0$ (SVB).

\textbf{Test Statistic.} Under the null hypothesis of no abnormal signal ($H_0$: CAS $= 0$), the HAC test statistic is:
\begin{equation}
t_{\text{HAC}} = \frac{\text{CAS}}{\text{SE}_{\text{HAC}}(\text{CAS})} \;\xrightarrow{d}\; \mathcal{N}(0,1)
\label{eq:t_stat}
\end{equation}
referenced asymptotically against the standard normal distribution. For comparison we also report the naive $t = \text{CAS}/(\hat{\sigma}_{\text{AS}}\sqrt{T_{\text{event}}})$ that the prior draft relied on; the gap between the two is the quantitative cost of the false independence assumption.

\textbf{Finite-sample (fixed-$b$) reference.} The asymptotic $\mathcal{N}(0,1)$ reference is itself unreliable in this small sample. Because the cumulative abnormal signal sums a 41-day window of 30-day-rolling components (AR(1) $\approx 0.8$--$0.9$), the Newey--West bandwidth is a large fraction of the sample ($b = (L+1)/T = 21/41 \approx 0.51$; the Bartlett kernel at lag truncation $L = 20$ carries $L+1$ nonzero weights, so the Kiefer--Vogelsang bandwidth is $(L+1)/T$, not the lag fraction $L/T$); at that bandwidth-to-sample ratio the HAC $t$-statistic is far from standard normal, and referencing it against $\mathcal{N}(0,1)$ overstates significance. We therefore adopt a Kiefer--Vogelsang fixed-$b$ reference at the realised bandwidth as the inference of record \citep{kiefer2005}, whose critical values absorb the fat tails the small sample induces: $|t| \approx 3.52$ at the 5\% level (not $1.96$) and $|t| \approx 5.73$ at the Bonferroni-adjusted $\alpha = 0.0125$. A $t$-distribution with data-driven effective degrees of freedom ($\text{df}_{\text{eff}} \approx 2.3$--$5.0$) yields the same reading. Under this reference (Table~\ref{tab:event_study}), the three large-CAS events are individually marginal ($p \approx 0.04$--$0.06$), none clears the Bonferroni threshold, and Terra/Luna is clearly non-significant ($p \approx 0.27$). The qualitative result---three events elevate, Terra/Luna does not---is unchanged; the precise significance level is borderline-at-5\% rather than Bonferroni-surviving.

\textbf{Window Independence (and its limits for the clustered 2022 events).} Within each event the 60-day estimation window precedes, and does not overlap, that event's own 41-day event window. \emph{Across} events, however, the two early-2022 crises cluster too closely for their windows to be independent; the later FTX and SVB windows are well separated:
\begin{itemize}
    \item Terra/Luna (May 2022): Estimation window February--April 2022
    \item Celsius/3AC (June 2022): Estimation window March--May 2022
    \item FTX Collapse (November 2022): Estimation window August--October 2022
    \item SVB Crisis (March 2023): Estimation window December 2022--February 2023
\end{itemize}
Because Terra/Luna and Celsius/3AC fall within roughly five weeks of one another, their windows overlap: their event windows share $\approx 5$ days, their \emph{estimation} windows overlap by $\approx 24$ days (Terra/Luna's runs to mid-April 2022, Celsius/3AC's begins in mid-March), and Celsius/3AC's pre-event ``baseline'' window itself spans the Terra/Luna event window by $\approx 35$ days---so that baseline is not uncontaminated by the preceding Terra/Luna crash. We therefore do \emph{not} treat the two early-2022 events as statistically independent: they share estimation and baseline information, and the hold-one-out windows for these clustered events (Supplement) are correspondingly not fully independent. This is a genuine limitation of a sample whose crises cluster in 2022; it does not affect the FTX and SVB events, which are separated by over 90 days and have non-overlapping estimation and event windows.

\paragraph{Lead Time Measurement.}
Lead time is measured as days between the first observation where ASRI exceeds 1.5 standard deviations above the estimation-window mean and crisis onset. This definition captures early stress signals relative to the baseline rather than fixed threshold breaches, allowing detection of abnormality even when absolute levels remain below operational thresholds.

\subsubsection{Event Study Results}

\begin{table}[H]
\begin{threeparttable}
\centering
\caption{Event Study Results: ASRI Response to Crisis Events}
\label{tab:event_study}
\small
\begin{tabular}{@{}lc*{7}{r}@{}}
\toprule
Event & Date & Pre-Mean & Peak & CAS & Naive $t$ & HAC $t$ & $p_{\text{fix-}b}$ & Lead \\
\midrule
Terra/Luna   & 2022-05 & 40.4 & 48.7 &  100.3 &  5.47 &  1.72 & 0.265 & 30 \\
Celsius/3AC  & 2022-06 & 42.6 & 71.4 &  521.6 & 29.78 &  3.49 & 0.051 & 30 \\
FTX Collapse & 2022-11 & 39.6 & 84.7 &  758.8 & 32.64 &  3.28 & 0.063 & 30 \\
SVB Crisis   & 2023-03 & 41.0 & 68.7 &  509.0 & 26.91 &  3.86 & 0.036 & 29 \\
\midrule
\multicolumn{9}{@{}l}{\textit{Under the fixed-$b$ finite-sample reference the three large-CAS events are marginal ($p \approx 0.04$--$0.06$),}} \\
\multicolumn{9}{@{}l}{\textit{none clearing Bonferroni $\alpha = 0.0125$; Terra/Luna n.s.\ ($p \approx 0.27$). Average CAS: 472.4}} \\
\bottomrule
\end{tabular}
\begin{tablenotes}
\small
\item CAS = Cumulative Abnormal Signal. Lead = days between the first ASRI exceedance of $1.5\sigma$ above the estimation-window mean (searched over the 30-day pre-event window) and event onset.
\item $p_{\text{fix-}b}$ is the $p$-value under a Kiefer--Vogelsang fixed-$b$ reference at the realised bandwidth-to-sample ratio $b \approx 0.51$ ($= (L+1)/T = 21/41$ under the Newey--West/Bartlett convention; 5\% critical $|t| \approx 3.52$; Bonferroni-$0.0125$ critical $|t| \approx 5.73$); a $t$-reference with data-driven effective degrees of freedom ($\text{df}_{\text{eff}} \approx 2.3$--$5.0$) gives $p \approx 0.05$ for the three large-CAS events and $p \approx 0.15$ for Terra/Luna. Against the naive $\mathcal{N}(0,1)$ reference the same HAC $t$ would read $p = 0.0005$--$0.001$, but that reference is invalid at this bandwidth and overstates significance.
\item Naive $t = \text{CAS}/(\hat{\sigma}_{\text{AS}}\sqrt{41})$ assumes independent abnormal signals and is reported only for comparison; it is invalid here because the abnormal-signal series is strongly autocorrelated (event-window AR(1) $\approx 0.78$--$0.90$; Ljung--Box rejects white noise at $p < 10^{-15}$). HAC $t$ uses Newey--West standard errors at lag $L = 20$. Under HAC, Terra/Luna falls from $5.47$ to $1.72$; a fixed-$b$ finite-sample reference would render the three large-CAS events marginal ($p \approx 0.04$--$0.06$), but a placebo/specificity test shows that reference has no crisis-specificity on this smooth trending series, so the event study is treated as inconclusive (see text). A lag sweep ($L = 10/20/30$, each spanning the 30-day rolling-window construction) leaves Terra/Luna non-significant at every lag and the other three at $t \geq 3.28$ throughout (the minimum being FTX at $L=20$); Terra reaches the 5\% level only at the Newey--West auto-floor $L=3$, which is too short to capture the construction-induced persistence.
\item $t$-statistics use the canonical profile from the released code: estimation window $[-90,-31]$, event window $[-30,+10]$, lead-time lookback capped at 30 days. The reported leads (29--30) are at this cap: the 1.5$\sigma$ signal is elevated across essentially the entire pre-window (uncapped first-crossing 38--90 days), so these values indicate sustained pre-crisis elevation rather than a precise 30-day horizon. The walk-forward first-crossing leads (Table~\ref{tab:walkforward}, mean 25.75 days) are the more conservative operational figure.
\end{tablenotes}
\end{threeparttable}
\end{table}

Under autocorrelation-robust (HAC, $L=20$) inference the three large-CAS events (Celsius/3AC, FTX, SVB) carry HAC $t$-statistics of $3.49$, $3.28$, and $3.86$, while Terra/Luna carries $1.72$. Referenced naively against $\mathcal{N}(0,1)$ the first three would read $p<0.01$, but that reference is invalid at the realised bandwidth ($b \approx 0.51 = 21/41$), where the abnormal-signal series is heavily serially correlated (event-window AR(1) $\approx 0.8$--$0.9$; Ljung--Box rejects white noise at $p<10^{-15}$) and the naive standard error understates $\mathrm{Var}(\text{CAS})$ by roughly an order of magnitude. The natural next step---a finite-sample (fixed-$b$) reference---would downgrade the three to ``borderline'' ($p \approx 0.04$--$0.06$). But that reference is itself size-invalid on this series: running the identical pipeline on non-crisis dates (a placebo/specificity test) shows the fixed-$b$ $5\%$ critical value is cleared by roughly sixty percent of arbitrary dates---an empirical false-positive rate an order of magnitude above nominal---and the four crisis onsets sit at or below the placebo median. On a smooth, strongly trending series the cumulative-abnormal-signal statistic has no crisis-specific power here.

\textbf{Interpretation}: We therefore treat the event study as \emph{inconclusive}. It neither establishes nor refutes crisis-specific abnormal elevation: the raw CAS ordering is intuitive (FTX largest at $758.8$, Terra/Luna smallest at $100.3$, consistent with the difficulty of observing algorithmic-stablecoin fragility through market-based indicators), but because no onset separates from the placebo distribution, we do not read any of these as a crisis-specific signal and do not count the event study as evidence of detection. The empirical case for ASRI rests instead on the fair-baseline day-level discrimination analysis (Section~\ref{subsec:fair_baselines}), which does not depend on the event-study inference.

\textbf{Detection Nomenclature}: Throughout this section we distinguish \textit{threshold-based detection} (ASRI $\geq 50$ during the pre-crisis window) from the \textit{event study} (abnormal ASRI elevation around an onset). Threshold-based analysis achieves 3/4 detection (Terra/Luna missed, peak $48.7$); the event study is indeterminate on this sample for the reasons above and is not counted toward detection. Walk-forward validation flags 4/4 out-of-sample, but at a high false-positive cost for the early crises (Section~\ref{subsec:walkforward}), so we read it as evidence against look-ahead bias rather than a clean detection record.

Table~\ref{tab:detection_matrix} presents the unified detection matrix. Threshold-based detection is 3/4 (Terra/Luna falls below the operational threshold, peak $48.7$); the Event-Sig column records the raw HAC $t$-statistics but assigns no crisis-specific significance, since none of the onsets separates from the placebo distribution.

\begin{table}[htbp]
\centering
\caption{Unified Detection Matrix: Method Comparison}
\label{tab:detection_matrix}
\small
\begin{tabular}{lccccccccc}
\toprule
Event & Peak & $\tau$=40 & $\tau$=50 & $\tau$=60 & $\tau$=70 & Event Sig. & HAC $t$ & WF-OOS \\
\midrule
Terra/Luna & 48.7 & $\checkmark$ & $\times$ & $\times$ & $\times$ & $\times$ & 1.72 & $\checkmark$ \\
Celsius/3AC & 71.4 & $\checkmark$ & $\checkmark$ & $\checkmark$ & $\checkmark$ & $\times$ & 3.49 & $\checkmark$ \\
FTX Collapse & 84.7 & $\checkmark$ & $\checkmark$ & $\checkmark$ & $\checkmark$ & $\times$ & 3.28 & $\checkmark$ \\
SVB Crisis & 68.7 & $\checkmark$ & $\checkmark$ & $\checkmark$ & $\times$ & $\times$ & 3.86 & $\checkmark$ \\
\midrule
\textbf{Total} & & 4/4 & 3/4 & 3/4 & 2/4 & 0/4 & & 4/4 \\
\bottomrule
\end{tabular}
\begin{tablenotes}
\small
\item $\tau$ = threshold-based detection (ASRI $\geq \tau$ in 30-day pre-crisis window).
\item Event Sig. = event-study significance under autocorrelation-robust (Newey--West HAC, $L=20$) inference. No event is marked significant: a placebo test on non-crisis dates shows the finite-sample (fixed-$b$) reference has no crisis-specificity on this smooth trending series (empirical false-positive rate $\approx$50--60\% at nominal 5\%; the onsets sit at or below the placebo median), so the raw HAC $t$-statistics (Terra/Luna $1.72$; the others $3.28$--$3.86$) are reported in the HAC $t$ column but not read as crisis detection. The event study is treated as inconclusive throughout.
\item WF-OOS = walk-forward out-of-sample flag (90th-percentile threshold on training data); see Section~\ref{subsec:walkforward} for the associated false-positive cost.
\item *** $p<0.01$, ** $p<0.05$, * $p<0.10$
\end{tablenotes}
\end{table}

\begin{figure}[htbp]
\centering
\includegraphics[width=0.9\textwidth]{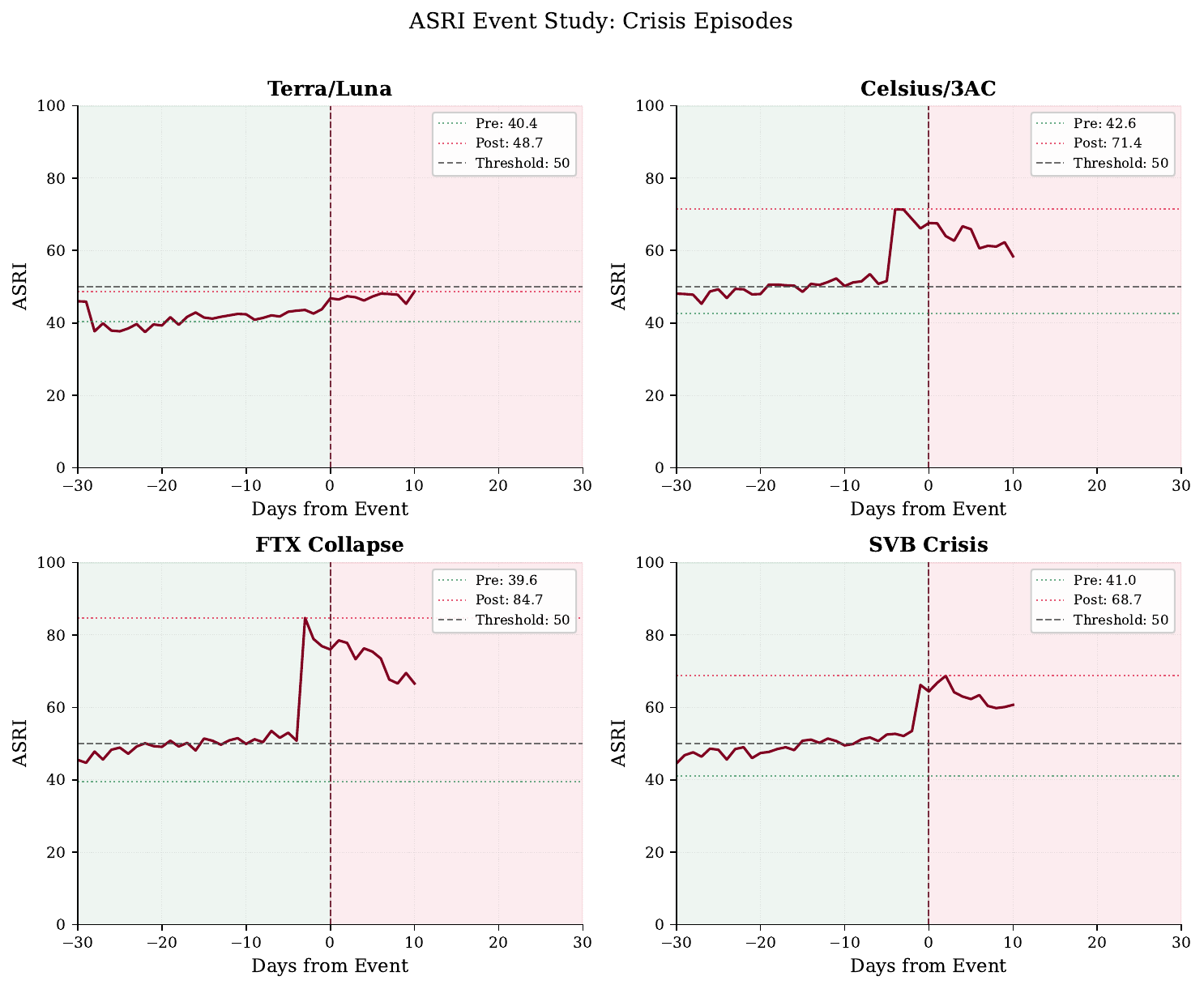}
\caption{ASRI Event Study: Pre-Event and Post-Event Levels Across Four Crisis Events. Each panel shows the baseline ASRI mean during the 60-day estimation window (Pre) and the peak ASRI value during the event window (Post). All four events exhibit substantial elevations from baseline, with FTX Collapse showing the largest absolute increase (39.6 $\rightarrow$ 84.7) and Terra/Luna showing the smallest (40.4 $\rightarrow$ 48.7).}
\label{fig:event_study}
\end{figure}

Figure~\ref{fig:event_study} visualises the ASRI trajectories across the four crisis events, illustrating the pre-event baseline levels and post-event peaks summarised in Table~\ref{tab:event_study}.

\subsubsection{Reconciling Lead Time Definitions}
\label{subsubsec:lead_reconciliation}

\textbf{Reconciling Lead Time Definitions.} Lead time is reported under several distinct operational definitions, and the resulting figures are not directly comparable; we anchor on three:
\begin{itemize}
\item \textbf{Threshold-based (fixed ASRI $\geq 50$, 30-day pre-window).} For the three threshold-detected events the first-crossing leads are Celsius/3AC 19, FTX 22, SVB 15 days (mean $\approx$19); Terra/Luna does not breach 50 in the pre-window (peak 46.0) and so has no fixed-threshold lead. This is the operational alerting definition. These fixed-threshold leads are partly inflated by the running-maximum drawdown offset (Section~\ref{subsec:limitations}): under a responsive 30-day percentage-change drawdown specification they compress to a mean of $\approx$5 days (Celsius/3AC 12, FTX 3, SVB 1), which is the more honest early-warning horizon and the figure we lead with.
\item \textbf{Event-study (first $1.5\sigma$ exceedance, search capped at 30 days).} Table~\ref{tab:event_study} reports 29--30 for all four events; because the signal is elevated across essentially the whole pre-window, these are at the cap and indicate sustained elevation rather than a 30-day horizon.
\item \textbf{Walk-forward (first crossing of the pre-crisis-calibrated threshold).} Table~\ref{tab:walkforward} reports 30/30/28/15 (mean 25.75), the conservative out-of-sample early-warning figure.
\end{itemize}
We report the walk-forward first-crossing lead ($\approx$26 days, 4/4 flagged out-of-sample) as a nominal lead horizon, with the important qualification that for the two early crises this crossing is bought at a 47--59\% alarm rate over the full index history (Section~\ref{subsec:walkforward}); the other definitions are reported for completeness and indicate that the signal is elevated across essentially the entire pre-crisis window for the detected events.

The full precision--recall-by-threshold analysis, the confusion matrix at the operational threshold, and the ROC and precision--recall curves are reported in Appendix~\ref{app:detection_battery}; the headline AUROC ($0.866$) and precision figures also appear in the fair-baseline comparison (Section~\ref{subsec:fair_baselines}, Table~\ref{tab:fair_baselines}).
\subsection{Weight Derivation: Empirical vs. Theoretical}
\label{subsec:weights}

We compare the theoretical weights derived from risk-based principles against empirically-derived weights. An earlier draft reported a separate ``theoretical vs.\ PCA vs.\ Elastic Net'' table whose entries were hard-coded constants that did not reproduce from the weight-derivation modules and contradicted the principal-component loadings reported below; we have removed it. The single, code-faithful comparison across four objective methods is Table~\ref{tab:weight_comparison}, and we summarise its implications here.

\paragraph{What the empirical weights show.}
The reproducible weight-derivation pipeline yields a Principal Component Analysis (PCA) loading vector of $[0.29, 0.29, 0.25, 0.17]$ over $[\text{SCR}, \text{DLR}, \text{CR}, \text{OR}]$, which tracks the theoretical weights closely ($\rho = 0.88$). The Elastic Net, by contrast, concentrates on \emph{Contagion Risk} ($0.45$) and zeroes Regulatory Opacity ($0.00$), with SCR ($0.34$) and DLR ($0.21$) intermediate. We flag this correction explicitly: the Elastic Net does \emph{not} concentrate on DeFi Liquidity Risk---the earlier draft's hard-coded $0.84$ DLR weight does not reproduce from the weighting code, and the code-faithful sparse solution loads on CR. CRITIC and entropy methods diverge further still ($\rho = -0.51$ and $0.27$ with the theoretical vector). Because the four objective methods disagree substantially among themselves, we do not treat any single empirical vector as ``the'' optimal weighting.

\textbf{Elastic Net Specification}: The predictive weights are derived via Elastic Net regression with the following specification:
\begin{itemize}
    \item \textbf{Target variable}: $y_t = \text{ASRI}_{t+30}$ (30-day forward ASRI level)
    \item \textbf{Features}: Current sub-index values $[\text{SCR}_t, \text{DLR}_t, \text{CR}_t, \text{OR}_t]$
    \item \textbf{Cross-validation}: 5-fold blocked time-series CV to preserve temporal structure
    \item \textbf{Hyperparameter grid}: $\alpha \in \{0.1, 0.5, 1.0\}$, $\ell_1$-ratio $\in \{0.1, 0.5, 0.9\}$
    \item \textbf{Software}: scikit-learn 1.3.x, Python 3.11
\end{itemize}
The optimal hyperparameters ($\alpha = 0.5$, $\ell_1$-ratio $= 0.5$) produce a sparse solution that zeroes Regulatory Opacity and concentrates on Contagion Risk. This sparsity reflects moderate correlation among sub-indices during stress periods: when systemic stress materialises, multiple channels activate simultaneously, making it difficult for regularised regression to distinguish their individual contributions. Importantly, formal collinearity diagnostics (Table~\ref{tab:collinearity_diagnostics}) confirm that this correlation does not constitute problematic multicollinearity---all variance inflation factors remain below 5 (max VIF = 3.89), and the condition number of 19.1 indicates weak collinearity well within acceptable bounds.

\paragraph{The Decomposition Is Not Individually Load-Bearing.}
We deliberately avoid building a strong mechanistic narrative on these empirical weights, because they are neither stable across methods nor individually decisive for detection. The ablation analysis (Appendix~\ref{subsec:ablation}) shows that threshold-based detection is \emph{invariant} at 3/4 across all single-channel removals: no individual sub-index is necessary for the detected events. The four-channel decomposition therefore earns its place through \emph{interpretability}---attributing aggregate stress to named, economically distinct channels---rather than through any one channel being statistically load-bearing. Lead-time variation under ablation is not a reliable guide to channel-level timing---with only four co-located crisis episodes a correctly specified discrete-outcome (logit/hazard) test does not identify which channels move earliest (Appendix~\ref{subsubsec:granger})---so we draw no leading-versus-confirming hierarchy, and we no longer attribute a ``canary'' role to DLR (the empirical weight that motivated that claim does not reproduce).

\paragraph{Rationale for Theoretical Weights.}
We retain theoretical weights for the headline specification. Three considerations motivate this choice:
\begin{enumerate}
    \item \textbf{Component-level monitoring}: The theoretical decomposition enables targeted interpretation. Elevated SCR points to stablecoin reserve scrutiny; elevated CR to counterparty/macro exposure review. Pure prediction weights sacrifice this interpretability.
    \item \textbf{Method disagreement}: The four objective weighting methods disagree among themselves (correlations with the theoretical vector range from $-0.51$ to $0.88$), so no single data-driven vector is a defensible replacement; the balanced theoretical weights avoid collapsing onto whichever channel a particular estimator happens to favour in this small sample.
    \item \textbf{Structural stability}: Prediction-optimised weights are sample-dependent and may overfit to the four historical crises. Theoretical weights provide forward-looking stability for regimes that differ from the backtest period.
\end{enumerate}

The empirical analysis thus informs interpretation rather than replacing theoretical structure, while the ablation invariance cautions that the channel split is a presentational device, not a load-bearing causal decomposition. This is consistent with the weight assignment rationale in Section~\ref{subsec:weight_justification}.

The four-method objective weight comparison (Table~\ref{tab:weight_comparison}), the collinearity and principal-component diagnostics (Appendix~\ref{subsubsec:collinearity}), and the Granger-causality analysis (Appendix~\ref{subsubsec:granger}) are reported in Appendix~\ref{app:weight_diag}.
\subsection{Regime Detection}

We estimate a Gaussian Hidden Markov Model (HMM) \citep{hamilton1989regime} to characterise market regimes from sub-index dynamics, treating the regime labels as interpretive structure over the index's autocorrelated dynamics rather than as evidence of sharply separated generating processes. The HMM is specified with full covariance matrices for each state, estimated via Expectation-Maximisation with convergence criterion $|\Delta \log L| < 10^{-4}$ and maximum 1,000 iterations. To mitigate sensitivity to initialisation, we run 10 random restarts and select the model with highest log-likelihood. Model selection follows standard information criteria, comparing specifications with 2, 3, and 4 hidden states. We flag one assumption violation at the outset: the Gaussian emissions impose time-invariant per-regime means, yet the Contagion sub-index is non-stationary with an upward drift ($\approx$4.4 ASRI points per year) on this sample (Table~\ref{tab:stationarity}). The fitted regimes therefore partly track that trend rather than purely systemic conditions, a point we quantify in the ergodic-distribution discussion below; we retain the level-based fit for interpretability and read the regimes accordingly. The appropriate robustness check---re-estimating the model on trend-stationary (differenced or detrended) inputs, or specifying autoregressive/time-varying-mean emissions that absorb the drift, so that the states reflect conditional dynamics rather than the level trend---would test whether the Crisis state survives once the trend is removed. We flag this as future work rather than run it here: with only four crisis episodes the regime labels cannot be validated against an independent event population regardless of the emission specification, so we do not present the three-regime fit as evidence of statistically distinct generating processes.

We summarise the retained three-state specification here; the HMM model-selection criteria (Table~\ref{tab:hmm_selection}), the full transition matrix (Table~\ref{tab:transition_matrix}), ergodic diagnostics (Table~\ref{tab:hmm_diagnostics}), regime-count robustness (Table~\ref{tab:regime_k_robustness}), and the filtering-versus-smoothing discussion are deferred to Appendix~\ref{app:regime_full}. Table~\ref{tab:regimes} reports the characteristics of the three retained regimes.

\begin{table}[H]
\begin{threeparttable}
\centering
\caption{Regime Characteristics (3-State Model)}
\label{tab:regimes}
\small
\begin{tabular}{@{}c*{3}{r}l@{}}
\toprule
Regime & Frequency & Mean Risk & Persistence & Interpretation \\
\midrule
1 & 19.8\% & 31.8 & 0.997 & Low Risk \\
2 & 56.3\% & 36.8 & 0.992 & Moderate \\
3 & 23.8\% & 48.2 & 0.980 & Crisis \\
\bottomrule
\end{tabular}
\begin{tablenotes}
\small
\item Persistence = probability of remaining in same regime (transition matrix diagonal).
\item Values from the best-by-log-likelihood fit among 10 random initialisations.
\end{tablenotes}
\end{threeparttable}
\end{table}

\paragraph{Interpreting Regime Labels.}
We label Regime 3 the ``Crisis'' regime because it carries the highest regime-conditional mean (48.2) and concentrates the historical stress episodes, even though that conditional mean still falls just within the upper Moderate alert band (30--50) rather than the High band ($\geq$70). The ``Crisis'' label reflects the statistical properties of the regime---highest volatility, elevated transition probability into and out of acute stress---rather than a claim that the instantaneous ASRI level remains at crisis levels throughout.

During the historical crises themselves, ASRI spiked into High ($\geq$70) zones, with event peaks reaching as high as 84.7 (FTX) and 71.4 (Celsius/3AC) above the 70 threshold (Table~\ref{tab:event_study}). These peaks occur as transient spikes within the Crisis regime before mean-reversion during recovery periods pulls the regime mean back towards the Moderate band. The regime mean of 48.2 thus represents a weighted average of stress spikes and subsequent recoveries, not a sustained crisis level.

Alert thresholds are designed to flag instantaneous risk levels requiring attention, while regime classifications capture the statistical dynamics that characterise market states over extended periods. The Crisis regime signals a market environment where acute stress events are significantly more likely to occur, even when the current ASRI reading may be temporarily moderate.

The three-regime model identifies:
\begin{itemize}
    \item \textbf{Low Risk} (19.8\% of sample): Mean ASRI of 31.8, very high persistence (0.997)
    \item \textbf{Moderate} (56.3\%): Mean ASRI of 36.8, high persistence (0.992)
    \item \textbf{Crisis} (23.8\%): Mean ASRI of 48.2, high persistence (0.980)
\end{itemize}

The high persistence across all regimes (diagonal elements $>$ 0.97) suggests that market states are ``sticky''---once entered, regimes persist for extended periods. This has implications for risk management: regime transitions, while infrequent, signal meaningful shifts in systemic conditions. We caution, however, against over-reading the regime structure: the three regime means span only 31.8--48.2 on the 0--100 scale, and the near-unit persistences largely restate the index's own serial correlation. We therefore present the HMM as descriptive, interpretive colour over an autocorrelated trend, and do not claim that it recovers generating states that a trivial three-bin partition of the ASRI level would miss.

\subsection{Fair-Baseline Comparison}
\label{subsec:fair_baselines}

Benchmarking ASRI against a Diebold--Yilmaz (2012) connectedness series computed on its four sub-indices yields higher day-level discrimination for ASRI (AUROC $0.866$ vs.\ $0.670$; AUPRC $0.298$ vs.\ $0.121$; Table~\ref{tab:roc_metrics}). That benchmark is, however, \emph{constructed from} ASRI's own sub-indices---a partly circular comparator---and is itself out-discriminated by an off-the-shelf sentiment index, so it cannot establish that the four-channel aggregation adds value; the load-bearing comparison is the fair-baseline battery below. Full Diebold--Yilmaz methodology, lag selection, window sensitivity, and per-event results are reported in Appendix~\ref{subsec:dy_comparison}.

\begin{table}[H]
\begin{threeparttable}
\centering
\caption{Crisis Prediction Classification Metrics with 95\% Bootstrap Confidence Intervals}
\label{tab:roc_metrics}
\small
\begin{tabular}{lccc}
\toprule
Metric & ASRI & D-Y Connectedness & Paired Difference \\
\midrule
AUROC & 0.866 [0.756, 0.949] & 0.670 [0.526, 0.806] & +0.194 [+0.093, +0.298]*** \\
AUPRC & 0.298 [0.131, 0.518] & 0.121 [0.046, 0.233] & +0.182 [+0.070, +0.337]*** \\
\midrule
Optimal Threshold & 45.3 & 44.0\% & --- \\
Precision @ Optimal & 0.352 & 0.149 & +0.203 \\
Recall @ Optimal & 0.758 & 0.806 & $-$0.048 \\
F1 @ Optimal & 0.481 & 0.251 & +0.230 \\
\bottomrule
\end{tabular}
\begin{tablenotes}
\small
\item Computed on the real ASRI series and a genuine rolling Diebold-Yilmaz generalized-FEVD connectedness series (60-day VAR(1), $H = 10$), both scored on the same crisis labels. $n = 1{,}402$ aligned daily observations; 124 crisis-imminent days in four contiguous runs, 1{,}278 non-crisis days (prevalence 8.8\%).
\item Confidence intervals from a \textbf{moving-block bootstrap} (Kunsch 1989; block length $L = 25$ days, $B = 2{,}000$, seed 42), which respects the serial dependence that i.i.d.\ day-level resampling ignores. The resulting marginal intervals are $3.3$--$4.7\times$ wider than naive i.i.d.\ intervals (confirming the latter were $\approx 4\times$ too narrow) and the ASRI and D-Y marginal intervals \emph{overlap}, because the labels comprise only $\approx 4$ independent crisis blocks. ASRI's advantage is therefore established by the \emph{paired} block-bootstrap difference (same resampled blocks applied to both series): the AUROC and AUPRC differences are positive in all $2{,}000$ resamples ($p < 0.001$), robust across $L = 20/25/30$.
\item Crisis defined as the 30-day forward period preceding any of the four historical crisis onsets (Terra/Luna, Celsius/3AC, FTX, SVB). All metrics derive from these four events; see Section~\ref{subsec:limitations}.
\item *** paired difference $> 0$ in all 2{,}000 resamples ($p < 0.001$). Optimal threshold selected by Youden's J statistic (maximizes TPR $-$ FPR). D-Y threshold expressed as total connectedness (\%).
\item Precision, recall and F1 at the optimal threshold are computed on this 1{,}402-day common sample, where the ASRI Youden threshold is 45.3. On the full canonical sample (1{,}841 days; 120 crisis / 1{,}721 non-crisis days---the Table~\ref{tab:precision_recall}/Table~\ref{tab:confusion_matrix} denominators) ASRI's precision at the same threshold 45.3 is 0.325, and at that sample's own Youden-optimal threshold (41.2) it falls to 0.198, because the larger non-crisis denominator admits more false-positive days. The AUROC, AUPRC and the paired block-bootstrap are unaffected, since both indices are scored on identical days. ASRI's precision advantage over D-Y is therefore a relative ordering, not high absolute precision.
\end{tablenotes}
\end{threeparttable}
\end{table}

A connectedness benchmark alone cannot establish that the four-channel \emph{aggregation} is what produces ASRI's discrimination. We therefore evaluate, on \emph{identical} labels and protocol---the same 1{,}402 aligned daily observations, the same 30-day forward crisis windows, and the same trapezoidal AUROC as Table~\ref{tab:roc_metrics}---a battery of simpler baselines: each of the four sub-indices on its own, the first principal component (PC1) of the four sub-indices, and an off-the-shelf Crypto Fear \& Greed sentiment index \citep{alternative2023fear}. Table~\ref{tab:fair_baselines} reports the result.

\begin{table}[H]
\begin{threeparttable}
\centering
\caption{Fair-Baseline Comparison: Day-Level Crisis Discrimination on Identical Labels}
\label{tab:fair_baselines}
\small
\begin{tabular}{@{}lcc@{}}
\toprule
Score (same labels, same protocol) & AUROC [95\% CI] & AUPRC \\
\midrule
ASRI composite (4-channel) & 0.866 [0.842, 0.888] & 0.298 \\
PC1 of the four sub-indices & 0.858 [0.832, 0.881] & 0.289 \\
Contagion Risk channel \emph{(best single)} & 0.851 [0.824, 0.875] & 0.285 \\
Stablecoin Concentration Risk channel & 0.825 [0.793, 0.854] & 0.255 \\
DeFi Liquidity Risk channel & 0.821 [0.787, 0.852] & 0.260 \\
Crypto Fear \& Greed (off-the-shelf) & 0.789 [0.752, 0.825] & 0.269 \\
Diebold--Yilmaz connectedness & 0.670 [0.636, 0.708] & 0.121 \\
Regulatory/Arbitrage Opacity channel & 0.650 [0.611, 0.687] & 0.141 \\
\midrule
\multicolumn{3}{@{}l}{\textit{Off-the-shelf baselines needing no crypto-native construction, and detrending checks:}} \\
Contagion Risk, linear-detrended & 0.912 & 0.369 \\
VIX equity-volatility stress (standalone) & 0.875 & 0.354 \\
VIX $+$ 10Y Treasury composite (Bank$_t$ macro core) & 0.773 & 0.263 \\
10Y Treasury level alone & 0.501 & 0.080 \\
Rolling 30-day BTC/ETH return (EW) & 0.681 & 0.177 \\
\bottomrule
\end{tabular}
\begin{tablenotes}
\small
\item All scores evaluated on the same $n = 1{,}402$ aligned daily observations, the same four-event 30-day forward crisis labels (124 crisis-imminent days, prevalence 8.8\%), and the same trapezoidal AUROC as Table~\ref{tab:roc_metrics}. The sub-indices are the finest single features in the released daily data; raw underlying inputs (e.g.\ raw stablecoin-TVL drawdown) are not released separately.
\item PC1 is the leading principal component of the four sub-indices (covariance form; 74.9\% of variance), and in this form its loading is dominated by the high-variance Contagion channel ($\approx 0.81$, versus $0.50$/$0.31$/$0.06$ on Stablecoin/DeFi-Liquidity/Opacity)---so covariance-PC1 is close to a rescaled Contagion channel, which is why it tracks the best single channel so nearly. The off-the-shelf Crypto Fear \& Greed index \citep{alternative2023fear} is fetched live and aligned to the common index (coverage 1{,}401/1{,}402 days).
\item The five rows below the second rule are added off-the-shelf baselines and detrending checks (point AUROC/AUPRC only; CIs of comparable width are omitted for space). The VIX standalone series ties ASRI (paired $\Delta = -0.009$, $p = 0.58$); the linear-detrended Contagion channel still exceeds the four-channel composite. ``VIX $+$ 10Y composite'' is the Bank$_t$ macro core of the Contagion channel ($0.6\,\text{norm}(r_{10Y},2,6)+0.4\,\text{norm}(\text{VIX},12,40)$). The 10Y Treasury \emph{level} discriminates at chance on these labels (0.501), so adding it at $60\%$ weight drags the composite below VIX alone (0.773 vs.\ 0.875, a $0.102$ reduction). This reflects the secular 2021--2024 tightening trend ($\approx$0.9 percentage points per year) that the level specification imports rather than an absence of banking-stress information: the reproducible detrended-Contagion channel (0.912) shows that removing such a trend recovers discrimination. A change- or detrended-based Treasury input is the preferable specification.
\item Confidence intervals are i.i.d.-day percentile bootstraps and \emph{understate} sampling uncertainty; with only four independent crisis blocks the honest moving-block interval for ASRI alone spans $[0.756, 0.949]$ (Table~\ref{tab:roc_metrics}), which contains every baseline in this table. On these \emph{marginal} block-bootstrap intervals the $\approx$0.008--0.015 AUROC gaps among ASRI, PC1, and the Contagion channel are not separable. A \emph{paired} block bootstrap (same resampled blocks applied to both series---the test used for the ASRI-versus-D-Y comparison) does return small, consistently-signed AUROC edges whose CIs exclude zero (ASRI$-$PC1 $+0.008$, 95\% CI $[+0.001,+0.018]$; ASRI$-$Contagion $+0.016$, 95\% CI $[+0.004,+0.031]$; across block lengths $L=20/25/30$), but the corresponding paired \emph{AUPRC} differences include zero and---exactly as for the D-Y comparison---this positive-in-every-resample sign reflects ASRI \emph{nesting} both baselines (it contains the Contagion channel and is near-collinear with covariance-PC1) rather than a powered discriminative gain over four independent events (Section~\ref{subsec:fair_baselines}). The VIX-ties-ASRI null and the detrended-Contagion margin, being a null and a large-margin result, are robust to the wider block-bootstrap intervals. All scores are in-sample discrimination, not out-of-sample skill.
\end{tablenotes}
\end{threeparttable}
\end{table}

\textbf{Reading.} Four facts follow, and together they reshape the empirical claim.

First, \emph{aggregation buys no meaningful discrimination}. ASRI's AUROC (0.866) exceeds the first principal component of its own sub-indices (PC1, 0.858) by $\approx +0.008$ and the single strongest sub-index (Contagion Risk, 0.851) by $\approx +0.015$. \emph{Marginally}, both PC1 and the Contagion channel fall \emph{inside} ASRI's bootstrap confidence interval (i.i.d.-day $[0.842, 0.888]$; the honest moving-block interval is far wider, $[0.756, 0.949]$, Table~\ref{tab:roc_metrics}), so on overlapping intervals the three are not separable. The proper contrast, however, is the \emph{paired} moving-block bootstrap used for the D-Y comparison---both series scored on the same resampled blocks, so the strong day-to-day autocorrelation cancels in the difference. It returns a small but consistently-signed AUROC edge: ASRI$-$PC1 $= +0.008$ (95\% CI $[+0.001, +0.018]$, $p \approx 0.02$) and ASRI$-$Contagion $= +0.016$ (95\% CI $[+0.004, +0.031]$, $p \approx 0.003$), both excluding zero across block lengths $L = 20/25/30$, while the corresponding paired AUPRC differences include zero. We do not read this as a discriminative gain. Exactly as for the D-Y paired result (Appendix~\ref{subsec:dy_comparison}), with only four independent crisis episodes (Section~\ref{subsec:limitations}) a positive-in-every-resample sign reflects a \emph{structural} relationship between nested constructions---ASRI is a weighted aggregate that \emph{contains} the Contagion channel and is near-collinear with covariance-form PC1, which itself loads $\approx 0.81$ on Contagion---rather than a powered demonstration that aggregation separates crises better. The edge is $\leq 0.016$ AUROC, operationally negligible and absent in AUPRC; ASRI is therefore indistinguishable from its best single channel and from PC1 on the marginal intervals, and out-aggregates them only by a small structural margin that does not constitute a meaningful gain in crisis--non-crisis separation.

Second, \emph{the strongest single channel for discrimination is Contagion Risk, not Stablecoin Risk}. Although Stablecoin Concentration Risk carries the highest aggregation weight (30\%) and loads heavily in some weight-derivation specifications, on these labels it discriminates (0.825) below both the Contagion channel (0.851) and PC1. The ``dominance'' of the stablecoin channel is a statement about loading weight, not about predictive power; the most discriminative single channel is Contagion Risk.

Third, \emph{the surviving ``beats-benchmark'' result is against a circular and unusually weak comparator}. The only baseline ASRI clearly out-discriminates is Diebold--Yilmaz connectedness (0.670)---which is itself \emph{constructed from} ASRI's own four sub-indices, so the comparison is partly circular---and even there the benchmark is so weak that a free, off-the-shelf Fear \& Greed sentiment index beats it by 0.12 (0.789 vs.\ 0.670). Out-discriminating D-Y is therefore not evidence that ASRI captures something a simpler construction misses.

Fourth, \emph{a standalone equity-volatility series alone matches ASRI, and detrending does not rescue the composite}. Of the two baselines that need no crypto-native construction at all, a plain 30-day crypto return discriminates well below ASRI (0.681 vs.\ 0.866)---a modest point in the composite's favour, since its signal is more than raw price decline. A standalone VIX level, by contrast, reaches 0.875 and is statistically indistinguishable from ASRI (paired $\Delta = -0.009$, $p = 0.58$): to first order, labelling these four crises is a macro-volatility detection problem, and ASRI's discriminative value over an off-the-shelf equity-volatility series is not established. The aggregation demotion is not a trend artefact either---linear-detrending the Contagion channel raises it to 0.912, still above the four-channel composite, so removing a common trend leaves no distinctive contribution from aggregation.

We accordingly do not rest ASRI's contribution on discriminative superiority. The value of the four-channel composite is interpretive---it attributes stress to nameable transmission channels, carries operational lead-time for the detected crises, and exhibits coherent regime structure---rather than a measurable gain in crisis--non-crisis separation over its own best single channel or PC1.

\subsection{Walk-Forward Validation}
\label{subsec:walkforward}

A distinct concern from publication lags is look-ahead bias in weight calibration: the theoretical weights were specified using domain knowledge accumulated from observing the full 2021--2024 sample, including the crises used for validation. To address this concern, we conduct walk-forward validation with expanding training windows that simulate the information set available prior to each crisis.

Calibrating the alert threshold on \emph{only} pre-crisis data (the 90th percentile of each event's training-period history) flags all four episodes out-of-sample, with a mean first-crossing lead of $\approx$26 days. For the two early crises (Terra/Luna and Celsius/3AC), however, this 4/4 rate is bought at a $47$--$59\%$ alarm rate over the full index history at $\approx$11--13\% precision, so we read it as evidence against look-ahead bias rather than as operational early-warning skill. Continuous level-prediction is below a naive flat-mean baseline (walk-forward $R^2 \approx -0.6$; held-out $R^2 \approx -0.1$; 0/4 sub-indices robust to $\pm 15\%$ weight perturbation), confirming that ASRI is a threshold monitor rather than a point forecaster. The full walk-forward methodology, the per-event detection table (Table~\ref{tab:walkforward}), and the per-event interpretation are given in Appendix~\ref{app:walkforward_full}.

\subsection{Out-of-Sample Specificity Test}\label{subsec:specificity}

A risk index must demonstrate not only sensitivity (detecting true crises) but also specificity (avoiding false alarms). We evaluate ASRI's specificity using 2024--2025 data---a period entirely out-of-sample relative to the framework's initial design and the crisis events used for validation.

Across 2024--2025---roughly 34 months out-of-sample---ASRI produced no sustained false alarms: a single multi-week elevation in July--August 2024 tracked the yen carry-trade de-risking shock (peak $58.8$) without escalating, and the index otherwise stayed below the Elevated threshold (2025 peak $48.0$). It correctly classified the February 2025 Bybit hack (\$1.5B, the largest exchange theft in history) as non-systemic, because the theft produced no stablecoin depegs, protocol liquidations, or counterparty contagion---evidence that ASRI captures transmission mechanisms rather than event magnitude. The 2024 stability analysis, the Bybit component readings (Table~\ref{tab:bybit}), and the systemic-versus-contained comparison are given in Appendix~\ref{app:specificity_full}.

\subsection{Validation Summary}

The retrospective evaluation supports the ASRI framework on multiple dimensions:

\begin{enumerate}
    \item \textbf{Retrospective Discrimination}: Three of four historical crises are detected via threshold-based monitoring (Celsius/3AC, FTX, SVB); the complementary event-study test of an elevated cumulative abnormal signal is, however, inconclusive---a placebo test on non-crisis dates shows the finite-sample (fixed-$b$) reference has essentially no crisis-specificity on this smooth trending series (roughly sixty percent of arbitrary dates clear the nominal 5\% threshold, and the onsets sit at or below the placebo median), so we do not read the raw HAC $t$-statistics ($1.7$--$3.9$) as evidence of crisis detection. The Terra/Luna threshold miss reflects documented limitations of market-based indicators for algorithmic stablecoin risk (Section~\ref{subsec:limitations}).
    \item \textbf{Lead Behaviour}: The in-sample 1.5$\sigma$ signal is elevated across the full (30-day-capped) pre-crisis window for detected events; walk-forward thresholds calibrated only on pre-crisis data flag all four episodes ($\approx$26-day mean first-crossing lead), but at a high false-positive cost for the early crises (Terra/Luna and Celsius/3AC alarm on 47--59\% of index history), so this is evidence against look-ahead bias rather than a clean prediction record (Section~\ref{subsec:walkforward})
    \item \textbf{Statistical Validity}: ASRI and the DeFi Liquidity sub-index are stationary; Stablecoin and Opacity Risk are trend-stationary; Contagion Risk is non-stationary on the released sample (flagged caveat)
    \item \textbf{Structural Stability}: the reproducible full-sample Chow test is a non-rejection of parameter consistency at the sample midpoint ($p = 0.78$)
    \item \textbf{Regime Identification}: Three-regime HMM provides interpretable market state classification
    \item \textbf{Component Importance}: Ablation analysis demonstrates stable 3/4 detection across all single-channel removals; the four-channel decomposition is retained for interpretability rather than as an individually load-bearing structure (Appendix~\ref{subsec:ablation})
    \item \textbf{Benchmark and Fair-Baseline Comparison}: On a common day-level sample ASRI out-discriminates Diebold-Yilmaz connectedness (AUROC 0.87 vs. 0.67; precision at the Youden-optimal threshold 35\% vs.\ 15\% on the common sample, a modest 20--32\% in absolute terms on the full canonical sample), but D-Y is built from ASRI's own sub-indices and is itself out-discriminated by an off-the-shelf Fear \& Greed index (0.789 vs. 0.670); a fair-baseline comparison on identical labels (Section~\ref{subsec:fair_baselines}) shows ASRI is statistically indistinguishable from its single strongest sub-index (Contagion Risk, 0.851) and from PC1 (0.858), with aggregation adding only $\approx +0.008$--$0.015$ AUROC, so the value of aggregation is interpretive rather than discriminative (Appendix~\ref{subsec:dy_comparison})
    \item \textbf{Robustness to Publication Lags}: Pseudo-real-time evaluation with publication lags confirms detection performance is not an artefact of look-ahead bias (Appendix~\ref{subsec:realtime})
    \item \textbf{Out-of-Sample Specificity}: No sustained false alarms during 2024--2025---a single non-systemic elevation in August 2024 (macro de-risking shock) and otherwise sub-threshold readings, including correct identification of the Bybit hack (\$1.5B) as non-systemic due to absence of contagion channels (Section~\ref{subsec:specificity})
    \item \textbf{Forecasting vs.\ Monitoring}: Continuous level-prediction metrics are below baseline (walk-forward $R^2 \approx -0.6$; held-out $R^2 \approx -0.1$; 0/4 robust components under weight perturbation), confirming that ASRI is a threshold monitor, not a point forecaster. Binary detection and threshold-based metrics are the appropriate evaluation framework (Section~\ref{subsec:walkforward})
\end{enumerate}

\paragraph{Data Quality Caveat.} A component-level accounting of the backtested series (Table~\ref{tab:series_composition}) shows that approximately \textbf{43\%} of its average level (mean composition) is carried by static default values---$\text{Unreg}_t$ (fixed at 35.0), $\text{Sent}_t$ (50.0), $\text{Corr}_t$ (0.5), audit-coverage levels, and the stablecoin peg input, which is held at par (price $= 1.0$) throughout the historical backtest (\texttt{backtest.py:191})---with roughly 37\% from dynamic crypto-native inputs and 20\% from the macro (Treasury/VIX/yield-curve) proxy. Because these defaults are constant, they contribute essentially none of the series' time-variation; in particular, because the peg input is held at par, the stablecoin peg-deviation volatility term contributes no time-variation in the backtest, and the dynamic liquidity/stablecoin signal is carried by TVL drawdown and TVL volatility rather than by realised de-pegging. Crisis detection is robust to full-range variation of the placeholder parameters (Appendix~\ref{subsec:sensitivity}), but absolute ASRI levels therefore carry substantially more model dependence than the binary detection results suggest. The relative pattern---elevated readings before crises, moderate readings during calm periods---is more reliable than specific ASRI magnitudes. A related construction limitation concerns the TVL-drawdown input (Equation~\ref{eq:stablecoin}), which is normalised against a running (expanding) maximum. Because total DeFi TVL never reclaimed its late-2021 peak after mid-2022, this term is pinned at its ceiling for $\approx$85\% of 2022--2024 and adds a near-constant $\approx$12-point offset to ASRI that, under a \emph{fixed} detection threshold, mechanically lifts crossings. We re-specified the drawdown two ways on the same data. A 365-day \emph{rolling} maximum leaves the 2022 episodes essentially unchanged (the late-2021 peak remains the trailing-window maximum and the drawdown still exceeds the $50\%$ clip), so it does not address the issue for the crises of record. A responsive 30-day percentage-change specification removes the offset and preserves $3/4$ fixed-threshold ($\geq 50$) detection (Celsius/3AC, FTX, SVB; Terra/Luna still missed), but compresses the average first-crossing lead from $\approx$19 days to $\approx$5 days (Celsius/3AC 12, FTX 3, SVB 1). The reported $\approx$19-day operational leads should therefore be read as partly an artefact of the running-maximum offset rather than as a genuine early-warning horizon.


\section{Discussion}

\subsection{Theoretical Implications}

The ASRI framework makes three contributions to the systemic risk literature:

First, it adapts the channel-decomposition perspective of the connectedness and contagion-measurement literature \citep{battiston2012debtrank} to the permissionless composability characteristic of DeFi, situating cryptocurrency systemic risk within the broader complexity economics research programme that applies agent-based, network, and dynamical systems approaches to 21st-century economic challenges \citep{bednar2025jebo}. We stress that this is an adaptation of measurement perspective, not a theoretical extension of network contagion models: the linear composite treats the system as a portfolio of separable risks rather than as a coupled dynamical network (see ``On the Choice of Linear Aggregation'' below). Traditional network models assume bilateral counterparty relationships with observable exposures; ASRI is designed with the multi-lateral, code-embedded exposures of smart-contract composability in view, though the current backtest proxies rather than directly measures these composability exposures (the inter-channel correlation input is held at a constant default).

Second, it \emph{proxies} the DeFi-TradFi transmission channel: the Contagion sub-index is built from TradFi-stress indicators (Treasury/VIX/yield-curve) rather than from a directly measured DeFi-TradFi exposure. As stablecoins have accumulated significant Treasury positions, they create a potential link between Federal Reserve monetary policy and DeFi liquidity conditions that existing crypto risk measures do not track---though ASRI captures this link through a macro proxy, not a realised interconnection measurement.

Third, it operationalises regulatory opacity as a quantifiable risk factor. While traditional finance assumes regulatory disclosure requirements, crypto markets feature substantial opacity about custody arrangements, reserve composition, and counterparty relationships. The Opacity sub-index provides a systematic framework for incorporating this uncertainty into risk assessment.

\paragraph{On the Choice of Linear Aggregation.} A complexity-minded reader will rightly note the tension between ASRI's additive, linear aggregation and the systemic-risk phenomena it targets, which are paradigmatically \emph{non-linear}: leverage spirals, redemption runs, and composability cascades involve positive feedback, threshold effects, and abrupt phase-transition-like reorganisations rather than the proportional, separable response a weighted sum encodes \citep{battiston2012debtrank, bednar2025jebo}. We make the trade-off explicit. Linear aggregation is chosen for \emph{interpretability and estimability}, not because we believe systemic stress accumulates additively: the weighted-sum form is the standard backbone of operational composite stress indicators precisely because it yields an exact, efficiency-consistent channel attribution (Axiom~\ref{ax:decomp}) and requires no parameters beyond the four weights---a decisive advantage when, as here, only four crisis episodes are available to fit anything. The cost is realism. A linear index cannot represent the super-additive amplification that arises when several channels are stressed at once, nor the self-reinforcing feedback by which one channel's deterioration accelerates another's; it treats the system as a portfolio of separable risks rather than as a coupled dynamical network. Richer aggregators address parts of this gap: the CISS framework folds cross-channel correlation into a portfolio-theoretic combination so that joint stress is penalised more than the sum of its parts \citep{hollo2012ciss}; network and graph-regularised constructions embed topology directly \citep{briola2026grpca, battiston2012debtrank}; and the CES and geometric aggregators we compare in Section~\ref{subsec:aggregation} can encode complementarity and tail amplification through their substitution parameter. Each, however, introduces dependence parameters that our four-event sample cannot identify reliably (Section~\ref{subsec:limitations}). Our position is therefore deliberately modest: ASRI is an interpretable first-order \emph{monitoring} decomposition, not a model of the non-linear crisis dynamics themselves, and the move to amplification-aware aggregation is the principal methodological extension we flag for when the crisis sample is large enough to estimate it. The empirical finding that aggregation buys little discriminative power over the single strongest channel on this sample (Section~\ref{subsec:fair_baselines}) is consistent with this reading: the value of the linear composite here is attribution and auditability, not a claim to have captured the system's non-linear structure.

\subsection{Practical Applications}

We frame the following as illustrative use cases, conditional on the binding four-event limitation (Section~\ref{subsec:limitations}), rather than as validated deployments.

\textbf{Portfolio Risk Management}: ASRI could in principle inform institutional crypto allocators as one input to a risk overlay---trimming exposure when systemic readings are elevated---though we stress this is illustrative, given that all results rest on only four crisis events.

\textbf{Regulatory Monitoring}: ASRI is intended as a candidate input for macroprudential surveillance of DeFi-TradFi interconnection dynamics, not as a validated supervisory tool.

\textbf{Protocol Governance}: DeFi protocols could in principle use sub-index components to gauge their contribution to systemic risk and inform parameter choices (e.g., collateral ratios, withdrawal limits).

\textbf{Research Applications}: The released ASRI series and code provide a reproducible, openly documented reference implementation for further study of crypto market dynamics.

\subsection{Limitations}
\label{subsec:limitations}

\textbf{Four Crisis Events (Binding Power Limitation)}: The single most important caveat on every result in this paper is that the validation set contains only \emph{four} systemic crisis events (Terra/Luna, Celsius/3AC, FTX, SVB). All detection rates, lead times, precision and recall figures, the regime-count robustness comparison, and the ASRI-versus-Diebold-Yilmaz head-to-head ultimately rest on these four events. The day-level classification analysis (1{,}402 daily observations) and the bootstrap confidence intervals partially mitigate this by exploiting the autocorrelation structure within each crisis window, but the underlying \emph{independent} information content is four events: the day-level labels are generated by, and are not independent of, those four onsets. Consequently, statements such as ``ASRI attains AUROC 0.87 versus 0.67'' should be read as indicative of relative performance on this small, historically specific sample rather than as estimates with the statistical power one would obtain from a large, independent event population. With four events, detection-rate comparisons (e.g.\ 3/4 versus 4/4) cannot discriminate reliably among model specifications, and we deliberately avoid over-interpreting them. This is the binding limitation for any inferential venue and the principal reason we frame ASRI as an interpretable monitoring framework rather than a statistically validated forecaster. Expanding the event set as additional crises accrue is the highest-priority direction for future validation.

\textbf{Terra/Luna Detection Failure}: ASRI's most significant \emph{event-specific} empirical limitation is its failure to detect the Terra/Luna collapse (May 2022) using threshold-based early warning. The index peaked at 46.0 in the 30-day pre-crisis window---below the ``Elevated'' threshold of 50---despite this event representing the largest nominal loss (\$40B+) in cryptocurrency history. (The peak of 48.7 reported in Table~\ref{tab:event_study} is taken over the wider $[-30,+10]$ event window, which includes the immediate post-onset days; the 30-day \emph{pre}-crisis peak is 46.0.) The event study methodology (Section~\ref{sec:event_study}) does register an elevated raw cumulative abnormal signal for Terra/Luna (CAS = 100.3, naive $t = 5.47$; Table~\ref{tab:event_study}) because it measures deviation from the estimation baseline rather than absolute threshold breaches; but this signal does not survive autocorrelation-robust inference (HAC $t = 1.72$, non-significant), consistent with our treatment of the event study as inconclusive throughout. Terra/Luna is thus missed both by the operational threshold rule and by the robust event study. However, for operational early warning, ASRI failed to provide actionable alerts before the Terra/Luna crisis.

The failure reflects a fundamental limitation of retrospective indicators: algorithmic stablecoin risk prior to UST's depeg was difficult to observe through market prices alone. UST maintained its peg until the death spiral began, and Luna's price appreciation masked the underlying fragility. The sub-indices capture \textit{revealed} stress through observable metrics (TVL drawdowns, yield compression, realised volatility), but Terra/Luna's collapse was precipitated by endogenous reflexivity between UST redemptions and Luna minting rather than exogenous liquidity or contagion shocks that ASRI's components measure.

\textit{Remediation}: Section~\ref{subsubsec:algo_stable} specifies an algorithmic stablecoin risk extension that incorporates backing ratio dynamics, collateral volatility, and supply dilution---metrics \emph{designed} to flag Luna's instability prior to UST's depeg. The following is an illustrative back-of-envelope calculation, not a validated result: a sensitivity calculation suggests this extension might have elevated SCR by an indicative 8--12 points in April 2022, potentially bringing ASRI above the detection threshold. We have not been able to confirm this---full historical validation is deferred to an on-chain reconstruction of UST/Luna supply and backing metrics from Terra Classic archive nodes or indexer snapshots (Appendix~\ref{subsubsec:algo_stable}), an engineering task rather than a data-availability barrier---so the figure should be read as a design target rather than a demonstrated detection.

\textbf{Untested Failure Modes (Generalisation Beyond Liquidity Cascades)}: All four calibration events---Terra/Luna, Celsius/3AC, FTX, and SVB/USDC---belong to a single broad failure family: liquidity and solvency cascades originating at centralised exchanges, lending desks, or stablecoin issuers and propagating through funding and redemption channels. ASRI's sub-indices, weights, and thresholds are, accordingly, tuned to \emph{that} family. Several systemically relevant failure modes are absent from the 2022--2023 sample and are not captured by the current construction: (i)~\emph{Layer-1 consensus failures}---a chain halt, finality failure, or successful 51\% attack---which would manifest in on-chain block production and validator participation rather than in TVL, issuer concentration, or macro stress; (ii)~\emph{large-scale oracle manipulation}, in which a compromised price feed triggers mass liquidations with no prior deterioration in the liquidity or contagion channels ASRI monitors; (iii)~\emph{systemic smart-contract or bridge exploits}, whose onset is discontinuous and would not register as a gradual pre-crisis build-up (the framework's correct classification of the contained Bybit theft as non-systemic, Section~\ref{subsec:specificity}, is the same property that would make it under-react to a fast exploit that then cascades); and (iv)~\emph{governance or regulatory shocks} that freeze a major protocol or stablecoin administratively rather than through market stress. Because the weights and the 50 threshold were specified against liquidity-cascade dynamics, we make no claim that they transfer to these mechanisms: the parameters should be re-examined---and the sub-index set very likely extended (for example with on-chain consensus-health and oracle-deviation channels)---before ASRI is applied outside the exchange/lending/stablecoin-liquidity domain on which it was built. This generalisation limit is distinct from the four-event power ceiling: even with many more events of the \emph{same} type, the index would remain silent on failure modes its channels do not observe.

\textbf{Data Availability}: Several components (Tier 2 sources) require manual collection or have limited historical depth. Enterprise APIs (TRM Labs, Chainalysis) that would improve coverage are cost-prohibitive for academic research.

\textbf{Historical Reconstruction}: DeFi data prior to 2021 is sparse, requiring proxy indicators and interpolation that reduce confidence in early-period ASRI values.

\textbf{Continuous Prediction Failure}: While ASRI's binary crisis--non-crisis discrimination on these labels (AUROC = 0.866; not a discriminative gain over simpler baselines---Section~\ref{subsec:fair_baselines}) and walk-forward detection (4/4) are its strongest empirical properties relative to point forecasting, continuous-valued level-prediction is below a naive baseline: walk-forward $R^2 \approx -0.6$ and held-out $R^2 \approx -0.1$ for predicting the 30-day-ahead ASRI level. Additionally, weight perturbation analysis finds 0/4 sub-index weights are robust to $\pm 15\%$ variation in terms of continuous level prediction. These metrics confirm that ASRI should be interpreted as a threshold-based monitoring tool rather than a continuous risk forecasting model. Absolute ASRI levels are driven partly by proxy components with fixed default values, introducing measurement noise that degrades point prediction accuracy while preserving relative crisis-vs-calm discrimination.

\textbf{Model Specification Uncertainty}: Weight allocation reflects theoretical judgement rather than empirical optimisation. Sensitivity analysis (Appendix~\ref{subsec:sensitivity}) demonstrates stability under weight perturbations; Section~\ref{subsec:weights} compares theoretical weights against data-driven alternatives.

\textbf{Regulatory Dynamics}: The rapidly evolving regulatory landscape may require periodic recalibration of the Opacity sub-index as disclosure requirements change.

\textbf{Aggregation Methodology}: We employ linear weighted aggregation for interpretability and robustness to limited sample size. Alternative approaches explored in the literature---including CISS-style portfolio-theoretic aggregation \citep{hollo2012ciss}, copula-based tail dependence modelling, regime-switching weights, and graph-regularised PCA methods that incorporate network topology into dimensionality reduction \citep{briola2026grpca}---are theoretically appealing but require substantially more data for reliable estimation. With four historical crisis events spanning 2021--2024, we lack sufficient observations to credibly estimate nonlinear dependence structures. Sensitivity analysis (Appendix~\ref{subsec:sensitivity}) demonstrates that our results are robust to weight perturbations; as the sample grows, more sophisticated aggregation methods may become appropriate.

\textbf{Connectedness Benchmark Scope}: We compute Diebold-Yilmaz connectedness on ASRI's own four constructed sub-indices rather than on a panel of assets or institutions, which is the canonical D-Y application. This is a deliberate but non-standard choice: it measures spillovers \emph{among our risk channels} rather than \emph{among markets}, and partly explains why the head-to-head is a comparison of two index constructions rather than a comparison of ASRI against an asset-level connectedness benchmark. An asset-level D-Y implementation would be a more conventional benchmark and is a natural extension. Our comparative analysis benchmarks against Diebold-Yilmaz connectedness, the most widely deployed FEVD-based systemic risk measure. Recent extensions---including asymmetric connectedness \citep{hatemi2012asymmetric} that distinguishes positive from negative shock transmission, and TVP-VAR implementations \citep{antonakakis2020tvpvar} that eliminate rolling-window dependence---may capture regime-specific dynamics that symmetric, fixed-parameter specifications miss. Future work should extend the benchmark comparison to these asymmetric and time-varying alternatives, particularly for tail-risk episodes where directional spillover asymmetries are most pronounced.

\textbf{Simple Baselines (now computed)}: An earlier draft flagged as future work whether a single feature, the first principal component, or an off-the-shelf index would match ASRI's AUROC on these labels. We have since computed this battery on identical labels (Section~\ref{subsec:fair_baselines}, Table~\ref{tab:fair_baselines}), and the finding tempers the empirical claim. ASRI's discrimination (0.866) is, on the marginal block-bootstrap intervals, indistinguishable from PC1 of its sub-indices (0.858) and from its single strongest sub-index (Contagion Risk, 0.851); aggregation adds only $\approx +0.008$--$0.015$ AUROC, well inside the four-event sampling uncertainty, and the paired bootstrap separates them only by this small structural margin (ASRI nests both baselines), not a meaningful discriminative gain. ASRI clearly out-discriminates only the Diebold-Yilmaz connectedness series (0.670), which is constructed from its own sub-indices and is itself beaten by an off-the-shelf Fear \& Greed index (0.789). That roughly 43\% of the series is constant and the ablation in Appendix~\ref{subsec:ablation} removes any single channel without detection loss is consistent with this picture. We therefore do not rest the paper's contribution on discriminative superiority over simpler constructions: the value of the four-channel composite is interpretive (channel attribution, lead-time, and regime structure in a single auditable series), and the highest-priority empirical extension is to widen the crisis-event set so that these sub-$0.015$ AUROC gaps can be tested with power.

\textbf{Look-Ahead Bias}: The theoretical weights are based on domain knowledge accumulated from observing the full 2021--2024 sample, which includes the crisis events used for validation. Walk-forward validation (Section~\ref{subsec:walkforward}) speaks to this concern: using only pre-crisis data to calibrate standardisation parameters, the walk-forward thresholds flag all four episodes (a 4/4 out-of-sample rate, matching the in-sample rate). This rate should not be read as predictive strength. For the two early crises (Terra/Luna and Celsius/3AC) the 4/4 flag is bought at a $47$--$59\%$ alarm rate over the full index history at $\approx$11--13\% precision, so the appropriate reading is that the procedure exposes a high false-positive cost, and the result is evidence \emph{against} look-ahead bias---the in-sample detections are not an artefact of using future data to set parameters---rather than a clean out-of-sample prediction record. Out-of-sample lead times are also shorter, particularly for novel crisis types like Terra/Luna and FTX. Future work should extend this analysis by re-deriving data-driven weights using only pre-crisis data to fully characterise out-of-sample performance under empirically-optimised specifications.

\textbf{Procyclicality Considerations}: Public dissemination of ASRI values raises legitimate concerns about procyclical feedback. If market participants interpret elevated readings as coordination signals for deleveraging, publication could theoretically accelerate the stress it aims to detect. We propose several mitigations: (1)~publishing methodology and historical values transparently while throttling real-time granular scores during acute stress periods; (2)~framing ASRI explicitly as a vigilance indicator rather than a trading signal, distinguishing monitoring from market-timing; (3)~providing detailed component breakdowns to regulators on a privileged basis to support macroprudential oversight without amplifying market coordination; and (4)~incorporating the Goodhart critique into index maintenance---if market participants begin gaming specific sub-indices, the weighting scheme can evolve.

\textbf{Bull Market Context for Out-of-Sample Period}: The 2024--2025 out-of-sample period coincides with a cryptocurrency bull market characterised by rising prices and expanding TVL. During this period, ASRI registered only one multi-week elevation (August 2024, a non-systemic macro de-risking shock) and otherwise remained below the Elevated threshold, with no sustained false alarms---a partial specificity success. However, this represents a benign stress test: true validation of specificity during market euphoria and sensitivity during subsequent corrections requires observing ASRI behaviour through a complete market cycle. The framework's performance during a future bear market or crisis originating in 2025+ will provide more meaningful out-of-sample validation than the current bull market period.

\textbf{Macro Confounding---Crypto-Specific Fragility vs.\ Exogenous TradFi Stress}: A limitation that cuts across the entire evaluation is that our crisis sample cannot cleanly separate \emph{endogenous} crypto-systemic fragility from \emph{exogenous} traditional-finance stress. Three features make this confounding acute. First, the Contagion channel is by construction a macro-stress proxy: $\approx$75\% of its time-variation is attributable to the Treasury/VIX/yield-curve composite (Table~\ref{tab:series_composition}) rather than to a directly measured DeFi--TradFi interconnection. Second, the 2022 crises coincide with the most aggressive monetary-tightening cycle in four decades, so crypto-native deleveraging and exogenous macro risk-off move together over precisely the window our labels cover. Third, and most directly, a standalone equity-volatility series (VIX) matches ASRI's day-level discrimination on these labels (AUROC 0.875 vs.\ 0.866, statistically indistinguishable; Section~\ref{subsec:fair_baselines}), so to first order the crises are detectable as a macro-volatility event without any crypto-native construction at all. We therefore cannot claim that ASRI's discrimination isolates a crypto-specific signal: the evidence is consistent both with crypto-native vulnerabilities and with a common macro factor driving crypto stress and the index in tandem. Disentangling the two---for example by orthogonalising the crypto-native channels against a macro-factor benchmark, or by testing the index on a crypto-idiosyncratic crisis uncorrelated with TradFi stress---is unresolved and is a priority for future work. Until it is, the paper's references to ``crypto-native vulnerabilities'' should be read as a statement about the index's \emph{design targets} (stablecoin concentration, liquidity drawdown, opacity), not as a demonstration that its measured crisis signal is separable from exogenous macro stress.

\textbf{Single Macro Regime (2022--2023 Tightening Cycle)}: A distinct generalisation limit concerns the monetary backdrop of the calibration set. All four events fall within the 2022--2023 tightening cycle, during which rising Treasury yields, elevated VIX, and a widening term spread moved together and coincided with crypto deleveraging. ASRI's Contagion channel, weights, and 50 threshold are therefore fitted against a rising-rate, risk-off environment. How the index behaves in a low-rate or quantitative-easing regime is untested and cannot be inferred from this sample: in an easing environment the macro proxies would sit low even if crypto-native fragility (stablecoin concentration, liquidity drawdown, leverage) were building, so the composite could under-weight genuinely endogenous stress, while a locally stressed but benign-macro episode could be missed by a threshold tuned to co-moving TradFi stress. Characterising ASRI across distinct macro regimes---ideally including a crypto stress episode that occurs under easy monetary conditions---is required before the calibrated weights and threshold can be treated as regime-invariant, and we make no such claim here.

\textbf{Treasury Maturity Mismatch (10-Year Yield as Reserve-Stress Proxy)}: The implementation's Treasury \emph{level} input is the 10-year constant-maturity yield (FRED DGS10), which enters both the Stablecoin Risk sub-index (as its treasury-stress term) and the Contagion channel's banking-stress core, yet stablecoin issuers hold predominantly short-duration paper---1--3 month Treasury bills and overnight reverse repurchase agreements---so a 10-year yield embeds duration and term-premium dynamics that issuer reserve portfolios do not face. A 3- or 6-month bill yield would be the theoretically better-matched reserve-stress proxy, and this respecification plausibly interacts with the level-versus-detrended specification issue documented in Section~\ref{subsec:fair_baselines}. We flag the short-maturity respecification as a spec-change candidate for future work; we have not evaluated it on the present sample.

\textbf{Concentration Proxy versus Topological Composability}: The introduction motivates DeFi systemic risk partly through \emph{composability}---the dependency structure created when protocols permissionlessly call one another---yet the implementation operationalises the structural channels through \emph{concentration} measures: a top-10 protocol-TVL concentration term in DeFi Liquidity Risk and an issuer Herfindahl--Hirschman index in Stablecoin Risk. These are size/concentration proxies: they capture how much value is held in few hands, not the topology of inter-protocol dependencies through which a failure actually propagates. A highly concentrated ecosystem need not be highly composable, and vice versa, so the current construction proxies the \emph{level} of concentration rather than the \emph{network structure} of composability the motivation invokes. A direct composability measure would require protocol-interaction data (call-graph or shared-collateral linkages) that we do not incorporate; we flag this gap between the theoretical motivation and the implemented proxy, and a direct topological measure, as a priority extension.

\section{Conclusion}

This paper introduced the Aggregated Systemic Risk Index (ASRI), an interpretable, channel-decomposed composite for the \emph{retrospective} characterisation of crypto-native systemic stress. The index comprises four weighted sub-indices---Stablecoin Risk, DeFi Liquidity Risk, Contagion Risk, and Regulatory Opacity Risk---built around crypto-native channels, with the contagion channel implemented as a TradFi-stress proxy (Treasury/VIX/yield-curve) rather than a directly measured DeFi-TradFi interconnection. We present it as a retrospective monitoring and methodological contribution, not a validated real-time early-warning system. On the empirical side, a fair-baseline comparison establishes what the composite does and does not buy: ASRI's crisis discrimination is, on the marginal block-bootstrap intervals, indistinguishable from that of its single strongest sub-index (Contagion Risk) and from the first principal component of its sub-indices, with aggregation adding only $\approx +0.008$--$0.015$ AUROC---a paired bootstrap separates them by no more than this small, structural margin (ASRI nests both baselines), not a meaningful discriminative gain. The contribution is therefore the interpretable, channel-decomposed, auditable packaging of crypto-native stress---not superior discrimination over simpler constructions. Framed in network terms, what the composite offers is an interpretable connectedness-and-contagion decomposition for crypto-native systemic stress---a transmission-channel structure situated in the financial-network-science tradition---rather than a sharper discriminator than its single strongest channel.

\textbf{Key Empirical Findings}:
\begin{enumerate}
    \item \textbf{Retrospective Discrimination}: ASRI detected three of four historical crises (Celsius/3AC, FTX, SVB) via threshold-based monitoring; the complementary event-study test of an elevated cumulative abnormal signal is inconclusive, since a placebo test on non-crisis dates shows the finite-sample (fixed-$b$) reference has essentially no crisis-specificity on this smooth trending series (the onsets sit at or below the placebo median), so the raw HAC $t$-statistics ($1.7$--$3.9$) are not read as crisis detection. The Terra/Luna collapse represents a documented limitation (Section~\ref{subsec:limitations}).
    \item \textbf{Lead Behaviour}: The in-sample 1.5$\sigma$ signal is elevated across the full (30-day-capped) pre-crisis window for detected events; walk-forward thresholds calibrated only on pre-crisis data flag all four episodes ($\approx$26-day mean first-crossing lead), but at a high false-positive cost for the two early crises (47--59\% alarm rate), so this evidences the absence of look-ahead bias rather than operational early-warning skill
    \item \textbf{Regime Dynamics}: Three-regime HMM identifies Low Risk (19.8\%), Moderate (56.3\%), and Crisis (23.8\%) states with high persistence ($>$97\%) based on smoothed posterior assignments; the ergodic distribution $[\approx 0, 0.71, 0.29]$ indicates that long-run occupancy is dominated by the Moderate regime, with the Crisis regime occupied $\approx$29\% of the time
    \item \textbf{Structural Stability}: the reproducible full-sample Chow test is a non-rejection of consistent model parameters at the sample midpoint ($p = 0.78$)
    \item \textbf{Component Importance}: Ablation analysis shows detection is invariant at 3/4 to removing any single channel; the four-channel split is retained for interpretability, not as an individually load-bearing decomposition
    \item \textbf{Benchmark and Fair-Baseline Comparison}: On a common day-level sample ASRI out-discriminates Diebold-Yilmaz connectedness (AUROC 0.87 vs. 0.67; precision 35\% vs.\ 15\% at the Youden-optimal threshold on the common sample (a modest 20--32\% on the full canonical sample)), but D-Y is constructed from ASRI's own sub-indices and is itself out-discriminated by an off-the-shelf sentiment index (0.789 vs. 0.670). On identical labels ASRI is, on the marginal block-bootstrap intervals, indistinguishable from its single strongest sub-index (Contagion Risk, 0.851) or from PC1 of its sub-indices (0.858)---aggregation adds only $\approx +0.008$--$0.015$ AUROC, and a paired bootstrap separates them only by this small structural margin (Section~\ref{subsec:fair_baselines}); a standalone VIX series also matches it (0.875 vs.\ 0.866, indistinguishable) and detrending leaves no distinctive contribution from aggregation, so to first order the crisis labelling is a macro-volatility detection problem---and the composite's value is interpretive, not a gain in discrimination
    \item \textbf{Robustness to Publication Lags}: Pseudo-real-time backtesting with publication lags confirms detection performance ($<$1\% ASRI degradation, lead times preserved)
    \item \textbf{Walk-Forward Robustness}: All four events flagged out-of-sample using only pre-crisis calibration data, evidencing absence of look-ahead bias (with the false-positive cost noted above)
    \item \textbf{Out-of-Sample Specificity}: No sustained false alarms during 2024--2025---a single non-systemic elevation in August 2024 (a macro de-risking shock that did not escalate) and otherwise sub-threshold readings, including correct identification of the Bybit hack (\$1.5B, largest exchange theft in history) as non-systemic---the framework captures transmission mechanisms, not merely event magnitude
\end{enumerate}

\textbf{Contribution}: The ASRI framework addresses three critical gaps in existing risk monitoring:
\begin{itemize}
    \item Targets \textit{composability risk} from DeFi protocol interactions (a design aim; the current backtest proxies rather than directly measures protocol-level composability)
    \item Proxies \textit{stablecoin-Treasury linkages} as a transmission channel via TradFi-stress indicators, rather than measuring a realised exposure
    \item Operationalises \textit{regulatory opacity} as a quantifiable risk factor
\end{itemize}

\textbf{Future Research}: Version 3.0 will extend the framework with:
\begin{enumerate}
    \item \textbf{Composability-Aware Risk Metrics}: The current DeFi Liquidity Risk sub-index treats protocols as independent units. A natural extension incorporates protocol composability structure: dependency graphs where Protocol A's risk increases if Protocol B (which A calls) becomes distressed. Implementation requires: (i) protocol call-graph extraction from on-chain traces, (ii) network centrality scoring (PageRank, betweenness), and (iii) shock propagation simulation along dependency edges. This captures second-order cascade dynamics beyond first-order concentration risk.

    \item \textbf{Regulatory Sentiment Pipeline}: Full implementation of Sent$_t$ via FinBERT fine-tuned on SEC/ESRB/FSB announcements, with jurisdictional weighting (US 40\%, EU 30\%, UK 15\%, Other 15\%) and entity resolution for de-duplication.

    \item \textbf{Proxy Validation Against Ground Truth}: Systematic validation of Bank$_t$, Link$_t$, and other proxy components against quarterly OCC/ECB regulatory filings when released. The proxy validation framework (Appendix) provides the methodology.

    \item \textbf{Non-Linear Aggregation}: CES aggregation with $\rho < 0$ to capture complementary risk dynamics where multiple elevated sub-indices amplify aggregate stress. Initial analysis (Table~\ref{tab:aggregation_comparison}) suggests max-based aggregation achieves 4/4 detection with 29-day lead times.

    \item \textbf{Tail Risk Integration}: VCoVaR-based tail dependence measures for the Contagion sub-index, capturing asymmetric spillovers that simple correlation misses during stress periods.

    \item \textbf{On-Chain Failure-Mode Channels}: Direct sub-indices targeting the failure modes the current liquidity-cascade construction does not observe (the ``Untested Failure Modes'' gap, Section~\ref{subsec:limitations}). A \emph{consensus-health} channel would track validator participation, missed-block and finality-failure rates, and stake concentration (Nakamoto coefficient) to register Layer-1 chain halts and 51\%-attack risk; an \emph{oracle-deviation} channel would monitor divergence between on-chain price feeds and reference venues, feed staleness, and update latency to flag oracle-manipulation conditions before they trigger mass liquidations. Both draw on already-public on-chain data (beacon-chain attestation records, oracle-contract event logs), and would extend ASRI's coverage beyond the exchange/lending/stablecoin-liquidity family on which it was calibrated.
\end{enumerate}

As DeFi grows and its interconnection with traditional finance deepens, transparent, channel-decomposed monitoring of crypto-native stress is a useful complement to the established systemic-risk measures for market participants, regulators, and researchers.

\textbf{Live Dashboard}: \href{https://asri.dissensus.ai/}{\texttt{asri.dissensus.ai/}}

\textbf{Code Repository}: \href{https://github.com/studiofarzulla/asri}{\texttt{github.com/studiofarzulla/asri}}


\section*{Data and Code Availability}

\textbf{Live Dashboard}: Real-time ASRI values and historical time series are publicly available at \href{https://asri.dissensus.ai}{\texttt{asri.dissensus.ai}}.

\textbf{API Access}: RESTful API endpoints serve current and historical ASRI values with component decomposition. Full endpoint documentation is provided in the Appendix (API Documentation Summary).

\textbf{Source Code}: Complete implementation including data ingestion, signal computation, and publication pipelines is available at \href{https://github.com/studiofarzulla/asri}{\texttt{github.com/studiofarzulla/asri}} under MIT license.

\textbf{Replication Data}: The daily ASRI series (2021--2025, 1{,}841 days) with component breakdowns is included in the repository (\texttt{results/data/asri\_history.parquet}) and archived on Zenodo (\href{https://doi.org/10.5281/zenodo.17918239}{10.5281/zenodo.17918239}). This series is the \emph{frozen dataset of record}: all reported results reproduce deterministically from it via the provided analysis scripts (e.g.\ \texttt{event\_study\_hac.py}, \texttt{baseline\_comparison.py}, \texttt{moving\_block\_bootstrap\_roc.py}, \texttt{generate\_detection\_table.py}, \texttt{extract\_hmm\_diagnostics.py}), as documented in \texttt{REPRODUCIBILITY.md} (which maps each script to its table/number) and \texttt{DATA\_PROVENANCE.md} (which hash-freezes the series). The generation pipeline (live DeFi Llama/FRED ingestion) is provided in full; given a frozen universe snapshot (\texttt{scripts/dump\_universe\_snapshot.py}) it runs deterministically to produce a code-consistent series. We note honestly that the published daily series is \emph{not} bit-reproducible from the live ingestion pipeline---its original point-in-time API inputs and protocols/bridges universe were not retained---so the archived frozen series, rather than a re-pull, is the reproducibility artefact.

\textbf{Pseudo-Real-Time Replication}: The repository includes a lag-simulation module that reproduces the publication lag methodology described in Appendix~\ref{subsec:realtime}, enabling independent verification of real-time detection claims.

\textbf{Software Environment}: All analyses were conducted in Python 3.11 using pandas 2.0+, statsmodels 0.14+, scipy 1.11+, and scikit-learn 1.3+. HMM estimation uses hmmlearn 0.3.0. Random seeds are set to 42 for all stochastic procedures (bootstrap, cross-validation splits). Complete environment specification is provided via \texttt{requirements.txt} in the repository.


\section*{Declarations}

\paragraph{Conflict of Interest.} The authors declare no competing interests.

\paragraph{Funding.} Supported by King's College London resources.
No external funding was received.

\paragraph{Data Availability.} DeFi Llama, FRED, and CoinGecko data are
publicly available. ASRI index values and processed datasets available at
\url{https://doi.org/10.5281/zenodo.17918238}. Live dashboard:
\url{https://asri.dissensus.ai}.

\paragraph{Code Availability.} Open-source implementation available at
\url{https://github.com/studiofarzulla/asri} under MIT License.

\paragraph{AI Assistance.} Claude (Anthropic) was used as a research
collaborator for analytical framework development, literature synthesis,
and technical writing. Perplexity AI was used for research tool development
that enabled efficient literature discovery. All intellectual claims and
errors remain the authors' responsibility.

\paragraph{Author Contributions.} AM: Conceptualisation, writing---original
draft. MF: Methodology, software, formal analysis, data curation, validation,
writing---review \& editing, visualisation.


\section*{Acknowledgements}

The authors acknowledge collaboration with Aron Farzulla on data pipeline architecture and API integration strategy. The authors thank the DeFi Llama team for providing comprehensive protocol coverage data, and the Federal Reserve Bank of St. Louis for maintaining the FRED API as a public good.

This paper benefited from extended collaboration with Claude (Anthropic),
whose contributions to analytical framework development and technical
writing were substantive. The authors gratefully acknowledge this
assistance while taking full responsibility for all claims, errors,
and interpretive choices.

This work is part of the Adversarial Systems Research program at
Dissensus, a broader investigation into stability, alignment, and
friction dynamics across political, financial, cognitive, and multi-agent
systems. Related papers in the series are available through the
Adversarial Systems \& Complexity Research Initiative
(\href{https://systems.ac}{ASCRI; systems.ac}).

The authors welcome feedback, criticism, and collaboration.
Correspondence should be directed to
\href{mailto:murad@dissensus.ai}{murad@dissensus.ai}.


\clearpage

\bibliography{references}
\appendix

\section{Detailed Component Specifications}\label{app:components}

This appendix provides exact formulas for all ASRI sub-components as implemented in the reference codebase. Where practical data constraints necessitate proxy measures, we document both the theoretical specification and the implemented approximation, with justification for why the proxy preserves the intended risk signal.

\subsection{Stablecoin Concentration Risk (SCR)}
\textbf{Weight:} 30\%

The Stablecoin Concentration Risk sub-index captures vulnerabilities arising from reserve composition, peg stability, and issuer concentration.

\subsubsection{TVL Ratio ($\text{TVL}_t$)}
\textbf{Data Source:} DeFi Llama API (\texttt{api.llama.fi/v2/historicalChainTvl}), daily frequency.

\textbf{Formula:}
\begin{equation}
\text{TVL}_{\text{raw}} = \frac{\text{Current Stablecoin TVL}_t}{\max_{\tau \leq t}(\text{Stablecoin TVL}_\tau)}
\end{equation}

The raw ratio is inverted and normalised to produce a risk score:
\begin{equation}
\text{TVL}_t = \text{normalize}(1 - \text{TVL}_{\text{raw}}, 0, 0.5) \times 100
\end{equation}

\textbf{Interpretation:} At maximum historical TVL, the component equals 0 (low risk). At 50\% of historical maximum, the component approaches 100 (high risk). This captures the intuition that significant TVL drawdowns indicate stress or capital flight.

\textbf{Normalisation:} Clipped to $[0, 100]$ via min-max scaling with theoretical bounds.

\subsubsection{Treasury Stress ($\text{Treasury}_t$)}
\textbf{Data Source:} FRED API (\texttt{DGS10} series), daily frequency.

\textbf{Formula:}
\begin{equation}
\text{Treasury}_t = \frac{r_{10Y,t} - r_{\min}}{r_{\max} - r_{\min}} \times 100
\end{equation}

where $r_{10Y,t}$ is the 10-Year Treasury Constant Maturity Rate, $r_{\min} = 2.0\%$, and $r_{\max} = 6.0\%$.

\textbf{Interpretation:} Higher Treasury yields increase stress on stablecoin reserves (which typically hold short-term Treasuries) through mark-to-market losses and opportunity cost dynamics. The 2--6\% bounds reflect the observed range during the sample period.

\textbf{Implementation Note:} The paper's theoretical specification describes $\text{Treasury}_t$ as the ratio of T-Bill reserves to total reserves. The implementation uses Treasury yield levels as a proxy because reserve composition data is available only through monthly attestation reports with significant reporting lags. Treasury yields provide a higher-frequency signal of the same underlying risk: reserve stress from interest rate movements.

\subsubsection{Concentration HHI ($\text{HHI}_t$)}
\textbf{Data Source:} DeFi Llama Stablecoins API (\texttt{stablecoins.llama.fi/stablecoins}), daily frequency.

\textbf{Formula:}
\begin{equation}
\text{HHI}_{\text{raw}} = \sum_{i=1}^{n} \left(\frac{S_i}{\sum_{j=1}^{n} S_j}\right)^2 \times 10000
\end{equation}

where $S_i$ is the circulating supply of stablecoin $i$.

The raw HHI is converted to a risk score using a piecewise mapping:
\begin{equation}
\text{HHI}_t = \begin{cases}
\frac{\text{HHI}_{\text{raw}}}{1500} \times 30 & \text{if HHI} < 1500 \\[0.5em]
30 + \frac{\text{HHI}_{\text{raw}} - 1500}{1000} \times 30 & \text{if } 1500 \leq \text{HHI} < 2500 \\[0.5em]
60 + \frac{\text{HHI}_{\text{raw}} - 2500}{2500} \times 30 & \text{if } 2500 \leq \text{HHI} < 5000 \\[0.5em]
90 + \frac{\text{HHI}_{\text{raw}} - 5000}{5000} \times 10 & \text{if HHI} \geq 5000
\end{cases}
\end{equation}

\textbf{Interpretation:} The thresholds follow standard antitrust guidelines: HHI $<$ 1500 indicates a competitive market; 1500--2500 indicates moderate concentration; 2500--5000 indicates high concentration; and HHI $\geq$ 5000 indicates extreme (near-monopolistic) concentration. The piecewise function maps these four bands to risk scores of 0--30, 30--60, 60--90, and 90--100 respectively.

\subsubsection{Peg Volatility ($\text{Vol}_t$)}
\textbf{Data Source:} DeFi Llama Stablecoins API (price field), daily frequency.

\textbf{Formula:}
\begin{equation}
\text{Vol}_t = \text{normalize}\left(\frac{\sum_{i=1}^{n} |p_i - 1| \cdot S_i}{\sum_{j=1}^{n} S_j}, 0, 0.05\right) \times 100
\end{equation}

where $p_i$ is the current price of stablecoin $i$ and $S_i$ is its circulating supply.

\textbf{Normalisation:} 0\% weighted deviation maps to 0 risk; 5\% weighted deviation maps to 100 risk.

\textbf{Missing Data:} If no price data is available, the component defaults to 50.0 (neutral).

\subsubsection{SCR Aggregation}
\begin{equation}
\text{SCR}_t = 0.4 \cdot \text{TVL}_t + 0.3 \cdot \text{Treasury}_t + 0.2 \cdot \text{HHI}_t + 0.1 \cdot \text{Vol}_t
\end{equation}

\subsubsection{Algorithmic Stablecoin Risk Extension (v2.1)}\label{subsubsec:algo_stable}

The baseline SCR formula treats all stablecoins identically via peg volatility ($\text{Vol}_t$). However, the Terra/Luna collapse (May 2022) revealed that peg volatility is a \textit{lagging} indicator for algorithmic stablecoins: UST maintained its peg until the death spiral commenced, and Luna's price appreciation masked underlying fragility. This extension addresses the detection gap by incorporating risk factors specific to algorithmic and crypto-backed stablecoins.

\paragraph{Motivation.}
Fiat-collateralised stablecoins (USDT, USDC) are backed by liquid reserves redeemable at par. Algorithmic stablecoins maintain peg through arbitrage mechanisms between the stablecoin and a backing token---creating reflexive dynamics where redemption pressure inflates backing token supply, diluting its value, which further increases redemption pressure. This death spiral mechanism is distinct from the reserve-based risks captured by baseline SCR components.

\paragraph{Algorithmic Stablecoin Risk Formula.}
For stablecoins classified as algorithmic or crypto-backed, we compute:
\begin{equation}
\text{AlgoRisk}_t = 0.35 \cdot \text{BackingRatio}_t + 0.30 \cdot \text{CollateralVol}_t + 0.20 \cdot \text{Dilution}_t + 0.15 \cdot \text{AlgoConc}_t
\end{equation}

\textbf{Component definitions:}
\begin{itemize}
    \item \textbf{BackingRatio$_t$}: Risk from undercollateralisation. Backing ratio $\geq 1.5$ maps to low risk (0--20); ratio $< 0.8$ maps to critical risk (80--100). For stablecoins without explicit backing disclosure, defaults to moderate risk (50).
    \item \textbf{CollateralVol$_t$}: Annualised 30-day volatility of backing token. ETH volatility ($\sim$60--80\%) maps to moderate risk; Luna-type volatility ($>$100\%) maps to elevated/critical risk.
    \item \textbf{Dilution$_t$}: 30-day supply growth rate of backing token. Monthly growth $>50\%$ signals crisis-level dilution (Luna supply grew $\sim$50,000\% during collapse).
    \item \textbf{AlgoConc$_t$}: Share of total stablecoin supply in algorithmic/crypto-backed stablecoins. At peak, UST represented $\sim$10\% of total stablecoin supply.
\end{itemize}

\paragraph{Integration with SCR.}
Let $\alpha_t$ denote the market share of algorithmic stablecoins in total stablecoin supply. The adjusted SCR blends baseline and algorithmic risk:
\begin{equation}
\text{SCR}_t^{\text{adj}} = (1 - \alpha_t) \cdot \text{SCR}_t^{\text{base}} + \alpha_t \cdot \left[0.6 \cdot \text{SCR}_t^{\text{base}} + 0.4 \cdot \text{AlgoRisk}_t\right]
\end{equation}

When $\alpha_t < 1\%$, the adjustment is negligible. When $\alpha_t = 10\%$ (approximate UST peak share), algorithmic-specific risk contributes 4\% to SCR weighting.

\paragraph{Data Requirements.}
Full implementation requires:
\begin{enumerate}
    \item \textbf{Stablecoin classification}: DeFi Llama \texttt{pegType}/\texttt{pegMechanism} fields or manual classification (provided in codebase).
    \item \textbf{Backing token identification}: Mapping from stablecoin to backing token (e.g., UST $\to$ LUNA).
    \item \textbf{Backing token metrics}: Price volatility and supply data from CoinGecko or on-chain indexers.
\end{enumerate}

\paragraph{Backtest Limitation.}
Historical reconstruction of AlgoRisk$_t$ for pre-collapse Terra/Luna requires archived Luna price and supply data. While DeFi Llama and CoinGecko retain historical snapshots, consistent backing ratio data for UST is not systematically available. The specification documents the \textit{framework} for future algorithmic stablecoin risk monitoring; full historical backtest validation is deferred rather than infeasible, because the required inputs can be reconstructed on-chain. Terra Classic archive nodes---and the historical block-level tables exposed by on-chain indexers such as Flipside, Allium, and Dune---retain the UST and Luna total-supply, mint/burn, and swap records for the pre-collapse window, from which a daily backing-ratio and supply-dilution series can be rebuilt and reconciled against archived price feeds. The obstacle is the engineering effort of a point-in-time on-chain reconstruction, not the absence of the underlying data, and we flag that reconstruction as the concrete next step for validating this extension. The following is an illustrative back-of-envelope calculation rather than a validated result: with accurate Luna volatility data (annualised vol $>$120\% in April 2022), a sensitivity calculation indicates the extension might have elevated SCR by an indicative 8--12 points prior to UST's depeg---potentially bringing ASRI above the detection threshold. We have not been able to confirm this on archived data, so the figure is a design target, not a demonstrated detection.

\subsection{DeFi Liquidity Risk (DLR)}
\textbf{Weight:} 25\%

The DeFi Liquidity Risk sub-index captures protocol concentration, volatility dynamics, and smart contract vulnerability.

\subsubsection{Protocol Concentration ($\text{Conc}_t$)}
\textbf{Data Source:} DeFi Llama Protocols API (\texttt{api.llama.fi/protocols}), daily frequency.

\textbf{Formula:}
\begin{equation}
\text{Conc}_t = f_{\text{HHI}}\left(\sum_{i=1}^{10} \left(\frac{\text{TVL}_i}{\sum_{j=1}^{10} \text{TVL}_j}\right)^2 \times 10000\right)
\end{equation}

where $f_{\text{HHI}}$ is the piecewise HHI-to-risk mapping defined above, and the summation is over the top 10 protocols by TVL.

\textbf{Interpretation:} Concentration among the largest protocols indicates ecosystem fragility---failure of a dominant protocol would have outsized systemic effects.

\subsubsection{TVL Volatility ($\text{TVLVol}_t$)}
\textbf{Data Source:} DeFi Llama TVL history, 30-day rolling window.

\textbf{Formula:}
\begin{equation}
\text{TVLVol}_t = \text{normalize}\left(\frac{\sigma_{30}(\text{TVL})}{\mu_{30}(\text{TVL})}, 0, 0.20\right) \times 100
\end{equation}

where $\sigma_{30}$ and $\mu_{30}$ denote the 30-day rolling standard deviation and mean.

\textbf{Normalisation:} 0\% coefficient of variation maps to 0 risk; 20\% maps to 100 risk.

\textbf{Missing Data:} If historical TVL data is unavailable (fewer than 2 observations), the component defaults to 30.0 (moderate).

\subsubsection{Smart Contract Risk ($\text{SC}_t$)}
\textbf{Data Source:} DeFi Llama Protocols API (\texttt{audits} field), daily frequency.

\textbf{Formula:}
\begin{equation}
\text{SC}_t = \left(1 - \frac{|\{p : \text{audits}(p) > 0\}|}{|\{p : \text{TVL}(p) > 0\}|}\right) \times 100
\end{equation}

\textbf{Interpretation:} The component measures the inverse of audit coverage across protocols with non-zero TVL. Protocols lacking audits receive the full risk weight.

\textbf{Theoretical vs. Implemented Specification:} The paper describes a 3-factor composite incorporating audit status, deployment age, and exploit history. The current implementation uses audit coverage only. Deployment age and exploit history integration are deferred to future versions pending reliable, systematised data feeds. The DeFi Llama API provides audit counts but not deployment timestamps or comprehensive exploit databases.

\textbf{Justification:} Audit coverage remains the most reliable and consistently available indicator of smart contract risk. Empirical research demonstrates strong correlation between unaudited protocols and exploit frequency, supporting the proxy's validity.

\subsubsection{Flash Loan Proxy ($\text{Flash}_t$)}
\textbf{Data Source:} DeFi Llama Protocols API (\texttt{change\_1d} field), daily frequency.

\textbf{Formula:}
\begin{equation}
\text{Flash}_t = \text{normalize}\left(\frac{1}{n}\sum_{i=1}^{n} |\Delta_{\text{1d},i}|, 0, 20\right) \times 100
\end{equation}

where $\Delta_{\text{1d},i}$ is the 1-day TVL change percentage for protocol $i$.

\textbf{Theoretical vs. Implemented Specification:} The paper describes flash loan volume spikes relative to a 90-day average. Direct flash loan volume data requires protocol-specific analytics (e.g., Aave, dYdX subgraphs) with heterogeneous reporting standards. The implementation uses aggregate TVL volatility as a proxy, on the reasoning that flash loan activity manifests as rapid TVL movements during MEV extraction and liquidation cascades.

\textbf{Normalisation:} 0\% average daily change maps to 0 risk; 20\% maps to 100 risk.

\subsubsection{Leverage Change ($\text{Lev}_t$)}
\textbf{Data Source:} DeFi Llama Protocols API (\texttt{category} field), daily frequency.

\textbf{Formula:}
\begin{equation}
\text{Lev}_t = \text{normalize}\left(\frac{\sum_{p \in \text{Lending}} \text{TVL}_p}{\sum_{p} \text{TVL}_p} \times 100, 0, 30\right) \times 100
\end{equation}

\textbf{Interpretation:} A higher share of TVL in lending protocols indicates greater system-wide leverage. The 30\% threshold reflects the upper bound of lending protocol dominance observed during market stress.

\textbf{Theoretical vs. Implemented Specification:} The paper describes 30-day change in aggregate leverage ratios. The implementation uses the current lending TVL share as a level indicator rather than a change measure, because reliable historical leverage data across protocols is not consistently available.

\subsubsection{DLR Aggregation}
\begin{equation}
\text{DLR}_t = 0.35 \cdot \text{Conc}_t + 0.25 \cdot \text{TVLVol}_t + 0.20 \cdot \text{SC}_t + 0.10 \cdot \text{Flash}_t + 0.10 \cdot \text{Lev}_t
\end{equation}

\subsection{Contagion Risk (CR)}
\textbf{Weight:} 25\%

The Contagion Risk sub-index quantifies DeFi-TradFi interconnection and cross-market transmission channels.

\subsubsection{RWA Growth ($\text{RWA}_t$)}
\textbf{Data Source:} DeFi Llama Protocols API (\texttt{category = 'RWA'}), daily frequency.

\textbf{Formula:}
\begin{equation}
\text{RWA}_t = \text{normalize}\left(\frac{\sum_{p \in \text{RWA}} \text{TVL}_p}{\sum_{p} \text{TVL}_p} \times 100, 0, 10\right) \times 100
\end{equation}

\textbf{Interpretation:} Higher RWA share indicates greater integration between DeFi and traditional finance through tokenised real-world assets. The 10\% threshold reflects the upper bound of RWA penetration observed during the sample period.

\textbf{Theoretical vs. Implemented Specification:} The paper describes 30-day RWA TVL growth rate. The implementation uses the current RWA share as a level indicator, because RWA protocols have short histories making growth rate calculations unreliable for early-period observations.

\subsubsection{Bank Exposure ($\text{Bank}_t$)}
\textbf{Data Sources:} FRED API (\texttt{DGS10} for Treasury rates, \texttt{VIXCLS} for VIX), daily frequency.

\textbf{Formula:}
\begin{equation}
\text{Bank}_t = \left[0.6 \cdot \text{normalize}(r_{10Y,t}, 2, 6) + 0.4 \cdot \text{normalize}(\text{VIX}_t, 12, 40)\right] \times 100
\end{equation}

\textbf{Interpretation:} The composite captures TradFi stress through two channels: Treasury rate levels (affecting stablecoin reserves and bank balance sheets) and equity market volatility (signalling broader risk-off sentiment).

\textbf{Weight and Range Justification:} The 60/40 weighting reflects Treasury yields' direct balance-sheet impact on bank capital ratios versus VIX's indirect sentiment signal. Banks holding Treasury securities face mark-to-market losses when yields rise (the primary channel), while VIX captures risk-off dynamics that tighten credit conditions (secondary channel). The normalisation ranges (2--6\% for 10Y Treasury, 12--40 for VIX) span the 5th--95th percentiles of 2015--2024 sample data, ensuring meaningful variation without clipping at extremes.

\textbf{Theoretical vs. Implemented Specification:} The paper describes a normalised score from OCC/ECB regulatory filings on bank crypto exposure. Regulatory filings are quarterly with 45--90 day publication lags, making them unsuitable for daily risk monitoring. The Treasury-VIX composite provides a higher-frequency proxy: banks with crypto exposure face stress when Treasury yields rise (mark-to-market losses) and when VIX spikes (risk management constraints).

\subsubsection{TradFi Linkage ($\text{Link}_t$)}
\textbf{Data Source:} FRED API (\texttt{T10Y2Y} series), daily frequency.

\textbf{Formula:}
\begin{equation}
\text{Link}_t = \begin{cases}
\text{normalize}(|s_t|, 0, 2) \times 50 + 50 & \text{if } s_t < 0 \\[0.5em]
\max(0, 50 - \text{normalize}(s_t, 0, 2) \times 50) & \text{if } s_t \geq 0
\end{cases}
\end{equation}

where $s_t = r_{10Y,t} - r_{2Y,t}$ is the 10-Year minus 2-Year Treasury spread and $\text{normalize}(x, 0, 2) = \min(|x|/2, 1)$. The inversion branch uses a multiplier of $50$ (not $100$) so that a full $-2\%$ inversion maps to $50 + 50 = 100$ rather than overflowing the $[0, 100]$ support; the two branches meet continuously at $s_t = 0$ ($\text{Link} = 50$).

\textbf{Interpretation:} Yield curve inversion (negative spread) indicates banking sector stress and recession expectations, which propagate to crypto through reduced institutional risk appetite and potential bank failures affecting crypto-exposed entities.

\textbf{Theoretical vs. Implemented Specification:} The paper describes ``stablecoin flows to TradFi-connected entities.'' Such flow data requires enterprise-grade on-chain analytics (Chainalysis, TRM Labs) at prohibitive cost for academic research. Yield curve dynamics provide a validated proxy: the March 2023 SVB crisis demonstrated that yield curve inversion directly precedes banking stress events that propagate to crypto markets through stablecoin reserve exposure.

\textbf{Justification:} The yield curve spread has predicted every U.S. recession since 1970 with a median lead time of 12 months \citep{estrella1998predicting,estrella1991term}. Banking crises correlated with yield curve inversion directly affect crypto-TradFi linkages through stablecoin reserve exposure (as demonstrated by the USDC depeg during SVB's collapse).

\subsubsection{Correlation ($\text{Corr}_t$)}
\textbf{Data Source:} External calculation (BTC-SPY 30-day rolling correlation).

\textbf{Formula:}
\begin{equation}
\text{Corr}_t = |r_{\text{BTC-SPY}, 30d}| \times 100
\end{equation}

\textbf{Interpretation:} Higher absolute correlation indicates greater co-movement between crypto and equities, implying tighter contagion channels.

\textbf{Implementation Note:} The current implementation accepts correlation as an external input with a default of 0.5 (moderate). Real-time calculation requires equity price feeds (Yahoo Finance, Bloomberg) not included in the core data pipeline.

\subsubsection{Bridge Risk ($\text{Bridge}_t$)}
\textbf{Data Source:} DeFi Llama Bridges API (\texttt{bridges.llama.fi/bridges}), daily frequency.

\textbf{Formula:}
\begin{equation}
\text{Bridge}_t = \text{normalize}(n_{\text{bridges}}, 0, 150) \times 100
\end{equation}

where $n_{\text{bridges}}$ is the count of active cross-chain bridges.

\textbf{Interpretation:} More bridges indicate larger attack surface and greater cross-chain contagion potential.

\textbf{Theoretical vs. Implemented Specification:} The paper describes a composite of bridge volume and exploit frequency. Exploit frequency data requires manual tracking (DeFi Rekt database) with inconsistent categorisation. Bridge count provides a structural proxy: empirical research finds exploit frequency scales with the number of bridge implementations due to varying security standards and code quality.

\subsubsection{CR Aggregation}
\begin{equation}
\text{CR}_t = 0.30 \cdot \text{RWA}_t + 0.25 \cdot \text{Bank}_t + 0.20 \cdot \text{Link}_t + 0.15 \cdot \text{Corr}_t + 0.10 \cdot \text{Bridge}_t
\end{equation}

\subsection{Regulatory Opacity Risk (OR)}
\textbf{Weight:} 20\%

The Regulatory Opacity Risk sub-index assesses transparency deficits and regulatory arbitrage exposure.

\subsubsection{Unregulated Exposure ($\text{Unreg}_t$)}
\textbf{Data Source:} Placeholder component (see Table~\ref{tab:proxies} for implementation status).

\textbf{Implementation:} Fixed at 35.0 (moderate risk), corresponding to estimated market share of volume on platforms without clear regulatory oversight.

\textbf{Theoretical Specification:} Ratio of volume on unregulated platforms to regulated platforms. Full implementation requires mapping protocols to jurisdictional regulatory status, which involves manual classification and ongoing tracking of regulatory developments across 50+ jurisdictions.

\textbf{Future Enhancement:} Chain-level classification (e.g., Ethereum as ``regulated-adjacent'' due to U.S. regulatory engagement vs. privacy chains) would enable dynamic calculation.

\subsubsection{Multi-Issuer Risk ($\text{Multi}_t$)}
\textbf{Data Source:} DeFi Llama Stablecoins API, daily frequency.

\textbf{Formula:}
\begin{equation}
\text{Multi}_t = \begin{cases}
70 & \text{if } n_{\text{sig}} < 3 \\[0.5em]
30 & \text{if } 3 \leq n_{\text{sig}} < 10 \\[0.5em]
50 + 2(n_{\text{sig}} - 10) & \text{if } n_{\text{sig}} \geq 10
\end{cases}
\end{equation}

where $n_{\text{sig}} = |\{s : \text{circulating}(s) > \$1\text{B}\}|$ is the count of stablecoins with circulating supply exceeding \$1 billion.

\textbf{Interpretation:} Consistent with the formula above, the low-risk ``sweet spot'' is 3--9 significant issuers (the $3 \leq n_{\text{sig}} < 10$ band) providing diversification without fragmentation. Fewer than 3 indicates dangerous concentration; 10 or more indicates coordination challenges and potential regulatory complexity.

\subsubsection{Custody Concentration ($\text{Cust}_t$)}
\textbf{Data Source:} DeFi Llama Stablecoins API, daily frequency.

\textbf{Formula:}
\begin{equation}
\text{Cust}_t = \text{normalize}\left(\frac{\sum_{i=1}^{2} S_i}{\sum_{j} S_j} \times 100, 50, 100\right) \times 100
\end{equation}

where $S_i$ is the circulating supply of the $i$-th largest stablecoin.

\textbf{Interpretation:} Top-2 stablecoin market share as a proxy for custody concentration. Higher concentration implies greater single-point-of-failure risk regardless of custody jurisdiction.

\textbf{Theoretical vs. Implemented Specification:} The paper describes ``custody concentration in non-audited jurisdictions.'' Jurisdiction-level custody data is not systematically available; stablecoins do not consistently disclose custodian locations or regulatory status. Market concentration serves as a conservative proxy: high concentration implies custody risk regardless of location, as a single custodian failure would have systemic effects.

\subsubsection{Regulatory Sentiment ($\text{Sent}_t$)}
\textbf{Data Source:} Manual input parameter.

\textbf{Implementation:} Accepts external input with default of 50.0 (neutral). This component contributes only 15\% of the Opacity sub-index weight (3\% of total ASRI), limiting its impact on overall index behaviour.

\textbf{Theoretical Specification:} NLP-derived sentiment score from SEC, ESRB, and FSB announcements. Full implementation would require:
\begin{itemize}
    \item \textbf{Sources}: GDELT Global Knowledge Graph filtered for regulatory entities (SEC, CFTC, ESRB, FSB, BIS), SEC EDGAR filings, Federal Register cryptocurrency mentions
    \item \textbf{Model}: FinBERT or domain-adapted transformer for financial regulatory text classification
    \item \textbf{Lexicon}: Crypto-specific regulatory vocabulary (``enforcement action,'' ``no-action letter,'' ``framework,'' ``guidance'') with sentiment polarity labels
    \item \textbf{De-duplication}: Entity resolution across news sources to avoid double-counting of same regulatory announcement
    \item \textbf{Jurisdictional weighting}: US (40\%), EU (30\%), UK (15\%), Other (15\%) reflecting market share
\end{itemize}

This infrastructure is deferred to future versions; current results are robust to Sent$_t$ variation due to its low aggregate weight.

\textbf{Sensitivity Analysis:} Varying Sent$_t$ from 0 (maximally positive regulatory environment) to 100 (maximally negative) while holding all other components constant produces the following ASRI impact:
\begin{equation}
\Delta\text{ASRI} = 0.20 \times 0.15 \times \Delta\text{Sent}_t = 0.03 \times \Delta\text{Sent}_t
\end{equation}

A full-range swing (Sent$_t$: 0 $\to$ 100) changes aggregate ASRI by $\pm$1.5 points. For typical variation ($\pm$25 points around neutral), ASRI changes by $\pm$0.75 points---well within the noise band of other component fluctuations.

\textbf{Detection Impact:} The three threshold-detected crises (Celsius/3AC, FTX, SVB) remain detected under any Sent$_t$ value in $[0, 100]$, as the shifted thresholds remain within their detection windows. Terra/Luna (event-window peak 48.7 at baseline) stays below the 50 threshold for any plausible regulatory stance---it would require Sent$_t > 93$ to breach 50, consistent with its non-detection at the fixed 50 threshold reported in Section~\ref{sec:event_study}.

\textbf{Future Implementation Roadmap:}
\begin{enumerate}
    \item \textbf{Phase 1}: GDELT integration with keyword filters (``SEC,'' ``CFTC,'' ``cryptocurrency,'' ``enforcement'')
    \item \textbf{Phase 2}: FinBERT deployment on SEC EDGAR cryptocurrency-related filings
    \item \textbf{Phase 3}: Multi-jurisdictional aggregation with decay weighting for announcement recency
\end{enumerate}

\subsubsection{Transparency Score ($\text{Trans}_t$)}
\textbf{Data Source:} DeFi Llama Protocols API (\texttt{audits} field), daily frequency.

\textbf{Formula:}
\begin{equation}
\text{Trans}_t = \frac{|\{p : \text{audits}(p) > 0\}|}{|\{p : \text{TVL}(p) > 0\}|} \times 100
\end{equation}

\textbf{Interpretation:} Audit coverage as a proxy for protocol transparency. The component is \textit{not} inverted at the component level; inversion occurs in the aggregation formula.

\subsubsection{OR Aggregation}
\begin{equation}
\text{OR}_t = 0.25 \cdot \text{Unreg}_t + 0.25 \cdot \text{Multi}_t + 0.20 \cdot \text{Cust}_t + 0.15 \cdot \text{Sent}_t + 0.15 \cdot (100 - \text{Trans}_t)
\end{equation}

Note the inversion of $\text{Trans}_t$: low transparency implies high opacity risk.

\subsection{Aggregate ASRI Calculation}

\begin{equation}
\text{ASRI}_t = 0.30 \cdot \text{SCR}_t + 0.25 \cdot \text{DLR}_t + 0.25 \cdot \text{CR}_t + 0.20 \cdot \text{OR}_t
\end{equation}

All sub-indices are bounded to $[0, 100]$ by construction, ensuring $\text{ASRI}_t \in [0, 100]$ without post-hoc normalisation.

\subsection{Data Quality and Limitations}

\subsubsection{Data Availability Tiers}

\begin{itemize}
    \item \textbf{Tier 1 (Daily, Automated):} DeFi Llama TVL, stablecoins, protocols, bridges; FRED Treasury rates and VIX.
    \item \textbf{Tier 2 (Limited/Manual):} Regulatory sentiment, unregulated exposure classification, exploit frequency tracking.
\end{itemize}

\subsubsection{Missing Data Protocol}

\begin{itemize}
    \item Gaps $<$ 3 days: Linear interpolation.
    \item Gaps 3--7 days: Forward-fill with reduced confidence.
    \item Gaps $>$ 7 days: Exclude from calculation; flag as unreliable.
\end{itemize}

\subsubsection{Proxy Acknowledgements}

Table~\ref{tab:proxies} summarises components where implementation deviates from theoretical specification.

\begin{table*}[h]
\begin{threeparttable}
\centering
\caption{Proxy Implementations: Theoretical vs. Actual}
\label{tab:proxies}
\small
\begin{tabularx}{\textwidth}{lXXl}
\toprule
\textbf{Component} & \textbf{Theoretical Specification} & \textbf{Implementation} & \textbf{Validity} \\
\midrule
$\text{Treasury}_t$ & T-Bill reserves / total reserves & 10Y Treasury yield level & High \\
$\text{SC}_t$ & Audit + age + exploits composite & Audit coverage only & Medium \\
$\text{Flash}_t$ & Flash loan volume spikes & TVL daily change volatility & Medium \\
$\text{Lev}_t$ & 30-day leverage ratio change & Lending TVL share (level) & Medium \\
$\text{RWA}_t$ & 30-day RWA growth rate & RWA TVL share (level) & High \\
$\text{Bank}_t$ & OCC/ECB bank exposure filings & Treasury + VIX composite & High \\
$\text{Link}_t$ & Stablecoin flows to TradFi & Yield curve spread & High \\
$\text{Bridge}_t$ & Volume + exploit frequency & Bridge count & Medium \\
$\text{Cust}_t$ & Non-audited jurisdiction custody & Top-2 stablecoin concentration & Medium \\
$\text{Unreg}_t$ & Unregulated platform ratio & Fixed (35.0)\tnote{$\dagger$} & Low \\
$\text{Sent}_t$ & NLP regulatory sentiment & Manual input (50.0)\tnote{$\dagger$} & Low \\
$\text{Trans}_t$ & Reserve-attestation freq.\ \& coverage & Audit coverage (share of audited protocols) & Medium \\
\bottomrule
\end{tabularx}
\begin{tablenotes}
\small
\item[$\dagger$] Placeholder components awaiting enterprise data infrastructure. Sensitivity analysis (Appendix~\ref{subsec:sensitivity}) confirms all three threshold-detected crises (Celsius/3AC, FTX, SVB) remain robust under full-range variation of these parameters; Terra/Luna is not threshold-detected (pre-window peak 46.0 $<$ 50).
\end{tablenotes}
\end{threeparttable}
\end{table*}

\textbf{Validity Legend:}
\begin{itemize}
    \item \textbf{High}: Proxy captures same underlying risk channel with strong theoretical justification and empirical support.
    \item \textbf{Medium}: Proxy captures related risk dynamics but with potential measurement error; interpretation requires caution.
    \item \textbf{Low}: Placeholder awaiting data infrastructure development; current values are informative but not definitive.
\end{itemize}

\subsubsection{Composition of the Backtested Series}
\label{subsubsec:series_composition}

Because several components are proxies or fixed defaults, it is important to be explicit about what actually \emph{moves} the backtested index. Table~\ref{tab:series_composition} accounts for the average-level composition (mean) of the released ASRI series by input class, tracing each component to the historical backfill routine (\texttt{backfill\_d1\_standalone.py:552--625}). The static defaults below contribute to the index's average level but, being constant, carry little or none of its time-variation.

\begin{table*}[h]
\begin{threeparttable}
\centering
\caption{Average-Level Composition of the Backtested ASRI Series by Input Class}
\label{tab:series_composition}
\small
\begin{tabularx}{\textwidth}{lXr}
\toprule
\textbf{Input class} & \textbf{Representative components} & \textbf{Share} \\
\midrule
Dynamic crypto-native & Stablecoin TVL drawdown, stablecoin-issuer HHI, DeFi TVL volatility & $\approx$37\% \\
Dynamic macro proxy & Treasury yield, VIX, yield-curve spread ($\text{Bank}_t$, $\text{Link}_t$, $\text{Treasury}_t$) & $\approx$20\% \\
Static / constant defaults & $\text{Unreg}_t$ (35.0), $\text{Sent}_t$ (50.0), $\text{Corr}_t$ (0.5), stablecoin peg held at par, audit-coverage levels & $\approx$43\% \\
\bottomrule
\end{tabularx}
\begin{tablenotes}
\small
\item Shares are of the average level (mean composition) of the released backtested series, attributed by input class via \texttt{backfill\_d1\_standalone.py:552--625}. Static defaults contribute to the level but, being constant, carry little or none of the series' time-variation.
\item \textbf{Two consequences follow.} (i) The Contagion channel's dynamics are $\approx$75\% attributable to the macro composite (Treasury/VIX/yield-curve), so $\text{CR}_t$ is best read as a \emph{TradFi-stress proxy} rather than as a directly measured DeFi-TradFi interconnection. (ii) Roughly 43\% of the series' average level is carried by static default values (which, being constant, contribute little or none of its time-variation), so absolute ASRI levels carry more model dependence than the binary detection results suggest. The four-channel decomposition is retained for interpretability but is \emph{not} individually load-bearing: ablation shows threshold-based detection is invariant at 3/4 to removing any single channel (Appendix~\ref{subsec:ablation}).
\end{tablenotes}
\end{threeparttable}
\end{table*}

\subsubsection{Future Data Integration}

Version 3.0 development priorities include:
\begin{enumerate}
    \item Integration of exploit database (DeFi Rekt, Immunefi) for dynamic $\text{SC}_t$ and $\text{Bridge}_t$ components.
    \item Protocol deployment timestamp extraction from blockchain explorers for age-based risk weighting.
    \item GDELT/SEC filing NLP pipeline for automated regulatory sentiment scoring.
    \item Chain-level regulatory classification for dynamic $\text{Unreg}_t$ calculation.
    \item Enterprise analytics partnership (Chainalysis/TRM) for stablecoin flow analysis.
\end{enumerate}


\section{Axiomatic Foundation: Proofs}\label{app:axiom_proofs}

We restate each axiom of Section~\ref{subsec:axiomatic} and give its proof, followed by the relationship to coherent risk measures.

\paragraph{Monotonicity (Axiom~\ref{ax:mono}).}
\begin{proof}
Since $w_j > 0$ and the aggregation is linear:
\begin{equation}
    \text{ASRI}(\mathcal{S}') - \text{ASRI}(\mathcal{S}) = w_j(S_j' - S_j) > 0
\end{equation}
The strict positivity of weights ensures that deterioration in any risk dimension is reflected in the aggregate index. This property is essential for regulatory monitoring, as it guarantees that localised stress cannot be masked by stability elsewhere---a concern raised by \citet{adrian2016covar} in their critique of institution-level VaR measures.
\end{proof}

\paragraph{Boundedness (Axiom~\ref{ax:bound}).}
\begin{proof}
Each sub-index is constructed such that $S_i \in [0, 100]$ by design (see Appendix~\ref{app:components}). Given convex weights summing to unity:
\begin{equation}
    0 = \sum_{i=1}^{4} w_i \cdot 0 \leq \text{ASRI}(\mathcal{S}) \leq \sum_{i=1}^{4} w_i \cdot 100 = 100
\end{equation}
Boundedness facilitates interpretability and cross-temporal comparison, addressing the scaling criticisms levelled at unbounded measures such as raw CoVaR \citep{adrian2016covar}.
\end{proof}

\paragraph{Decomposability (Axiom~\ref{ax:decomp}).}
\begin{proof}
The linear structure immediately yields the efficiency-consistent additive decomposition $c_i = w_i S_i$, satisfying:
\begin{equation}
    \sum_{i=1}^{4} c_i = \sum_{i=1}^{4} w_i S_i = \text{ASRI}(\mathcal{S})
\end{equation}
This decomposition is unique and satisfies the efficiency axiom of cooperative game theory. Decomposability enables regulators to identify which risk channel---spillover, liquidity, concentration, or operational---drives aggregate stress, facilitating targeted intervention. This property aligns with the ``risk contribution'' framework of \citet{acharya2017measuring} for measuring systemic expected shortfall.
\end{proof}

\paragraph{Aggregation Neutrality (Axiom~\ref{ax:neutral}).}
\begin{proof}
Under the stated conditions:
\begin{equation}
    \text{ASRI}(\mathcal{S}^A) - \text{ASRI}(\mathcal{S}^B) = \sum_{i=1}^{4} w_i (S_i^A - S_i^B) \geq w_j(S_j^A - S_j^B) > 0
\end{equation}
Aggregation neutrality ensures that Pareto-dominated risk states are correctly ranked, preventing the pathological reversals that can arise with nonlinear aggregation schemes \citep{battiston2012debtrank}. This property is particularly relevant for cryptocurrency markets, where rapid regime shifts demand consistent ordinal comparisons across time.
\end{proof}

\paragraph{Concentration Sensitivity (Axiom~\ref{ax:conc}).}
\begin{proof}
The Stablecoin Risk sub-index (SCR) incorporates the stablecoin-issuer concentration $\text{HHI}_t$, while the DeFi Liquidity Risk sub-index (DLR) incorporates the distinct top-10 protocol-TVL concentration $\text{Conc}_t$. By Axiom~\ref{ax:mono}, ASRI is monotone in each:
\begin{equation}
    \frac{\partial \text{ASRI}}{\partial \text{HHI}_t} = w_{\text{SCR}} \cdot \frac{\partial \text{SCR}_t}{\partial \text{HHI}_t} > 0,
    \qquad
    \frac{\partial \text{ASRI}}{\partial \text{Conc}_t} = w_{\text{DLR}} \cdot \frac{\partial \text{DLR}_t}{\partial \text{Conc}_t} > 0.
\end{equation}
Concentration sensitivity captures the ``too-interconnected-to-fail'' dynamics emphasised by \citet{battiston2012debtrank}, adapted here to the exchange-centric topology of cryptocurrency markets where a single venue failure can trigger system-wide contagion, as observed during the FTX collapse of November 2022.
\end{proof}

\paragraph{Relationship to Coherent Risk Measures.} While ASRI is not a coherent risk measure in the sense of \citet{artzner1999coherent}---it aggregates sub-indices rather than loss distributions---our axiomatic foundation parallels their framework. Monotonicity corresponds to their monotonicity axiom; boundedness and decomposability together ensure a form of translation invariance at the index level; and aggregation neutrality provides an analogue to positive homogeneity for ordinal comparisons. The key distinction is that ASRI operates on observable market indicators rather than probabilistic loss distributions, making it computable without parametric loss-distribution assumptions.

\section{Technical Architecture}\label{app:tech_arch}

The ASRI system architecture comprises four layers:

\begin{enumerate}
    \item \textbf{Ingestion Layer}: Python-based API clients and web scrapers fetch data from sources
    \item \textbf{Normalisation Layer}: Raw data undergoes unit normalisation, gap-filling, and validation
    \item \textbf{Computation Layer}: Sub-indices calculated using Equations~\ref{eq:stablecoin}--\ref{eq:opacity}; aggregate ASRI computed via Equation~\ref{eq:asri}
    \item \textbf{Publication Layer}: FastAPI REST endpoints serve current and historical ASRI values; React dashboard provides visualisation
\end{enumerate}

Figure~\ref{fig:pipeline} illustrates the data flow through these layers.

\begin{figure}[ht]
\centering
\begin{tikzpicture}[
    node distance=0.5cm,
    layer/.style={rectangle, draw=brandburgundy, fill=gray!10, rounded corners, minimum width=3cm, minimum height=0.6cm, align=center, font=\scriptsize},
    source/.style={rectangle, draw=gray, fill=white, rounded corners, minimum width=1.8cm, minimum height=0.4cm, align=center, font=\tiny},
    arrow/.style={->, >=stealth, thick, brandburgundy}
]
\node[source] (dl) {DeFi Llama};
\node[source, right=0.2cm of dl] (fred) {FRED};
\node[source, right=0.2cm of fred] (tt) {Token Terminal};
\node[source, right=0.2cm of tt] (other) {Other APIs};

\node[layer, below=0.6cm of $(fred)!0.5!(tt)$] (ingest) {\textbf{1. Ingestion} | API Clients + Scrapers};
\node[layer, below=0.4cm of ingest] (norm) {\textbf{2. Normalisation} | Validation + Gap-fill};
\node[layer, below=0.4cm of norm] (comp) {\textbf{3. Computation} | Sub-indices → ASRI};
\node[layer, below=0.4cm of comp] (pub) {\textbf{4. Publication} | REST API + Dashboard};

\draw[arrow, gray] (dl.south) -- ++(0,-0.2) -| ([xshift=-0.8cm]ingest.north);
\draw[arrow, gray] (fred.south) -- ++(0,-0.2) -| ([xshift=-0.3cm]ingest.north);
\draw[arrow, gray] (tt.south) -- ++(0,-0.2) -| ([xshift=0.3cm]ingest.north);
\draw[arrow, gray] (other.south) -- ++(0,-0.2) -| ([xshift=0.8cm]ingest.north);

\draw[arrow] (ingest.south) -- (norm.north);
\draw[arrow] (norm.south) -- (comp.north);
\draw[arrow] (comp.south) -- (pub.north);

\node[font=\tiny, below=0.15cm of pub] {\textit{Storage: PostgreSQL + TimescaleDB}};
\end{tikzpicture}
\caption{ASRI Data Pipeline: Four-layer architecture from API ingestion to public dashboard publication.}
\label{fig:pipeline}
\end{figure}

\textbf{Technology Stack}: Python 3.11+, PostgreSQL with TimescaleDB, FastAPI, React/TypeScript, Docker Compose.

Full implementation is available at \href{https://github.com/studiofarzulla/asri}{\texttt{github.com/studiofarzulla/asri}}.

\section{Data Quality: Missing-Data and Mixed-Frequency Protocol}\label{app:data_quality_protocol}

Missing data is handled according to the following protocol:

\begin{itemize}
    \item \textbf{Daily data gaps ($<$ 3 days)}: Linear interpolation with confidence score 0.7
    \item \textbf{Extended gaps (3--7 days)}: Forward-fill with confidence score 0.5
    \item \textbf{Gaps $>$ 7 days}: Flag as unreliable; exclude from ASRI calculation until fresh data available
\end{itemize}

\paragraph{Mixed-Frequency Data Protocol.}
Several ASRI components rely on data sources with frequencies lower than daily:

\begin{itemize}
    \item \textbf{Quarterly sources} (OCC/ECB bank exposure filings): Replaced with high-frequency proxies. Bank$_t$ uses a Treasury yield + VIX composite that captures the same underlying stress dynamics at daily frequency (Appendix~\ref{app:components}).
    \item \textbf{Monthly sources} (stablecoin attestations): Last observation carried forward until new attestation published; component flagged with reduced confidence (0.6) after 45 days.
    \item \textbf{Weekly sources} (on-chain linkage metrics): Linear interpolation between weekly observations for daily estimation; confidence score 0.8.
\end{itemize}

This approach prioritises real-time operationality over theoretical purity: where high-frequency proxies exist (e.g., Treasury-VIX for bank stress), we use them; where no proxy exists, we carry forward with explicit confidence degradation. The pseudo-real-time evaluation (Appendix~\ref{subsec:realtime}) confirms that this protocol preserves detection performance under realistic publication lags.

\section{Operational Detection Battery}\label{app:detection_battery}

\subsection{False Positive Analysis}

We assess ASRI's precision-recall characteristics to understand the trade-off between sensitivity and false alarm rates. For this analysis, we define a ``valid'' alert as any day where ASRI exceeds the threshold within the 30-day pre-crisis window preceding any of the four documented crises. Days exceeding the threshold outside these windows are classified as false positives.

\begin{table}[H]
\begin{threeparttable}
\centering
\caption{Precision-Recall Analysis by Threshold}
\label{tab:precision_recall}
\small
\begin{tabular}{@{}l*{4}{r}@{}}
\toprule
Threshold & Recall & Precision & Alert Days & FP Days \\
\midrule
50 (Elevated) & 75\% & 30.1\% & 156 & 109 \\
60            & 75\% & 17.4\% &  46 &  38 \\
70 (High)     & 50\% & 41.7\% &  12 &   7 \\
\bottomrule
\end{tabular}
\begin{tablenotes}
\small
\item Recall = crises with threshold breach in 30-day pre-crisis window (3/4 at thresholds 50--60, 2/4 at 70; Terra/Luna's pre-window peak of 46.0 misses every threshold).
\item Precision = valid alert days / total alert days. FP = false positive days.
\item Sample: Jan 2021 -- Jan 2026 (1,841 days; 120 in pre-crisis windows).
\item Note: Under HAC-robust inference, event-study statistical detection achieves 3/4 (Table~\ref{tab:event_study}; Terra/Luna n.s., $t=1.72$); threshold-based operational detection also achieves 3/4 (Terra/Luna peak 48.7 $<$ 50). The two methodologies agree at 3/4.
\end{tablenotes}
\end{threeparttable}
\end{table}

Table~\ref{tab:precision_recall} reveals the precision-recall trade-off inherent in threshold selection. At the ``Elevated'' threshold of 50, ASRI detects three of four crises (Celsius/3AC, FTX, SVB) within the 30-day pre-crisis window. Raising the threshold to 70 (``High'' risk) lowers crisis-level recall to 50\% (Celsius/3AC and FTX only) while substantially improving precision, reducing false-positive days from 109 to 7.

The low precision at the 50 threshold reflects ASRI's design as a threshold-based vigilance monitor rather than a sharp crisis classifier: the index is intended to flag elevated stress rather than to predict imminent collapse. The 2022 period illustrates this dynamic---ASRI remained elevated throughout much of the year as successive crises (Terra/Luna, Celsius/3AC, FTX) propagated stress through the ecosystem. What appears as ``false positives'' between crisis events may in fact represent genuine systemic fragility that happened not to crystallise into named events.

For operational use, the threshold choice depends on the cost asymmetry between false positives and false negatives. Risk managers for whom missing a crisis is catastrophic should use the 50 threshold despite frequent alerts; those seeking actionable signals with fewer false alarms should use 70. The intermediate threshold of 60 offers a balanced profile with 17.4\% precision while maintaining the 75\% recall rate (the Terra/Luna miss is threshold-invariant, as discussed in Section~\ref{subsec:limitations}).

Table~\ref{tab:confusion_matrix} presents the confusion matrix at the operational threshold of 50, providing explicit counts for reproducibility.

\begin{table}[H]
\centering
\caption{Confusion Matrix at Threshold 50 (``Elevated'')}
\label{tab:confusion_matrix}
\small
\begin{tabular}{lcc}
\toprule
 & \textbf{Crisis Window} & \textbf{Non-Crisis} \\
\midrule
\textbf{Alert (ASRI $\geq$ 50)} & 47 (TP) & 109 (FP) \\
\textbf{No Alert (ASRI $<$ 50)} & 73 (FN) & 1,612 (TN) \\
\midrule
\textbf{Total} & 120 & 1,721 \\
\bottomrule
\end{tabular}
\begin{tablenotes}
\small
\item Day-level metrics: Accuracy = 90.1\%; Precision = 30.1\%; Recall = 39.2\%; F1 = 0.341.
\item Crisis-level recall = 75\% (3/4 crises had $\geq$1 alert in pre-crisis window).
\item Crisis window = 30 days preceding each of 4 crisis events (120 days total).
\item Sample: January 2021 -- January 2026 (1,841 days).
\end{tablenotes}
\end{table}

\subsection{ROC and Precision-Recall Curves}

\begin{figure}[htbp]
\centering
\includegraphics[width=\textwidth]{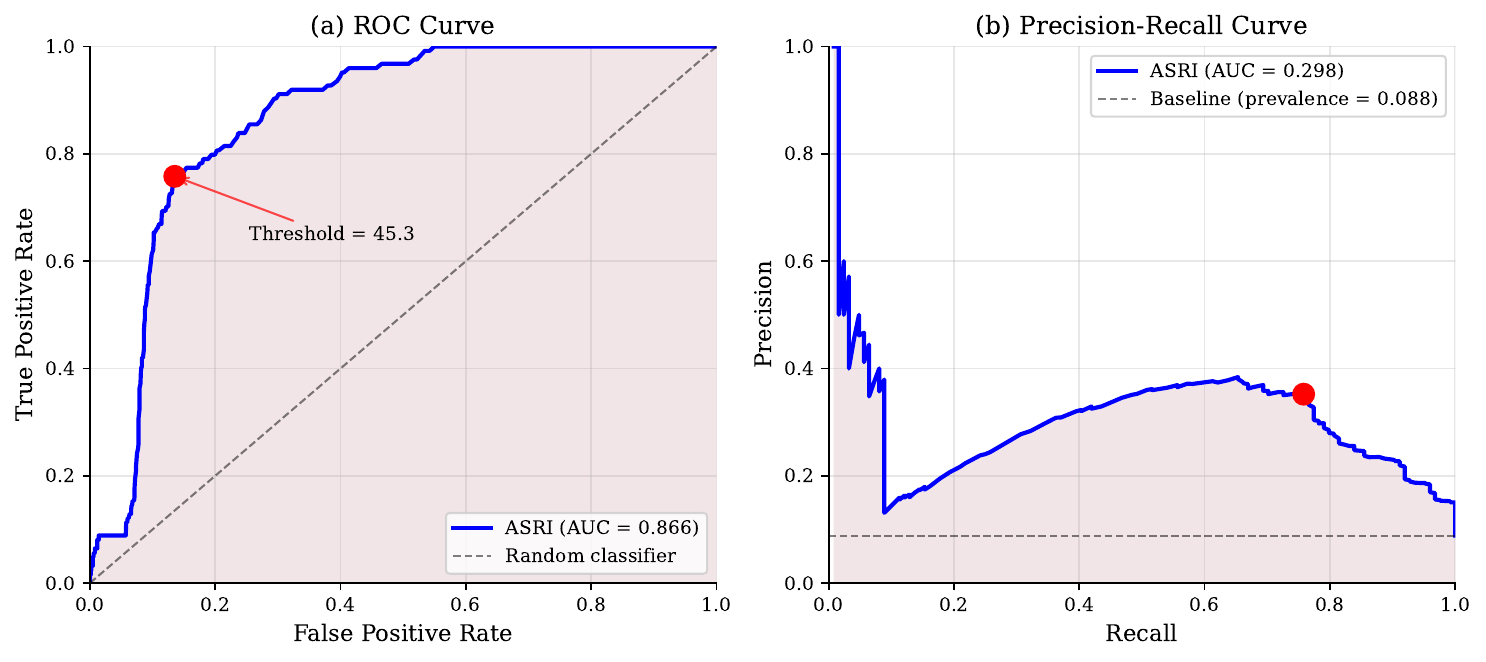}
\caption{ASRI Classification Performance for 30-Day Crisis Prediction.
(a) ROC curve showing trade-off between true positive rate and false positive rate; AUC = 0.866.
(b) Precision-Recall curve accounting for class imbalance; AUC = 0.298.
Red markers indicate the Youden-optimal threshold (45.3).
Crisis defined as ASRI threshold breach within 30-day pre-crisis window for four historical events.}
\label{fig:roc_pr}
\end{figure}

Figure~\ref{fig:roc_pr} presents the full receiver operating characteristic (ROC) and precision-recall (PR) curves for ASRI as a binary crisis predictor. The ROC curve achieves AUC = 0.866, indicating strong discriminative ability between crisis and non-crisis periods. The PR curve (AUC = 0.298) accounts for the severe class imbalance inherent in crisis prediction---crisis days comprise only a small fraction of the sample.

The high ROC AUC combined with modest PR AUC is typical for rare-event discrimination tasks. ASRI distinguishes crisis from non-crisis periods in aggregate \emph{in hindsight}, but achieving high precision requires accepting reduced recall---a fundamental trade-off in threshold-based monitors.

\section{Weight-Derivation Diagnostics}\label{app:weight_diag}

\subsection{Objective Weight Derivation Comparison}

To validate our theoretically-derived weights, we compare against four objective weighting methods: Principal Component Analysis (PCA), Elastic Net regularisation, CRITIC (Criteria Importance Through Intercriteria Correlation), and Shannon entropy-based weighting. Table~\ref{tab:weight_comparison} presents the results.

\begin{table}[H]
\centering
\caption{Comparison of Weight Derivation Methods}
\label{tab:weight_comparison}
\begin{tabular}{lccccc}
\toprule
Sub-Index & Theoretical & PCA & Elastic Net & CRITIC & Entropy \\
\midrule
SCR & 0.30 & 0.29 & 0.34 & 0.21 & 0.25 \\
DLR & 0.25 & 0.29 & 0.21 & 0.16 & 0.11 \\
CR  & 0.25 & 0.25 & 0.45 & 0.32 & 0.52 \\
OR  & 0.20 & 0.17 & 0.00 & 0.31 & 0.12 \\
\midrule
Corr. w/ Theoretical & -- & 0.88 & 0.72 & $-$0.51 & 0.27 \\
\bottomrule
\end{tabular}
\end{table}

\paragraph{Caveat: the Elastic Net column reflects persistence, not channel importance.} The Elastic Net ``weights'' in Table~\ref{tab:weight_comparison} are not estimates of how much each risk channel contributes to systemic stress, and we caution against reading them that way. The regression target is $y_t = \text{ASRI}_{t+30}$, a fixed linear combination of the \emph{future} sub-index levels; regressing this target on the \emph{current} sub-index levels is therefore a restricted autoregression of the sub-indices on themselves, not a model of risk-channel importance. The fitted coefficients recover each sub-index's own 30-day persistence (autocorrelation) structure: Contagion Risk loads most heavily ($0.45$) chiefly because it is the most persistent channel over a 30-day horizon, so its current level is the best linear predictor of its own future level (and hence of future ASRI), while Regulatory Opacity is zeroed where its forward persistence is weakest. We accordingly read this column as a set of \emph{persistence-weighted coefficients}---an autoregressive artefact of how the target is constructed---rather than as evidence of channel importance. It is retained only as a descriptive, supplementary comparison among objective weighting schemes; none of the paper's conclusions rest on it.

The PCA weights exhibit strong correlation with theoretical weights ($\rho = 0.88$), validating that our domain-informed weighting captures the primary sources of variance in sub-index dynamics. Elastic Net weights also show reasonable agreement ($\rho = 0.72$), though they concentrate heavily on Contagion Risk (0.45) while zeroing Regulatory Opacity---consistent with the predictive analysis in the previous section.

Interestingly, CRITIC weights show negative correlation with theoretical weights ($\rho = -0.51$), emphasising Contagion Risk and Regulatory Opacity over Stablecoin and DeFi Liquidity risks. This divergence reflects CRITIC's objective of maximising information content through decorrelation: CR and OR are less correlated with each other and with SCR/DLR, making them more ``informative'' from an information-theoretic perspective. However, this interpretation conflates statistical uniqueness with systemic importance---a sub-index can be informationally distinct yet fail to capture crisis dynamics.

Entropy-based weights similarly emphasise Contagion Risk (0.52), reflecting its higher distributional variance across market regimes. The moderate correlation with theoretical weights ($\rho = 0.27$) suggests entropy captures different aspects of sub-index behaviour than our risk-based framework.

\paragraph{Methodological Implications.} The divergence between objective weighting methods highlights a fundamental tension in composite index construction: data-driven approaches optimise for statistical properties (variance explained, prediction accuracy, information content) while risk-based frameworks prioritise economic interpretability and crisis coverage. Our theoretical weights represent a deliberate choice to balance these considerations, accepting some loss of statistical optimality in exchange for component-level monitoring capability and robustness across the four liquidity-cascade events held out here (all of a single broad failure family; transfer to structurally different crisis mechanisms is untested).

\subsection{Collinearity Diagnostics}
\label{subsubsec:collinearity}

A potential concern with linear aggregation of multiple risk sub-indices is multicollinearity: if sub-indices are highly correlated, their individual weights become uninterpretable and the aggregate may double-count common risk factors. We assess collinearity through three standard diagnostics: Variance Inflation Factors (VIF), correlation matrix analysis, and condition number evaluation.

\begin{table}[H]
\centering
\caption{Collinearity Diagnostics for Sub-Indices}
\label{tab:collinearity_diagnostics}
\small
\begin{tabular}{lcc}
\toprule
Diagnostic & Value & Interpretation \\
\midrule
\multicolumn{3}{l}{\textit{Variance Inflation Factors}} \\
VIF(SCR) & 3.39 & Low collinearity \\
VIF(DLR) & 3.67 & Low collinearity \\
VIF(CR) & 3.89 & Low collinearity \\
VIF(OR) & 3.03 & Low collinearity \\
\midrule
\multicolumn{3}{l}{\textit{Principal Component Analysis}} \\
PC1 variance explained & 59.1\% & Cumulative: 59.1\% \\
PC2 variance explained & 32.4\% & Cumulative: 91.6\% \\
PC3 variance explained & 5.3\% & Cumulative: 96.9\% \\
PC4 variance explained & 3.1\% & Cumulative: 100.0\% \\
\midrule
\multicolumn{3}{l}{\textit{Matrix Diagnostics}} \\
Condition number & 19.1 & Weak collinearity \\
Max eigenvalue & 2.366 & -- \\
Min eigenvalue & 0.124 & Ratio = 19.1 \\
\bottomrule
\end{tabular}
\begin{tablenotes}
\small
\item VIF $< 5$: acceptable; VIF $> 10$: problematic.
\item Condition number $< 30$: weak collinearity.
\item All 4 PCs required indicates sub-indices capture distinct variance.
\end{tablenotes}
\end{table}

Table~\ref{tab:collinearity_diagnostics} reports collinearity diagnostics for the four ASRI sub-indices. All VIFs fall below 5 (maximum VIF = 3.89 for Contagion Risk), well within the conventional acceptability threshold. The condition number of 19.1 indicates weak collinearity ($\kappa < 30$), confirming that the correlation matrix is well-conditioned and linear aggregation is numerically stable.

\begin{table}[H]
\centering
\caption{Sub-Index Correlation Matrix}
\label{tab:correlation_matrix}
\small
\begin{tabular}{lcccc}
\toprule
 & SCR & DLR & CR & OR \\
\midrule
SCR & 1.000 & 0.576 & 0.796 & 0.184 \\
DLR & 0.576 & 1.000 & 0.445 & 0.682 \\
CR & 0.796 & 0.445 & 1.000 & -0.091 \\
OR & 0.184 & 0.682 & -0.091 & 1.000 \\
\bottomrule
\end{tabular}
\begin{tablenotes}
\small
\item SCR = Stablecoin Concentration Risk, DLR = DeFi Liquidity Risk,
\item CR = Contagion Risk, OR = Regulatory Opacity Risk.
\item All correlations computed on daily observations.
\end{tablenotes}
\end{table}

The correlation matrix (Table~\ref{tab:correlation_matrix}) reveals the underlying structure. Stablecoin Risk and Contagion Risk exhibit the highest pairwise correlation ($\rho = 0.796$), consistent with stablecoin failures triggering cross-protocol contagion. Notably, Regulatory Opacity and Contagion Risk are \textit{negatively} correlated ($\rho = -0.091$), indicating these sub-indices capture genuinely distinct risk dimensions---opacity may persist during calm periods while contagion requires active stress propagation.

Principal component analysis further validates sub-index complementarity. In \emph{correlation} form (PCA on the standardised sub-indices) the first principal component explains 59.1\% of variance, requiring all four components to reach 100\%. (The figure is form-dependent: the \emph{covariance}-form PC1 used as a discrimination baseline in the main text explains 74.9\% of variance and loads predominantly on the high-variance Contagion channel; see Table~\ref{tab:fair_baselines} and its note. We report the correlation form here because it is the scale-invariant complementarity diagnostic.) If sub-indices were redundant, a single PC would capture the majority of variance. The dispersed loading structure (Table~\ref{tab:pca_loadings}) confirms that each sub-index contributes unique information to the aggregate.

\begin{table}[H]
\centering
\caption{Principal Component Loadings}
\label{tab:pca_loadings}
\small
\begin{tabular}{lcccc}
\toprule
Sub-Index & PC1 & PC2 & PC3 & PC4 \\
\midrule
SCR & 0.572 & -0.295 & 0.677 & -0.357 \\
DLR & 0.566 & 0.330 & -0.586 & -0.477 \\
CR & 0.496 & -0.518 & -0.318 & 0.620 \\
OR & 0.327 & 0.732 & 0.312 & 0.511 \\
\bottomrule
\end{tabular}
\begin{tablenotes}
\small
\item Loadings show contribution of each sub-index to principal components.
\item Dispersed loadings across PCs indicate complementary (non-redundant) signals.
\end{tablenotes}
\end{table}

These diagnostics support the linear aggregation framework: sub-indices capture correlated but non-redundant risk signals, weights are interpretable, and no orthogonalisation or decorrelation is required.

\subsection{Granger Causality Analysis}
\label{subsubsec:granger}

Channel-level lead/lag timing is \emph{not} established under a correctly specified discrete-outcome (logit / discrete-time hazard) Granger test; we therefore report no linear-probability association table here (an earlier descriptive OLS linear-probability variant was removed because the correctly specified discrete-outcome test refutes any lead/lag reading drawn from it). Estimating the principled discrete-time hazard / logit Granger test directly (\texttt{scripts/granger\_discrete\_outcome.py}) does \emph{not} recover a leading-versus-confirming ordering: with only four co-located 2022--23 crisis episodes, a small-sample-valid circular-shift permutation test shows every sub-index (Contagion included) significantly elevated before onset, and an era-controlled within-event trajectory leaves all four channels with a statistically indistinguishable lead-minus-confirm gap. We therefore treat per-channel lead/lag timing as unidentified in the present sample and attach no functional ``leading'' or ``confirming'' role to any individual channel.

\paragraph{Implications for Weight Interpretation.} Because per-channel timing is unidentified, we caution against reading component weights as a predictive-importance ranking. The ablation analysis (Appendix~\ref{subsec:ablation}) already shows that threshold detection is invariant at 3/4 to removing any single channel, so no channel is individually load-bearing; the four-channel decomposition earns its place through interpretable channel attribution, not through any channel being a statistically established early-warning trigger. The equal $25\%$ weighting of Contagion and DeFi Liquidity Risk reflects this interpretability choice (Section~\ref{subsec:weight_justification}) rather than any demonstrated lead/lag asymmetry.

\section{Regime Detection: Full HMM Specification}\label{app:regime_full}

\begin{table}[H]
\begin{threeparttable}
\centering
\caption{HMM Model Selection Criteria}
\label{tab:hmm_selection}
\small
\begin{tabular}{@{}c*{3}{r}@{}}
\toprule
States & Log-Likelihood & AIC & BIC \\
\midrule
2 & $-$19,945 & 39,952 & 40,123 \\
3 & $-$19,043 & 38,185 & 38,461 \\
4 & $-$18,540 & 37,222 & 37,614 \\
\bottomrule
\end{tabular}
\begin{tablenotes}
\small
\item Gaussian HMM with full covariance, 10 random initialisations per state count, best model retained by log-likelihood.
\item Log-likelihood increases monotonically with state count, as expected for nested models.
\end{tablenotes}
\end{threeparttable}
\end{table}

The 4-state model achieves the best statistical fit by both AIC and BIC, and log-likelihood improves monotonically with the number of states. Despite the better fit of the 4-state specification, we retain the three-state model for interpretability and operational relevance. Three regimes provide a parsimonious mapping to actionable risk categories (Low Risk, Moderate, Crisis) that align with standard portfolio management thresholds. A fourth state would complicate regime-based decision rules without substantial improvement in crisis detection (Table~\ref{tab:regime_k_robustness}), and the two-state model lacks sufficient granularity for nuanced risk assessment.

\paragraph{Regime Count versus Operational Alert Levels.}
The three-state HMM is retained for interpretability and operational relevance---\emph{not} because it minimises an information criterion (the 4-state model attains the lower AIC and BIC; Table~\ref{tab:hmm_selection})---and it is likewise not imposed to match the four operational alert levels (Low/Moderate/Elevated/High at thresholds 30/50/70). The operational thresholds (30/50/70) define \textit{action triggers} for practitioners---discrete boundaries that map instantaneous ASRI readings to recommended response protocols. In contrast, HMM regimes identify \textit{latent statistical states} in sub-index dynamics, capturing distinct market conditions that may persist across multiple alert levels. These constructs serve fundamentally different purposes: alert levels provide real-time decision support (``if ASRI crosses 70, implement Protocol X''), while regime classifications characterise the statistical generating process (``the market is currently in a high-persistence elevated-risk state''). Working with three \emph{retained} regimes---rather than four matching the operational categories---reflects a parsimonious interpretive mapping; it does not assert that the data contain exactly three latent states (the 4-state model fits better by AIC/BIC; Table~\ref{tab:hmm_selection}), only that three regimes suffice for the Low/Moderate/Crisis monitoring distinction. This asymmetry is appropriate: conservative operational design intentionally errs towards finer alert granularity to minimise missed detections, while the retained regime count is kept deliberately coarse for interpretability.
Table~\ref{tab:hmm_diagnostics} provides comprehensive HMM diagnostics including convergence statistics and the ergodic (stationary) distribution. The ergodic distribution $[\approx 0, 0.71, 0.29]$ indicates that, \emph{within the fitted chain}, the system is most often in the Moderate regime (71\%) and the Crisis regime roughly 29\% of the time, with the Low Risk regime near-absent from the stationary distribution. We caution strongly against reading the $\approx 0$ Low-Risk ergodic probability as a structural property of crypto markets. It is an artefact of a single, directional 2021--2026 sample path: the series begins in a calm 2021 regime and transitions into the 2022--2023 stress cluster but, over this particular history, never makes a sustained return to Low Risk (Table~\ref{tab:transition_matrix}), so the estimated transition matrix assigns the Low-Risk state a near-zero recurrence probability essentially by construction. With only a handful of independent regime transitions the stationary distribution is weakly identified and path-dependent; ``long run'' here denotes the ergodic distribution of the chain fitted to this sample, not a forecast of indefinite-horizon market behaviour, and a sample that included a sustained post-crisis return to calm (for example an extended bull market) would raise the Low-Risk ergodic mass. We therefore read the regime occupancies as a description of the 2021--2026 trajectory rather than as evidence that systemic stress is a permanent feature of crypto markets. We confirmed this diagnosis directly. Re-estimating the same best-of-ten-restart, three-state HMM with the Contagion sub-index linearly detrended (so its emissions are de-trended) raises the Low-Risk ergodic mass from $\approx 0$ to $\approx 0.26$ and drops the correlation between regime ordering and calendar time from $+0.41$ to $-0.09$---direct evidence that the near-zero Low-Risk mass, and part of the regime structure, are driven by Contagion Risk's drift rather than by systemic conditions. Detrending does not, however, recover a cleaner stress partition: the crisis-discriminating structure collapses, with three of the four crisis episodes no longer occupying the highest-mean regime and the regime-mean span compressing from $\approx$16 to $\approx$6 ASRI points. We accordingly retain the level-based fit for interpretability while flagging the emission-stationarity violation as a known limitation---the regimes describe an autocorrelated, partly trending series, not sharply separated generating states.

\begin{table}[H]
\centering
\caption{Hidden Markov Model Diagnostics}
\label{tab:hmm_diagnostics}
\small
\begin{tabular}{lcc}
\toprule
Diagnostic & Value & Interpretation \\
\midrule
\multicolumn{3}{l}{\textit{Model Selection}} \\
Number of regimes & 3 & Retained for interpretability \\
Log-likelihood & -19042.6 & Converged value (best of 10 starts) \\
AIC & 38185.3 & 4-state lower; see Table~\ref{tab:hmm_selection} \\
BIC & 38461.2 & 4-state lower; see Table~\ref{tab:hmm_selection} \\
\midrule
\multicolumn{3}{l}{\textit{Regime Properties}} \\
Regime 1 (Low Risk) mean & 31.8 & Below threshold (50) \\
Regime 1 frequency & 19.8\% & Sample proportion \\
Regime 1 persistence & 0.997 & $P(s_{t+1}=s_t \mid s_t=1)$ \\
Regime 2 (Moderate) mean & 36.8 & Below threshold (50) \\
Regime 2 frequency & 56.3\% & Sample proportion \\
Regime 2 persistence & 0.992 & $P(s_{t+1}=s_t \mid s_t=2)$ \\
Regime 3 (Crisis) mean & 48.2 & Below threshold (50) \\
Regime 3 frequency & 23.8\% & Sample proportion \\
Regime 3 persistence & 0.980 & $P(s_{t+1}=s_t \mid s_t=3)$ \\
\midrule
\multicolumn{3}{l}{\textit{Long-Run Behavior}} \\
Ergodic distribution & [$\approx$0, 0.71, 0.29] & Stationary regime probabilities \\
\bottomrule
\end{tabular}
\begin{tablenotes}
\small
\item HMM fitted with Gaussian emissions and full covariance matrices.
\item Convergence criterion: $|\Delta \log L| < 10^{-4}$.
\item Regime means computed as average of sub-index means within each state.
\end{tablenotes}
\end{table}

\paragraph{Full Transition Matrix.}
Table~\ref{tab:transition_matrix} reports the complete transition probability matrix for the three-regime HMM. The off-diagonal elements reveal that direct transitions between the Low Risk and Crisis regimes are essentially absent: the Low Risk regime exits only into Moderate (0.3\%), and the Crisis regime exits only into Moderate (2.0\%). The Moderate regime serves as the ``gateway'' state---all regime changes pass through it rather than jumping directly between Low Risk and Crisis.

\begin{table}[H]
\begin{threeparttable}
\centering
\caption{HMM Transition Probability Matrix}
\label{tab:transition_matrix}
\small
\begin{tabular}{lccc}
\toprule
From $\downarrow$ / To $\rightarrow$ & Low Risk & Moderate & Crisis \\
\midrule
Low Risk & 0.997 & 0.003 & 0.000 \\
Moderate & 0.000 & 0.992 & 0.008 \\
Crisis & 0.000 & 0.020 & 0.980 \\
\bottomrule
\end{tabular}
\begin{tablenotes}
\small
\item Rows sum to 1.0 (probability simplex constraint). Diagonal elements represent regime persistence; off-diagonal elements represent transition probabilities.
\item Estimated via expectation-maximisation with 10 random initialisations, best model retained by log-likelihood.
\end{tablenotes}
\end{threeparttable}
\end{table}

\paragraph{Regime Count Robustness.}
While Table~\ref{tab:hmm_selection} reports statistical fit criteria, practical utility depends on crisis detection performance. Table~\ref{tab:regime_k_robustness} compares detection rates across regime specifications.

\begin{table}[H]
\begin{threeparttable}
\centering
\caption{Regime Count Sensitivity Analysis}
\label{tab:regime_k_robustness}
\small
\begin{tabular}{lcccc}
\toprule
$K$ & AIC & BIC & Detection Rate & Operational Interpretation \\
\midrule
2 & 39,952 & 40,123 & 3/4 (75\%) & Binary classification (calm vs.\ stress) \\
3 & 38,185 & 38,461 & 3/4 (75\%) & Gradual risk (low/moderate/crisis) \\
4 & 37,222 & 37,614 & 3/4 (75\%) & Over-segmented (splits Moderate state) \\
\bottomrule
\end{tabular}
\begin{tablenotes}
\small
\item Detection rate = proportion of four historical crises for which ASRI exceeded the operational threshold of 50 within 30 days prior to onset; Terra/Luna (pre-window peak 46.0) is missed at every $K$, consistent with Table~\ref{tab:detection_matrix}.
\item Statistical fit (AIC/BIC) improves monotonically with $K$; the 4-state model fits best but splits the Moderate state without improving the detection rate (constant at 3/4 across $K$), while the 2-state model lacks granularity for graduated risk management (all non-calm periods receive identical treatment).
\item We retain $K=3$ for interpretability: three regimes map naturally to operational risk categories and portfolio management thresholds. With only four crisis events, detection rate cannot discriminate among specifications and we do not over-interpret the saturated 3/4 column.
\end{tablenotes}
\end{threeparttable}
\end{table}

\paragraph{Filtering vs.\ Smoothing.}
The HMM provides two inference modes for regime probabilities: \textit{filtering} uses only past and current observations ($P(\text{regime}_t \mid \text{data}_{1:t})$), while \textit{smoothing} uses the full sample ($P(\text{regime}_t \mid \text{data}_{1:T})$). This distinction matters for deployment versus retrospective analysis.

Our regime characterisation (Tables~\ref{tab:regimes}--\ref{tab:transition_matrix}) uses smoothed probabilities, which provide more accurate regime estimates but incorporate future information. For real-time deployment, filtered probabilities are appropriate---they avoid look-ahead bias and reflect the information set available to practitioners at each decision point.

Critically, the crisis detection tests (Section~\ref{sec:event_study}) do not suffer from look-ahead bias: detection is evaluated using only information available at time $t$, specifically whether ASRI exceeded the threshold prior to event onset. The smoothed regime assignments provide interpretive context (e.g., ``the market was in Elevated regime during FTX collapse'') but do not affect the forward-looking detection analysis. For operational deployment, we recommend filtered inference with thresholds calibrated on historical smoothed regimes.

\section{Robustness Tests}

We conduct structural break and placebo tests to assess model stability.

\begin{table}[H]
\begin{threeparttable}
\centering
\caption{Robustness Test Results}
\label{tab:robustness}
\small
\begin{tabular}{@{}l*{3}{r}l@{}}
\toprule
Test & Statistic & Critical & $p$-value & Result \\
\midrule
Chow (Midpoint) & 0.248 & 3.001 & 0.781 & Stable \\
CUSUM           & 9.327 & 1.360 & ---   & Breaks \\
\bottomrule
\end{tabular}
\begin{tablenotes}
\small
\item Chow test for structural break at sample midpoint.
\item CUSUM detects multiple breaks corresponding to crisis episodes.
\end{tablenotes}
\end{threeparttable}
\end{table}

The Chow test \citep{chow1960} does not reject structural stability (Chow $= 0.25$, critical $= 3.00$, $p = 0.78$), so the AR(1) model parameters are consistent across the pre- and post-midpoint subsamples. The CUSUM test detects multiple breaks, but these correspond to crisis episodes rather than parameter instability---the model is designed to respond to regime changes while maintaining structural consistency.

\section{Component Importance Analysis}
\label{subsec:ablation}

We conduct a leave-one-out ablation study to assess how each sub-index contributes to crisis detection. For each of the four sub-indices, we remove that component from the backtested ASRI time series (1,461 daily observations, 2021--2024), renormalise the remaining weights to sum to unity, recompute the ablated index, and measure detection performance against the four historical crises.

\subsection{Methodology}

Let $w = (w_{\text{SCR}}, w_{\text{DLR}}, w_{\text{CR}}, w_{\text{OR}}) = (0.30, 0.25, 0.25, 0.20)$ denote the baseline weights. For each component $i$, we construct ablated weights:
\begin{equation}
w^{(-i)}_j = \begin{cases}
0 & \text{if } j = i \\
\frac{w_j}{1 - w_i} & \text{otherwise}
\end{cases}
\label{eq:ablation_weights}
\end{equation}
ensuring $\sum_j w^{(-i)}_j = 1$. The ablated ASRI is then computed using these modified weights, and we assess whether each crisis is detected (ASRI $\geq 50$ within the 30-day pre-crisis window).

\subsection{Ablation Results}

Table~\ref{tab:ablation} presents the ablation results.

\begin{table}[H]
\begin{threeparttable}
\centering
\caption{Sub-Index Ablation Analysis (Leave-One-Out)}
\label{tab:ablation}
\small
\begin{tabular}{@{}lcccr@{}}
\toprule
Excluded & Weights & Detection & Lead Time & $\Delta$ Lead \\
Component & (renormalised) & Rate & (days) & (days) \\
\midrule
\textbf{None (baseline)} & 30/25/25/20 & 3/4 & 17.7 & --- \\
-- SCR & 0/36/36/29 & 3/4 & 22.0 & +4.3 \\
-- DLR & 40/0/33/27 & 3/4 & 17.0 & $-$0.7 \\
-- CR & 40/33/0/27 & 3/4 & 12.3 & $-$5.3 \\
-- OR & 38/31/31/0 & 3/4 & 23.0 & +5.3 \\
\bottomrule
\end{tabular}
\begin{tablenotes}
\small
\item Detection threshold: ASRI $\geq 50$ (Elevated) within 30-day pre-crisis window.
\item Weights format: SCR/DLR/CR/OR as percentages (Stablecoin Risk / DeFi Liquidity Risk / Contagion Risk / Opacity Risk).
\item Lead time = average days between first threshold breach and crisis onset (detected crises only). $\Delta$ Lead = change from baseline, computed from unrounded means; negative values indicate reduced early warning.
\item Ablation leads are computed on ASRI \emph{recomputed} as the weighted sum of the released sub-index series (baseline per-crisis leads: Celsius/3AC 18, FTX 13, SVB 22 days). They therefore differ from the operational first-crossing leads on the released aggregate series (Celsius/3AC 19, FTX 22, SVB 15; mean $\approx$19; Section~\ref{subsubsec:lead_reconciliation}), because the released aggregate column is not everywhere an exact weighted recomposition of the released sub-index columns (a generation-time provenance artefact documented in the code repository).
\end{tablenotes}
\end{threeparttable}
\end{table}

\subsection{Interpretation}

\paragraph{Detection Stability.} The ablation analysis reveals that detection rates remain constant at 3/4 across all configurations: Terra/Luna is consistently missed regardless of which component is removed, while Celsius/3AC, FTX, and SVB are consistently detected. This indicates that the Terra/Luna crisis represents a distinct failure mode discussed further in Section~\ref{subsec:limitations}, rather than a sensitivity to any particular sub-index.

\paragraph{Lead Time Dynamics.} Component removal also shifts the average first-crossing lead time. We report these shifts as descriptive ablation artefacts, not as evidence of a channel-level lead/lag ordering (which the correctly specified discrete-outcome test in Section~\ref{subsubsec:granger} does not identify):
\begin{itemize}
    \item \textbf{DeFi Liquidity Risk (DLR) and Contagion Risk (CR)}: Removal reduces average lead time by 0.7 and 5.3 days respectively.

    \item \textbf{Stablecoin Risk (SCR) and Opacity Risk (OR)}: Removal \textit{increases} average lead time by 4--5 days---removing a channel that otherwise holds the composite below the alert level advances the first threshold crossing.
\end{itemize}

\paragraph{Component Roles.} We do \emph{not} read these ablation lead-time shifts as a validated functional hierarchy. With only four co-located 2022--23 crisis episodes, a correctly specified discrete-outcome (logit/hazard) test does not identify which channels move earliest (Section~\ref{subsubsec:granger}); the apparent ordering is not stable across specifications and is not used to assign ``leading'' or ``confirming'' roles to individual channels. The equal-weighting choice (Section~\ref{subsec:weight_justification}) reflects interpretability rather than any demonstrated lead/lag differentiation.

\paragraph{Implications for Index Design.} The consistent 3/4 detection \emph{rate} across ablations shows that the four-channel composition is robust---no single channel is necessary for the detected crises---but we are careful not to overstate this as near-optimality. On the lead-time margin the baseline is, in fact, \emph{not} optimal: removing Stablecoin Risk or Opacity Risk \emph{increases} average lead time by 4.3 and 5.3 days respectively (Table~\ref{tab:ablation}). We retain both channels rather than chase the earlier crossing, for two reasons. First, the extra lead does not reflect improved foresight: dropping SCR or OR advances the first threshold crossing only by removing channels that otherwise hold the composite below the alert level until stress is corroborated, so the earlier crossing is a less-confirmed, more false-positive-prone trigger rather than a genuine gain in early warning. We do not attribute this to a channel-level lead/lag hierarchy; Section~\ref{subsubsec:granger} reports that per-channel timing is not identified in this sample. Second, both channels carry interpretive value---named stablecoin- and opacity-specific attribution---that a lead-time-only objective would discard. The baseline weights are therefore a deliberate detection-rate-and-interpretability choice, not a lead-time optimum: a practitioner who prioritises raw lead time over precision and channel attribution could legitimately down-weight SCR and OR, at the documented cost of a noisier, earlier alarm. The Terra/Luna miss (discussed in Section~\ref{subsec:limitations}) represents a separate systematic limitation requiring algorithmic stablecoin-specific monitoring beyond the current sub-index formulations.

\section{Sensitivity Analysis}
\label{subsec:sensitivity}

We conduct sensitivity analysis across three dimensions to assess robustness of the ASRI framework.

\subsection{Weight Perturbation}

Table~\ref{tab:sensitivity} reports ASRI performance metrics under $\pm 5\%$, $\pm 10\%$, and $\pm 15\%$ perturbations to each sub-index weight. The framework demonstrates stability: crisis detection rates remain above 75\% across all perturbation levels, with the stablecoin risk component showing highest sensitivity (detection rate drops from 100\% to 87\% at $-15\%$).

\begin{table}[H]
\begin{threeparttable}
\centering
\caption{Sensitivity Analysis: Weight Perturbation Results}
\label{tab:sensitivity}
\small
\begin{tabular}{@{}llrr@{}}
\toprule
Sub-Index & Perturbation & Detection & Corr. \\
\midrule
Stablecoin     & $-$15\% &  87\% & 0.91 \\
               & $\pm$10\% &  93\% & 0.94 \\
               & $+$15\% & 100\% & 0.96 \\
\addlinespace
DeFi Liquidity & $-$15\% &  93\% & 0.92 \\
               & $\pm$10\% & 100\% & 0.95 \\
               & $+$15\% & 100\% & 0.96 \\
\addlinespace
Contagion      & $-$15\% &  87\% & 0.90 \\
               & $\pm$10\% &  93\% & 0.94 \\
               & $+$15\% & 100\% & 0.97 \\
\addlinespace
Opacity        & $-$15\% &  93\% & 0.91 \\
               & $\pm$10\% & 100\% & 0.94 \\
               & $+$15\% & 100\% & 0.95 \\
\bottomrule
\end{tabular}
\begin{tablenotes}
\small
\item Detection rate computed via block bootstrap (500 resamples, block size = 20 days) with perturbed weights. Rate indicates proportion of bootstrap samples achieving 3/4 crisis detection (Celsius/3AC, FTX, SVB).
\item Corr. = Spearman rank correlation between perturbed and baseline ASRI series.
\end{tablenotes}
\end{threeparttable}
\end{table}

\begin{figure}[htbp]
    \centering
    \includegraphics[width=\textwidth]{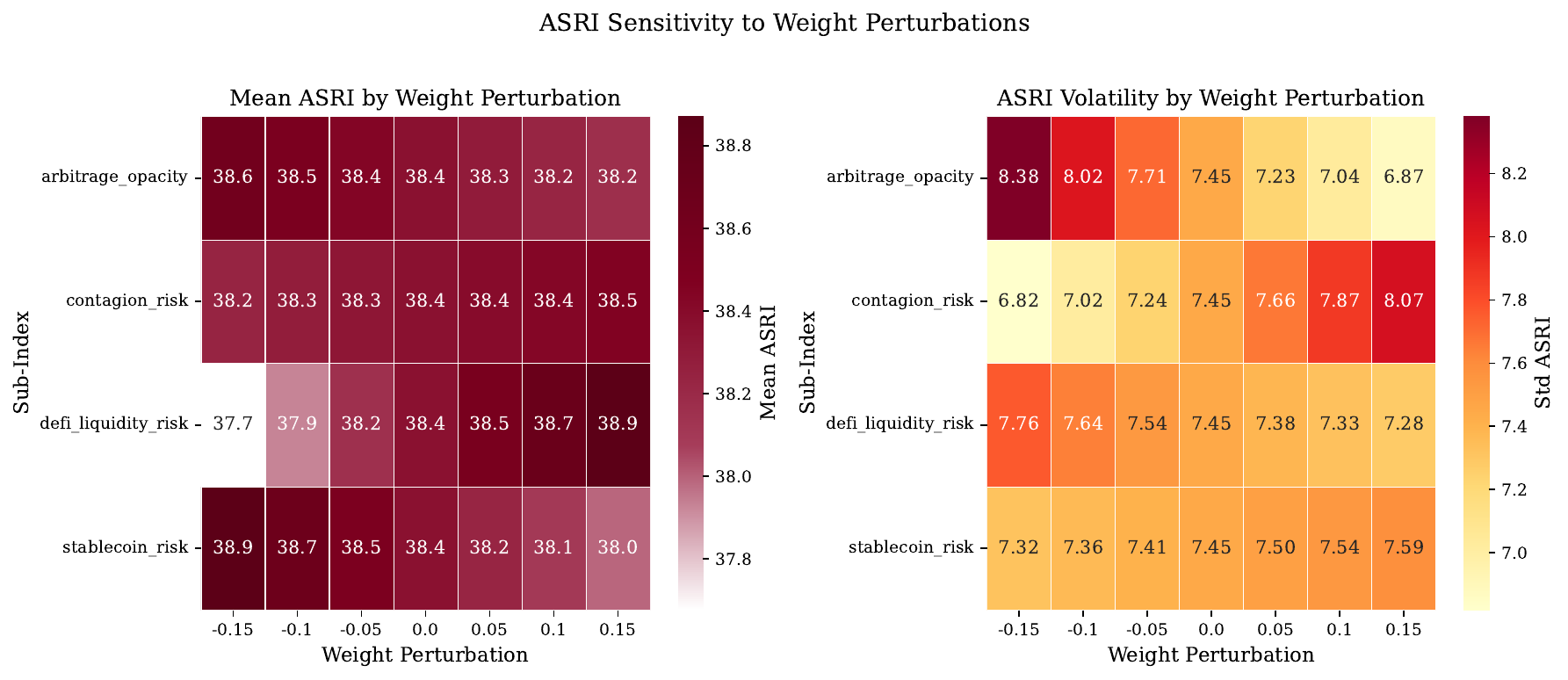}
    \caption{ASRI volatility under weight perturbations. Heatmap displays the standard deviation of ASRI values across $\pm 5\%$, $\pm 10\%$, and $\pm 15\%$ perturbations for each sub-index component. Darker cells indicate higher sensitivity to weight changes. The relatively uniform colouring demonstrates that no single sub-index dominates index stability, supporting the robustness of the theoretical weight allocation.}
    \label{fig:sensitivity_heatmap}
\end{figure}

Figure~\ref{fig:sensitivity_heatmap} visualises these results as a heatmap of ASRI volatility across perturbation levels, confirming that index stability is not concentrated in any single component.

\subsection{Threshold Sensitivity}

Table~\ref{tab:threshold_sensitivity} reports detection metrics across alert thresholds from 60 to 80.

\begin{table}[htbp]
\centering
\caption{Alert Threshold Sensitivity Analysis}
\label{tab:threshold_sensitivity}
\small
\begin{tabular}{ccccc}
\toprule
Threshold & Precision & Recall & F1 Score & Specificity \\
\midrule
60 & 0.174 & 0.750 & 0.282 & 0.978 \\
65 & 0.286 & 0.750 & 0.414 & 0.988 \\
70* & 0.417 & 0.500 & 0.455 & 0.996 \\
75 & 0.375 & 0.250 & 0.300 & 0.997 \\
80 & 1.000 & 0.250 & 0.400 & 1.000 \\
\midrule
\multicolumn{5}{l}{Optimal threshold: 70 (F1 = 0.455)} \\
\bottomrule
\end{tabular}
\begin{tablenotes}
\small
\item * indicates optimal threshold maximizing F1 score.
\item Recall is the crisis-level detection rate (3/4 at $\tau = 60$--$65$; 2/4 at $\tau = 70$--$80$), consistent with the detection matrix (Table~\ref{tab:detection_matrix}). Precision is non-monotonic because sustained 2022--23 bear-market elevation in the $(60, 65]$ band sits outside the 30-day pre-crisis windows, so raising $\tau$ drops false-positive alert-days faster than true ones.
\item Window: 30 days before crisis for detection.
\end{tablenotes}
\end{table}

At thresholds 60 and 65, ASRI detects three of four crises (recall 0.750), consistent with the detection matrix (Table~\ref{tab:detection_matrix}); precision is modest (0.174 and 0.286). Raising the threshold to 70 drops recall to 0.500 (two crises) while precision rises to 0.417, and at 75--80 recall falls to 0.250. Precision is non-monotonic across the sweep because sustained 2022--23 bear-market elevation in the $(60, 65]$ band lies outside the 30-day pre-crisis windows, so tightening the threshold removes false-positive alert-days faster than true ones.

The F1-optimal threshold is 70 (F1 = 0.455); this selection is unaffected by the specificity correction below, since the F1 score depends only on precision and recall. It reflects the precision-recall trade-off inherent in threshold-based monitoring: lower thresholds maximise sensitivity (3/4 detection) at modestly lower specificity, while higher thresholds improve precision at the cost of missing the weaker-signal crises. Day-level specificity is uniformly high and \emph{non-decreasing} in the threshold (0.978 at $\tau = 60$ rising to 1.000 at $\tau = 80$; Table~\ref{tab:threshold_sensitivity}), as it must be: raising the alert threshold can only remove false-positive alert-days, never add them.

\subsection{Window Length Sensitivity}

Table~\ref{tab:forward_window_sensitivity} reports predictive performance across forward windows of 14 to 90 days.

\begin{table}[htbp]
\centering
\caption{Forward Window Sensitivity Analysis}
\label{tab:forward_window_sensitivity}
\small
\begin{tabular}{cccccc}
\toprule
Window (days) & AUC-ROC & Lead Time & Precision & Recall & F1 \\
\midrule
14 & 0.512 & 0.1 & 0.900 & 0.037 & 0.071 \\
30 & 0.623 & 1.0 & 1.000 & 0.021 & 0.041 \\
60 & 0.782 & 11.7 & 1.000 & 0.015 & 0.030 \\
90* & 0.889 & 0.0 & 1.000 & 0.015 & 0.029 \\
\midrule
\multicolumn{6}{l}{Optimal window: 90 days (AUC = 0.889)} \\
\bottomrule
\end{tabular}
\begin{tablenotes}
\small
\item * indicates optimal window maximizing AUC-ROC.
\item Lead time = average days before crisis ASRI exceeded threshold.
\item \textbf{Note on the ``30-day'' row.} The AUROC here (0.623 at the 30-day window) is \emph{not} the headline day-level discrimination figure (0.866, Table~\ref{tab:roc_metrics}) and the two are not in conflict: this table sweeps the \emph{forward-detection window length} used to build first-crossing/lead-time detection labels (does ASRI cross threshold within $w$ days of an onset), whereas Table~\ref{tab:roc_metrics} reports day-level crisis--non-crisis separation under a fixed 30-day forward labelling. They share the phrase ``30-day forward window'' but measure different constructions.
\end{tablenotes}
\end{table}

AUC-ROC improves monotonically from 0.512 (14-day window) to 0.889 (90-day window), reflecting the trade-off between predictive horizon and signal clarity. Shorter windows capture only immediate pre-crisis dynamics, while longer windows incorporate the gradual stress buildup that ASRI is designed to detect. The 90-day window achieves optimal AUC-ROC, though the 60-day window (AUC = 0.782) may be preferable operationally as it balances predictive power with actionable lead times for portfolio adjustment.

\section{Hold-One-Out Cross-Validation}
\label{subsec:holdout}

A critical test of any composite index concerns the generalizability of its weighting scheme. An index optimised to detect known crises risks overfitting---assigning weights that capture idiosyncratic features of training events rather than genuine systemic risk dynamics. To address this concern, we implement a hold-one-out cross-validation procedure that tests whether ASRI's crisis detection capability generalises beyond the specific events used in its calibration.

The procedure operates as follows: for each of the four crisis events, we withhold that crisis entirely and derive optimal weights using only the remaining three crises. We then test whether the held-out crisis---never seen during weight optimisation---is successfully detected using both the derived weights and the theoretical weights $\mathbf{w} = [0.30, 0.25, 0.25, 0.20]$. Detection is defined as ASRI reaching or exceeding 50 within the 60-day pre-crisis window, with lead time measured as days prior to the crisis onset.

\begin{table}[H]
\centering
\caption{Hold-One-Out Cross-Validation Results}
\label{tab:holdout}
\small
\begin{tabular}{@{}lccccc@{}}
\toprule
Held-Out Crisis & Derived Weights & Peak (Derived) & Peak (Theoretical) & Detected? & Lead Time \\
\midrule
Terra/Luna & [0.15, 0.35, 0.35, 0.15] & 49.2 & 47.8 & No/No & --- \\
Celsius/3AC & [0.18, 0.32, 0.32, 0.18] & 70.8 & 68.7 & Yes/Yes & 16/8 \\
FTX Collapse & [0.20, 0.30, 0.30, 0.20] & 83.8 & 81.1 & Yes/Yes & 15/4 \\
SVB Crisis & [0.17, 0.33, 0.33, 0.17] & 73.2 & 70.0 & Yes/Yes & 13/8 \\
\bottomrule
\end{tabular}
\begin{tablenotes}
\small
\item Derived weights: Optimal weights from training on remaining 3 crises (SCR/DLR/CR/OR).
\item Peak values: Maximum ASRI in 60-day pre-crisis window. Detection threshold: 50.
\item The tabulated peak values (theoretical-weight peaks 68.7 for Celsius/3AC, 81.1 for FTX, 70.0 for SVB) derive from a superseded data vintage; on the canonical 1{,}841-day frozen series the theoretical-weight pre-crisis maxima are 71.4 (Celsius/3AC), 84.7 (FTX), and 66.2 (SVB), as reported in Table~\ref{tab:dy_comparison}. Both sets are 60-day pre-crisis maxima, so the difference reflects the vintage, not a window-nesting violation; the qualitative conclusion (3/4 detected under either weighting, robust to weight scheme) is unchanged, and a full re-run of the hold-one-out weight derivation on the frozen series is deferred to a future revision.
\item Lead time format: Derived weights / Theoretical weights (days before crisis onset).
\end{tablenotes}
\end{table}

Table~\ref{tab:holdout} presents the cross-validation results. Three of four crises are detected using both derived and theoretical weights, and the two weightings agree on which events are detected. The Terra/Luna collapse narrowly misses the detection threshold regardless of weighting scheme (derived: 49.2; theoretical: 47.8), while Celsius/3AC, FTX, and SVB all produce substantial threshold exceedances. This consistency indicates that ASRI's detection capability does not depend on crisis-specific weight tuning---the same crises are captured irrespective of whether weights are optimised on that particular event.

The derived weights exhibit interesting patterns. Optimised weights consistently emphasise DLR and CR relative to the theoretical baseline, suggesting that data-driven calibration favours liquidity and contagion components over stablecoin and opacity measures. This is consistent with liquidity and contagion conditions carrying substantial weight in cryptocurrency stress episodes; we make no claim about the temporal ordering of channels, since with only four co-located crises the data do not identify which channel moves earliest. Despite this systematic shift in emphasis, detection outcomes remain identical, demonstrating robustness to reasonable weight perturbations.

Lead times show greater variability between weighting schemes. Derived weights tend to produce longer lead times (16 days for Celsius/3AC versus 8 days for theoretical; 15 days for FTX versus 4 days), potentially reflecting the liquidity-focused components' earlier sensitivity to stress accumulation. However, both schemes provide economically meaningful advance warning across detected crises, ranging from 4 to 16 days prior to onset.

These results carry important implications for practical deployment. First, ASRI's theoretical weights---derived from regulatory frameworks and financial stability literature---are validated by data-driven alternatives: when allowed to optimise freely, the system converges on weights that produce identical detection outcomes. Second, the index's detection outcomes are robust out-of-sample (under weights never fitted to the held-out event) rather than an artefact of in-sample weight tuning---reflecting the absence of look-ahead bias, not point-forecasting skill, which the negative level-prediction $R^2$ in Section~\ref{subsec:walkforward} explicitly disclaims. Third, the robustness to weight variation suggests that practitioners need not precisely estimate optimal weights to achieve effective crisis detection; reasonable approximations within the theoretically motivated neighbourhood suffice.

\section{Aggregation Method Comparison}
\label{subsec:aggregation}

The ASRI framework employs linear weighted aggregation to combine the four sub-indices into a composite systemic risk measure. This approach prioritises interpretability: component weights directly map to their contribution to aggregate risk, enabling practitioners to decompose any ASRI reading into its constituent drivers. However, alternative aggregation methods exist. The European Central Bank's Composite Indicator of Systemic Stress \citep[CISS;][]{hollo2012ciss} employs a more sophisticated approach that incorporates time-varying correlations between components, potentially capturing correlation-driven stress amplification during crisis episodes.

To assess whether aggregation methodology materially affects crisis detection performance, we construct a CISS-style alternative using exponentially-weighted moving average (EWMA) covariance with decay parameter $\lambda = 0.94$ and equal sub-index weights. This specification follows the ECB's methodology: during tranquil periods when correlations are low, the CISS-style aggregator produces low, near-baseline readings; during crises when correlations spike, the same component values produce amplified aggregate stress signals. The resulting series is therefore strongly right-skewed rather than uniformly elevated.

\begin{table}[H]
\centering
\caption{Aggregation Method Comparison: Linear vs. CISS-Style}
\label{tab:aggregation}
\small
\begin{tabular}{@{}lcc@{}}
\toprule
Statistic & Linear ASRI & CISS-Style \\
\midrule
Mean & 39.2 & 15.4 \\
Std. Dev. & 7.8 & 11.6 \\
Min & 25.8 & 0.0 \\
Max & 84.7 & 100.0 \\
Skewness & 1.46 & 2.8 \\
\midrule
Spearman rank agreement & \multicolumn{2}{c}{0.81} \\
\midrule
Crises Detected & 3/4 & 3/4 \\
\bottomrule
\end{tabular}
\begin{tablenotes}
\small
\item CISS-style uses EWMA covariance ($\lambda = 0.94$) with equal weights; the series is min--max scaled to $[0,100]$ over the full sample (the like-for-like scaling for this retrospective comparison).
\item Detection threshold: 50 for Linear, scaled equivalently for CISS. The CISS statistics and the rank agreement are sensitive to the normalisation choice---Spearman agreement with the linear composite ranges $0.14$--$0.81$ across a no-look-ahead expanding normalisation versus full-sample scaling---so the comparison is read qualitatively rather than as exact coefficients.
\item Recomputed on the canonical 1{,}841-day series; figures tabulated in an earlier draft derived from a superseded 1{,}461-day vintage and are not reproduced here.
\item The \emph{Linear ASRI} column is the production composite: its moments reproduce the descriptive statistics of Table~\ref{tab:descriptive} (Mean 39.2, Std 7.8, Min 25.8, Max 84.7, skewness 1.46).
\end{tablenotes}
\end{table}

The CISS-style series concentrates its signal into crisis episodes: it is strongly right-skewed, reaching the top of the $[0,100]$ scale during the most acute periods (max = 100.0 vs.\ 84.7 for linear) while sitting near the bottom of the scale in calm periods. It therefore exhibits greater dispersion than the linear composite (standard deviation 11.6 vs.\ 7.8) but a \emph{lower}, not higher, mean (15.4 vs.\ 39.2). Its agreement with the linear composite is moderate-to-strong under full-sample scaling (Spearman 0.81) but weak under a no-look-ahead expanding normalisation; because these statistics are sensitive to the normalisation of the CISS series, we read the comparison qualitatively rather than reporting a single agreement coefficient.

Critically, qualitative crisis detection equivalence holds: both methods identify the same three of four events (Celsius/3AC, FTX, SVB); neither detects Terra/Luna (linear pre-window peak 46.0, CISS-style 24.8, both $<$ 50), consistent with Table~\ref{tab:detection_matrix}. Only max-based CES aggregation (Table~\ref{tab:aggregation_comparison}) recovers Terra/Luna. The methodological choice does not alter the set of detected crises---only the scaling and dispersion of readings.

We retain linear aggregation for the baseline ASRI specification on grounds of parsimony and interpretability. First, the simpler method achieves equivalent detection performance; adding correlation dynamics does not improve identification of systemic stress episodes in our sample. Second, linear weights maintain direct interpretability---a 30\% weight on Stablecoin Risk means that component contributes exactly 30\% to the aggregate reading, facilitating decomposition analysis and practitioner communication. Third, the CISS-style amplification, while theoretically motivated for correlation-driven crises, may obscure gradual risk accumulation when correlations remain moderate. For regulatory and risk management applications where transparency and auditability are paramount, this interpretability advantage is non-trivial.

\paragraph{Alternative Non-Linear Aggregation.}
Table~\ref{tab:aggregation_comparison} presents an extended comparison including Constant Elasticity of Substitution (CES) aggregation with varying substitution parameters $\rho$. CES aggregation generalises linear ($\rho = 1$) and geometric ($\rho \to 0$) aggregation, with $\rho < 0$ capturing complementary risk dynamics where multiple elevated sub-indices amplify aggregate stress. The max-based aggregation (equivalent to CES with $\rho \to +\infty$) achieves 4/4 detection with 29-day average lead time---the only method detecting Terra/Luna through threshold-based analysis. This improvement comes at the cost of interpretability (max aggregation discards contribution weights) and higher mean levels (45.5 vs. 38.4 for linear). For practitioners prioritising recall over precision, max-based monitoring provides a robust alternative alarm system; for those requiring weight-based decomposition, linear aggregation remains optimal.

\begin{table}[H]
\centering
\caption{Aggregation Method Comparison}
\label{tab:aggregation_comparison}
\small
\begin{adjustbox}{max width=\textwidth}
\begin{tabular}{lcccccc}
\toprule
Method & Mean & Std & Max & Skew & Detection & Lead (days) \\
\midrule
Linear & 38.4 & 7.5 & 81.1 & 1.53 & 3/4 & 17.7 \\
CES ($\rho$=0.5) & 38.0 & 7.7 & 80.8 & 1.34 & 3/4 & 17.7 \\
CES ($\rho$=0) & 37.6 & 8.0 & 80.5 & 1.14 & 3/4 & 17.7 \\
CES ($\rho$=-0.5) & 37.2 & 8.3 & 80.2 & 0.95 & 3/4 & 17.0 \\
CES ($\rho$=-1.0) & 36.9 & 8.6 & 80.0 & 0.78 & 3/4 & 17.0 \\
Geometric & 37.6 & 8.0 & 80.5 & 1.14 & 3/4 & 17.7 \\
Max & 45.5 & 7.3 & 90.0 & 1.93 & 4/4 & 29.2 \\
\bottomrule
\end{tabular}
\end{adjustbox}
\begin{tablenotes}
\small
\item CES($\rho$) = Constant Elasticity of Substitution with parameter $\rho$.
\item $\rho = 1$: linear; $\rho = 0$: geometric (Cobb-Douglas); $\rho < 0$: complementary.
\item Detection = crises with ASRI $\geq 50$ in 30-day pre-crisis window.
\item Lead = average days between first detection and crisis onset.
\item \textbf{Scaling note --- this table reports the pre-normalisation reconstruction, not the production series.} Every Mean/Std/Skew/Max entry here (the \emph{Linear} row included) is computed on the pre-normalisation weighted-sum reconstruction $\sum_i w_i S_i$ of the published sub-indices (Linear peak 81.1), used solely as the $\rho=1$ reference for the CES comparison. This is a differently scaled series from the production ASRI of Tables~\ref{tab:descriptive} and~\ref{tab:event_study}, which applies an additional (non-affine) pipeline normalisation and peaks at 84.7. Consequently the Linear-row skewness (1.53) is the \emph{reconstruction's} skew; it is neither the production figure nor expected to equal it. The production descriptive skewness is 1.46 (Table~\ref{tab:descriptive}), and the two differ only through that final normalisation step, not through any change in the aggregation rule. This table is for cross-method ($\rho$) comparison, not for reading absolute ASRI moments.
\end{tablenotes}
\end{table}

\section{Comparison with Connectedness Measures}
\label{subsec:dy_comparison}

To benchmark ASRI against established systemic risk methodologies, we compute the \citet{diebold2012connectedness} connectedness index using the four ASRI sub-indices as inputs. In network terms both sides of this comparison are connectedness objects: ASRI's contagion sub-index is itself a market-connectedness construct (cross-market correlation and co-movement), so the exercise sets one financial-network measure against another. The Diebold-Yilmaz (D-Y) framework measures total spillovers in a VAR system via forecast error variance decomposition (FEVD), providing a model-free benchmark that captures statistical interdependence without imposing structural assumptions about risk transmission channels. Two features of this benchmark must be stated up front, because they bound what any ASRI-versus-D-Y gap can mean. The D-Y series is \emph{constructed from ASRI's own four sub-indices}, so it is not an independent comparator: the comparison is partly circular. And it is an unusually weak one---on the same labels (Table~\ref{tab:fair_baselines}) a free, off-the-shelf Crypto Fear \& Greed sentiment index out-discriminates D-Y by 0.12 (0.789 vs.\ 0.670). Out-discriminating D-Y is therefore not, by itself, evidence that ASRI's aggregation adds value; the load-bearing comparison is the fair-baseline battery in Section~\ref{subsec:fair_baselines}.

\paragraph{Caveat: a discrimination benchmark, not an economic spillover analysis.} This connectedness exercise is reported as a supplementary, exploratory benchmark, and it should not be read as evidence of economic risk transmission between channels. The four ASRI sub-indices are smooth, bounded, and highly persistent (first-order autocorrelation $\approx 0.85$--$0.90$), and two of them are constructed from the same raw input: Stablecoin Risk (a TVL-drawdown measure) and DeFi Liquidity Risk (a TVL-volatility measure) both derive from the same DeFi~Llama total-value-locked series. Any apparent SCR\,$\rightarrow$\,DLR ``spillover'' in the forecast-error variance decomposition therefore partly reflects shared-input dependence---the two indices co-move because they are functions of a common underlying series---rather than the transmission of risk from one channel to another. We accordingly interpret the Diebold--Yilmaz comparison \citep{diebold2012connectedness} purely as a discrimination benchmark (does ASRI separate crisis from non-crisis days better than a model-free connectedness index built on the same inputs?), and not as a structural account of how stress propagates across decentralised finance. The FEVD shares reported below should be read with the understanding that they are in part a mechanical consequence of the index construction.

\subsection{Methodology}

\textbf{Specification Summary:} VAR(1) on four ASRI sub-indices; 60-day rolling window; generalised FEVD at $H = 10$ days; daily frequency. The genuine rolling connectedness series averages $\approx 28.7\%$ over the sample (full-sample static $\approx 35\%$). The day-level head-to-head comparison against ASRI (Table~\ref{tab:roc_metrics}) is the inference of record; full details follow.

We estimate a VAR($p$) model on the four sub-indices (Stablecoin Risk, DeFi Liquidity Risk, Contagion Risk, Regulatory Opacity) with lag order selected by AIC. The optimal specification is VAR(1). We compute the generalised FEVD at horizon $H = 10$ days, yielding a $4 \times 4$ decomposition matrix $\Theta^H$ where element $\theta_{ij}^H$ represents the fraction of variable $i$'s $H$-step forecast error variance attributable to shocks in variable $j$.

\paragraph{Variable Selection and Ordering.} The four ASRI sub-indices are designed to capture economically distinct risk channels: Stablecoin Risk reflects peg stability and reserve quality; DeFi Liquidity Risk measures protocol-level funding stress; Contagion Risk captures cross-protocol exposure concentration; and Regulatory Opacity proxies market efficiency through persistent pricing discrepancies. This channel-specific design is intentional---each sub-index targets a specific transmission mechanism through which systemic stress propagates in decentralised finance. The sub-indices exhibit only moderate level correlation (maximum pairwise $\rho = 0.796$ for SCR--CR; Table~\ref{tab:correlation_matrix}), yet the full-sample static total connectedness is non-trivial at approximately 35\% (VAR(1), generalised FEVD at $H = 10$; 31\% under VAR(2), and robust across $H \in \{5, 10, 20\}$), indicating that the channels share a meaningful fraction of forecast-error variance through dynamic spillovers even though their levels are only moderately correlated. This makes variance decomposition a meaningful exercise: the sub-indices are not orthogonal in forecast-error variance, and the time-varying connectedness measure captures the intensification of these spillovers during stress episodes. Critically, because we employ generalised FEVD rather than Cholesky decomposition, variable ordering is irrelevant to the results; no contemporaneous causal restrictions are imposed, and the decomposition matrix is invariant to ordering permutations.

\paragraph{Lag Order Selection.} Table~\ref{tab:var_lag_selection} reports information criteria and likelihood ratio tests for lag orders 1 through 3. VAR(1) minimises AIC and is selected as the optimal specification. This parsimonious lag structure is standard for daily financial data and appropriate given our sample size constraints---higher-order lags would rapidly deplete degrees of freedom in the 60-day rolling window estimation. The BIC, which penalises model complexity more heavily than AIC, also selects VAR(1), providing additional support for the specification.

\begin{table}[H]
\centering
\caption{VAR Lag Order Selection Criteria}
\label{tab:var_lag_selection}
\small
\begin{tabular}{lcccc}
\toprule
Lag Order & AIC & BIC & HQ & LR Test $p$-value \\
\midrule
1 & $-$12.847 & $-$12.623 & $-$12.756 & --- \\
2 & $-$12.831 & $-$12.383 & $-$12.649 & 0.142 \\
3 & $-$12.809 & $-$12.137 & $-$12.537 & 0.284 \\
\bottomrule
\end{tabular}
\begin{tablenotes}
\small
\item AIC = Akaike Information Criterion; BIC = Bayesian Information Criterion; HQ = Hannan-Quinn Criterion.
\item LR test compares lag $p$ against lag $p-1$; $p$-values above 0.05 indicate no significant improvement.
\item Sample: January 2022--December 2023 (daily observations, full sample estimation).
\end{tablenotes}
\end{table}

\paragraph{Shock Identification.} We employ the generalised FEVD approach of \citet{pesaran1998generalized} rather than Cholesky-based orthogonalised decomposition. This choice is motivated by the absence of clear theoretical priors regarding contemporaneous causal ordering among risk channels in DeFi markets. Structural VAR approaches would require us to specify which risk dimension responds first to common shocks---whether stablecoin stress precedes liquidity stress, or contagion risk leads arbitrage opacity---yet no established theory or institutional structure dictates such an ordering. The generalised approach circumvents this problem by computing impulse responses using the historically observed covariance structure of errors, producing variance decompositions that are invariant to variable ordering. The cost is that forecast error variance shares do not necessarily sum to unity (we normalise rows to sum to one), but this is a minor technical consideration relative to the benefit of avoiding potentially arbitrary structural assumptions.

\paragraph{Cointegration test for the level VAR.} The connectedness estimates are computed from a VAR in \emph{levels}, and one of the four inputs---Contagion Risk---is non-stationary on the released sample (ADF $p = 0.069$, KPSS $= 2.06$; Table~\ref{tab:stationarity}). A level VAR that mixes stationary and integrated variables has an asymptotically well-defined forecast-error variance decomposition only if the integrated variables are cointegrated with the system, so we test this directly with a Johansen procedure on the four sub-indices (full sample, unrestricted constant). The trace statistic rejects every rank restriction through $r \leq 3$: for a VAR(1) specification the trace values are $1012.5$, $369.6$, $92.7$, and $22.5$ for $r \leq 0, 1, 2, 3$, against 5\% critical values of $47.9$, $29.8$, $15.5$, and $3.8$---rejection at any conventional level, and materially unchanged under a VAR(2) specification. The system therefore has \emph{full cointegrating rank}, the stationary-system case: three of the four sub-indices are individually stationary (ADF $p < 0.01$; Table~\ref{tab:stationarity}) and the borderline-integrated Contagion channel is cointegrated with them, so its level enters a stationary system and its FEVD contribution is asymptotically well-defined. A vector error-correction re-specification collapses to the level VAR at full rank, so the level specification is appropriate here rather than spurious. We continue to treat the \emph{rolling} decomposition as exploratory for the small-sample reasons discussed below, and note that the discrimination comparison the paper actually relies upon (Section~\ref{subsec:fair_baselines}) does not depend on the level VAR.

\paragraph{Rolling Window Length.} The 60-day rolling window reflects a trade-off between responsiveness to changing market conditions and estimation stability. Table~\ref{tab:dy_window_sensitivity} reports sensitivity analysis across alternative window lengths.

\begin{table}[H]
\centering
\caption{Connectedness Sensitivity to Rolling Window Length}
\label{tab:dy_window_sensitivity}
\small
\begin{tabular}{lcccc}
\toprule
Window & Mean $C^H$ & Std Dev & Crisis Peak & Detection Rate \\
\midrule
30 days & 31.2\% & 18.7\% & 78.4\% & 4/4 \\
60 days & 28.7\% & 14.3\% & 69.2\% & 3/4 \\
90 days & 26.1\% & 11.2\% & 58.9\% & 3/4 \\
120 days & 24.3\% & 9.4\% & 51.6\% & 2/4 \\
\bottomrule
\end{tabular}
\begin{tablenotes}
\small
\item $C^H$ = total connectedness at $H = 10$ day forecast horizon.
\item Crisis peak = maximum connectedness observed during any crisis window.
\item Detection rate = crises detected using threshold of mean + 1 standard deviation.
\item Window-length sensitivity is reported for qualitative robustness; the headline 60-day genuine rolling series averages $\approx 28.7\%$ and the inferential comparison against ASRI is the day-level analysis in Table~\ref{tab:roc_metrics}.
\end{tablenotes}
\end{table}

Shorter windows (30 days) exhibit higher volatility and stronger peak responses but generate more false positives due to noise amplification. Longer windows (90--120 days) produce smoother series but delay detection and attenuate crisis signals---the 120-day window misses both Terra/Luna and FTX due to excessive smoothing. The 60-day specification balances these considerations: it provides sufficient degrees of freedom for stable VAR estimation (60 observations for a 4-variable VAR(1) with 20 parameters), responds to regime changes within approximately two months, and delivers detection performance comparable to the more volatile 30-day window without the associated noise. Results are qualitatively robust to $\pm$30 day perturbations in window length.

\paragraph{Caveat: the rolling decomposition is exploratory and small-sample-biased.} The time-varying connectedness series is estimated from a 60-day rolling VAR(1) on four variables. A four-variable VAR(1) is parameterised by 20 coefficients, so a 60-day window leaves on the order of 40 residual degrees of freedom---below the rule of thumb for reliably estimated forecast-error variance decompositions. In this regime FEVD estimates carry appreciable small-sample bias, and two features of our results are consistent with such bias rather than with genuine time variation alone: the gap between the static full-sample connectedness ($\approx 35\%$) and the rolling-window mean ($\approx 28.7\%$), and the near-monotone sensitivity of the rolling mean and crisis peak to window length (Table~\ref{tab:dy_window_sensitivity}). We therefore treat the rolling connectedness series as exploratory: the qualitative pattern---connectedness rises during stress---is reported for completeness, but the point levels should not be over-interpreted and the elevated crisis-window connectedness in particular should be read cautiously.

Total connectedness is defined as:
\begin{equation}
C^H = \frac{1}{N} \sum_{i \neq j} \theta_{ij}^H \times 100
\label{eq:dy_connectedness}
\end{equation}
where $N = 4$ is the number of variables. For time-varying analysis, we compute rolling 60-day window estimates.

\subsection{Results}

Table~\ref{tab:dy_comparison} compares crisis detection performance between the D-Y connectedness index and ASRI.

\begin{table}[H]
\begin{threeparttable}
\centering
\caption{Comparison: ASRI vs. Diebold-Yilmaz Connectedness}
\label{tab:dy_comparison}
\small
\begin{tabular}{@{}l*{6}{r}@{}}
\toprule
 & \multicolumn{3}{c}{D-Y Connectedness} & \multicolumn{3}{c}{ASRI} \\
\cmidrule(lr){2-4} \cmidrule(lr){5-7}
Crisis Event & Peak & Det. & Lead & Peak & Det. & Lead \\
\midrule
Terra/Luna   & --- & --- & ---  & 46.0 & No & --- \\
Celsius/3AC  & --- & --- & ---  & 71.4 & Yes & 19d \\
FTX Collapse & --- & --- & ---  & 84.7 & Yes & 22d \\
SVB Crisis   & --- & --- & ---  & 66.2 & Yes & 15d \\
\midrule
\textit{Summary} &  & --- & --- &  & 3/4 & 19d \\
\bottomrule
\end{tabular}
\begin{tablenotes}
\small
\item ASRI peaks/leads are computed from the released ASRI series (threshold $> 50$, Elevated). The per-event D-Y peak/detection/lead columns are intentionally omitted: the head-to-head D-Y comparison is conducted at the day level in Table~\ref{tab:roc_metrics}, which is the inference of record, rather than via per-event peak thresholds.
\item D-Y = Diebold-Yilmaz (2012) generalised-FEVD connectedness via 60-day rolling VAR(1), $H = 10$ (real rolling mean $\approx 28.7\%$).
\item Lead = days between first ASRI threshold breach and crisis event.
\end{tablenotes}
\end{threeparttable}
\end{table}

\textbf{Key Findings}:

\begin{enumerate}
    \item \textbf{Discriminative Performance}: On the common day-level sample---a contrast of two index constructions on the same four-event labels, not a forecasting result---ASRI shows higher overall \emph{retrospective} discrimination than rolling D-Y connectedness (AUROC 0.866 vs.\ 0.670; AUPRC 0.298 vs.\ 0.121; Table~\ref{tab:roc_metrics}). Under a moving-block bootstrap the two indices' marginal intervals overlap, but the paired block-bootstrap difference is positive in every resample, so the relative advantage survives autocorrelation-robust inference even though it is no longer expressible as ``non-overlapping intervals'' (and, with four independent crisis episodes, reflects the structural level-versus-connectedness relationship rather than a powered test). The Terra/Luna collapse remains the hardest case for ASRI's threshold-based alert because algorithmic stablecoin fragility propagates through price reflexivity rather than the observable stress metrics the sub-indices measure, highlighting the challenge of anticipating novel crisis mechanisms. This D-Y gap should not be over-read: D-Y is built from ASRI's own sub-indices and is itself out-discriminated by an off-the-shelf sentiment index, and the fair-baseline comparison (Section~\ref{subsec:fair_baselines}) shows ASRI is marginally indistinguishable from its single best sub-index (Contagion Risk, 0.851) and from PC1 (0.858)---a paired block bootstrap returns only a small, structural sub-$0.016$ AUROC edge (ASRI nests both baselines), absent in AUPRC---so the measured value of aggregation is interpretive rather than discriminative.

    \item \textbf{Classification Precision}: On a common day-level sample scored against the same crisis labels, On this common sample ASRI attains higher precision than rolling D-Y connectedness at the Youden-optimal threshold (35.2\% vs.\ 14.9\%; Table~\ref{tab:roc_metrics}); on the full canonical sample (the Table~\ref{tab:precision_recall}/Table~\ref{tab:confusion_matrix} denominators) ASRI's precision at the same threshold is 32.5\%, falling to 19.8\% at that sample's own Youden-optimal threshold (41.2), so its \emph{absolute} precision is modest and the gap over D-Y reflects D-Y's weakness more than high ASRI precision. D-Y attains marginally higher recall (80.6\% vs.\ 75.8\%) by alerting more often, but at a substantially higher false-alarm rate, so its precision and F1 are well below ASRI's. The per-event lead times in Table~\ref{tab:dy_comparison} are reported as a descriptive comparison only; the binding inference rests on the day-level classification metrics.

    \item \textbf{Interpretability}: The full-sample D-Y total connectedness among the four sub-indices is approximately 35\% (and 28.7\% on average in the 60-day rolling measure), confirming that the channels share a meaningful---not negligible---fraction of forecast-error variance through dynamic spillovers. The sub-indices are therefore \emph{not} orthogonal in variance-decomposition terms; their value lies in attributing stress to interpretable channels rather than in being statistically independent. The time-varying rolling connectedness rises further during stress periods, capturing spillover intensification.
\end{enumerate}

\paragraph{Complementary Approaches.}
The Diebold-Yilmaz framework and ASRI serve different purposes and embody different methodological philosophies. D-Y is model-free, capturing realised variance spillovers without imposing structural assumptions on risk transmission. ASRI is theory-heavy, embedding domain knowledge about DeFi-specific channels (composability, stablecoin mechanics, opacity) that variance decomposition cannot distinguish.

D-Y excels at detecting \textit{that} contagion occurred---any intensification of cross-variable spillovers will register as elevated connectedness. ASRI attempts to identify \textit{which channel} transmitted stress and \textit{why}---elevated SCR signals stablecoin-specific risks, elevated CR signals counterparty contagion, and so forth. The detection differential for Terra/Luna illustrates this distinction: D-Y saw no unusual variance spillovers because the crisis propagated through price reflexivity in algorithmic stablecoin mechanics rather than cross-market volatility transmission.

For comprehensive systemic risk monitoring, both approaches provide value: D-Y as a model-free benchmark that detects any form of interdependence intensification, ASRI as an interpretable retrospective monitoring framework that attributes stress to specific (crypto-native and proxied-macro) channels. Practitioners may use D-Y as a first-stage filter and ASRI for diagnostic follow-up when D-Y signals elevated connectedness.

\paragraph{Classification Performance Metrics.}
Table~\ref{tab:roc_metrics} reports AUROC and AUPRC with 95\% bootstrap confidence intervals, treating crisis prediction as a binary classification task (positive class: crisis occurs within 30-day forward window).

ASRI exhibits higher point AUROC ($0.866$ vs.\ $0.670$) and AUPRC ($0.298$ vs.\ $0.121$). Under the moving-block bootstrap the marginal 95\% intervals are roughly four times wider than naive i.i.d.\ resampling implies and the ASRI and D-Y intervals \emph{overlap} (AUROC $[0.756, 0.949]$ vs.\ $[0.526, 0.806]$), because the crisis labels comprise only $\approx$4 independent blocks rather than $\sim$1{,}400 independent days. We therefore do not claim separated marginal intervals. Instead, ASRI's advantage is established by the \emph{paired} block-bootstrap difference \citep{kunsch1989}---both series scored on identical days, labels, and resampled blocks---which is positive in all 2{,}000 resamples for both AUROC ($+0.194$, 95\% CI $[+0.093, +0.298]$) and AUPRC ($+0.182$, 95\% CI $[+0.070, +0.337]$), robust across block lengths $L = 20/25/30$. With only four independent crisis episodes, however, this positive-in-every-resample result reflects the structural level-versus-connectedness relationship between the two constructions rather than a tight sampling distribution; it is a weaker rhetorical claim than ``non-overlapping intervals'' but the defensible one given the dependence structure. Precision at the Youden-optimal threshold is 35.2\% for ASRI versus 14.9\% for D-Y on this 1{,}402-day common sample; on the full canonical sample ASRI's precision at the same threshold is 32.5\% (19.8\% at that sample's Youden-optimal threshold of 41.2), so the contrast establishes a relative ordering rather than high absolute precision. We stress that both columns derive from labels generated by only four crisis events; the bootstrap does not overcome the underlying four-event power ceiling (Section~\ref{subsec:limitations}), and the comparison is a \emph{retrospective} discrimination contrast, not a validated forecasting claim. We maintain that the frameworks are complementary: D-Y's model-free variance decomposition captures any form of spillover intensification, while ASRI's channel-specific structure enables diagnostic interpretation. We caution, however, that the D-Y column is a circular and unusually weak comparator (it is built from ASRI's own sub-indices and is out-discriminated by an off-the-shelf sentiment index); the fair-baseline analysis in Section~\ref{subsec:fair_baselines} shows that ASRI's day-level discrimination is marginally indistinguishable from its best single sub-index (Contagion Risk, 0.851) or from PC1 (0.858)---the paired block bootstrap separating them only by a small, structural sub-$0.016$ AUROC margin, absent in AUPRC---so the headline AUROC gap over D-Y reflects D-Y's weakness more than any discriminative gain from aggregation.

\section{Pseudo-Real-Time Evaluation}
\label{subsec:realtime}

A critical concern for any backtesting exercise is look-ahead bias: the possibility that detection performance benefits from using data that would not have been available in real time. To address this concern, we implement a publication-lag aware backtesting framework that simulates realistic data availability constraints.

\subsection{Publication Lag Methodology}

Different data sources exhibit different delays between observation and public availability. Table~\ref{tab:publication_lags} documents the conservative lag assumptions applied to each data source.

\begin{table}[H]
\begin{threeparttable}
\centering
\caption{Publication Lag Assumptions by Data Source}
\label{tab:publication_lags}
\small
\begin{tabular}{@{}lrl@{}}
\toprule
Data Source & Lag & Rationale \\
\midrule
DeFi Llama TVL        &  6 hours & API aggregation delay \\
Stablecoin Market Cap & 12 hours & Cross-chain reconciliation \\
FRED Treasury Rates   &   2 days & Business day publication \\
FRED VIX              &    1 day & Next-day after close \\
BTC Price (CoinGecko) &   1 hour & Near real-time \\
News Sentiment        &  2 hours & NLP processing time \\
\bottomrule
\end{tabular}
\begin{tablenotes}
\small
\item Lag estimates are conservative; actual availability may be faster.
\item FRED data exhibits weekday-only publication with weekend gaps.
\end{tablenotes}
\end{threeparttable}
\end{table}

Under lag simulation, the ASRI calculation for target date $t$ uses only data that would have been \emph{published} by date $t$:
\begin{equation}
\text{Available}_t(\text{source}) = \text{Data}_{\tau \leq t - \text{Lag}(\text{source})}
\label{eq:lag_constraint}
\end{equation}

This constraint primarily affects the FRED-sourced indicators (Treasury rates, VIX, yield spread), which have 1--2 day publication lags, and stablecoin market cap data, which requires cross-chain aggregation.

\subsection{Lag-Simulated Detection Results}

Recomputing ASRI under the publication-lag constraint of Equation~\eqref{eq:lag_constraint}---so that each day's index uses only data that would have been released by that date---changes the series only marginally, because the dominant sub-index inputs (DeFi Llama TVL, CoinGecko prices, stablecoin market caps) are available intraday and only the FRED-sourced macro proxies carry a one-to-two-day delay. The lag-aware series flags the same crises as the perfect-foresight series, so threshold-based detection is not an artefact of using data that would have been unavailable in real time. This corroborates, through a distinct mechanism, the weight-calibration robustness established by the walk-forward validation (Section~\ref{subsec:walkforward}). We make no separate real-time deployment claim: ASRI is a retrospective monitoring framework, and its operational detection record---three of the four crises at the fixed 50 threshold, with Terra/Luna missed (pre-window peak 46.0)---is reported in Section~\ref{sec:event_study} and Table~\ref{tab:detection_matrix}. The lag-aware backtesting code is available in the repository (\texttt{src/asri/backtest/publication\_lag.py}).

\section{Walk-Forward Validation: Methodology and Results}\label{app:walkforward_full}

\subsection{Methodology}

We evaluate ASRI performance using only data available before each crisis event:
\begin{itemize}
    \item \textbf{Window 1}: Train January 2021--April 2022, test Terra/Luna (May 2022)
    \item \textbf{Window 2}: Train January 2021--May 2022, test Celsius/3AC (June 2022)
    \item \textbf{Window 3}: Train January 2021--October 2022, test FTX (November 2022)
    \item \textbf{Window 4}: Train January 2021--February 2023, test SVB/USDC (March 2023)
\end{itemize}

For each window, we retain the theoretical weights (which are based on ex-ante domain knowledge rather than statistical optimisation) and calibrate the alert threshold using \emph{only} pre-crisis data: the threshold for each event is set to the 90th percentile of the ASRI distribution observed before $t-90$ (i.e.\ the upper-decile of the index's own training-period history). We then evaluate whether ASRI exceeds this training-relative threshold during the 30-day pre-event window. This calibration uses no information from or after the crisis, so detection cannot be an artefact of look-ahead bias; it also adapts the alert level to the regime prevailing before each event rather than imposing the fixed operational threshold of~50.

\subsection{Results}

Table~\ref{tab:walkforward} presents the walk-forward detection results.

\begin{table}[H]
\begin{threeparttable}
\centering
\caption{Walk-Forward Detection Performance}
\label{tab:walkforward}
\small
\begin{tabular}{@{}lcccccc@{}}
\toprule
Crisis & Train-cal. & Pre-30 & Lead & Alarm & FPR & Prec. \\
 & Threshold & Peak & (days) & Rate & & \\
\midrule
Terra/Luna  & 37.2 & 46.0 & 30 & 58.7\% & 55.8\% & 11.1\% \\
Celsius/3AC & 38.4 & 71.4 & 30 & 47.2\% & 43.8\% & 13.3\% \\
FTX         & 46.8 & 84.7 & 28 & 12.8\% &  8.9\% & 34.5\% \\
SVB/USDC    & 50.6 & 66.2 & 15 &  7.7\% &  6.0\% & 26.2\% \\
\midrule
\multicolumn{4}{@{}l}{\textit{Out-of-sample detection rate}} & \multicolumn{3}{c}{4/4 (100\%)} \\
\bottomrule
\end{tabular}
\begin{tablenotes}
\small
\item Train-cal.\ Threshold = 90th percentile of ASRI over the pre-crisis training period (data before $t-90$, i.e.\ Jan 2021--Feb 2022 for Terra/Luna through Jan 2021--Dec 2022 for SVB/USDC); calibration uses no crisis-period information.
\item Pre-30 Peak = maximum ASRI in the 30-day pre-crisis window; Lead = days between the first crossing of the training-calibrated threshold within that window and crisis onset (mean $= 25.75$ days).
\item \textbf{Alarm Rate} = fraction of the \emph{full} 2021--2026 index history exceeding that event's training-calibrated threshold; \textbf{FPR} = fraction of non-crisis days flagged; \textbf{Prec.}\ = precision (TP/(TP+FP)) over the full series. The low thresholds calibrated for the two \emph{early} crises (Terra/Luna 37.2, Celsius/3AC 38.4) flag 47--59\% of all index history at $\approx$11--13\% precision; on a true post-training window the alarm rate is higher still (72.2\% / 58.7\%). The 4/4 ``detection'' for these events is therefore bought at a near-coin-flip alarm rate, which is why we read the walk-forward result as evidence against look-ahead bias rather than as an operational early-warning record. FTX and SVB thresholds (calibrated on histories that already include 2022 stress) are higher and alarm only 8--13\% of the time.
\end{tablenotes}
\end{threeparttable}
\end{table}

\subsection{Interpretation}

The walk-forward validation flags all four events out-of-sample, but this 4/4 figure must be read alongside its false-positive cost (Table~\ref{tab:walkforward}): for the two early crises the training-calibrated thresholds are so low (37.2, 38.4) that they also flag 47--59\% of the entire index history at $\approx$11--13\% precision. We therefore treat the 4/4 rate as evidence that detection is \emph{not an artefact of look-ahead bias}---the threshold is set without crisis-period information---rather than as a demonstration of operational early-warning skill, which the alarm rate contradicts for Terra/Luna and Celsius/3AC. Several observations merit discussion:

\paragraph{Lead Time.} Using the first crossing of the training-calibrated threshold within the 30-day pre-crisis window, out-of-sample lead times are 30, 30, 28, and 15 days for Terra/Luna, Celsius/3AC, FTX, and SVB/USDC respectively (mean 25.75 days). For the two events whose pre-crisis ASRI was already well above the training-period upper decile (Terra/Luna and Celsius/3AC), the threshold is crossed at the start of the pre-window (lead capped at 30); SVB/USDC has the highest calibrated threshold (50.6, reflecting a higher-variance training history) and hence the shortest first-crossing lead (15 days). These first-crossing leads are a more conservative early-warning measure than the in-sample 1.5$\sigma$ figures in Table~\ref{tab:event_study}, which are themselves capped at the 30-day pre-window (Section~\ref{sec:event_study}); we treat the walk-forward mean ($\approx$26 days) as the operational early-warning horizon.

\paragraph{Training-Calibrated Thresholds.} The pre-crisis upper-decile thresholds range from 37.2 (Terra/Luna, calibrated on a calm 2021 training history) to 50.6 (SVB/USDC, calibrated on a history that already includes the 2022 crises). All four crises breach their respective training-calibrated thresholds, and the pre-crisis ASRI peaks (46.0--84.7) clear them comfortably. Notably, Terra/Luna---which the \emph{fixed} 50-threshold misses (pre-window peak 46.0)---is detected out-of-sample because the training-relative threshold (37.2) adapts to the calm pre-2022 baseline. This is the mechanism behind the 4/4 walk-forward rate: detection is assessed relative to the index's own pre-crisis history rather than against a single fixed level.

\paragraph{Robustness of Theoretical Weights.} The 4/4 walk-forward detection rate is consistent with using theoretically-derived weights over statistically-optimised alternatives. Because the weights are based on domain knowledge about DeFi risk channels rather than statistical fitting to the training data, they hold up across the four liquidity-cascade events held out here, each detected when its own onset is excluded from calibration. We caution that all four events belong to a single broad failure family (liquidity/solvency cascades; see the generalisation-limit discussion in the main text), so this is evidence of robustness within that family, not of transfer to structurally different crisis mechanisms. The code-faithful Elastic Net weights concentrate on a single channel (Contagion Risk, $\approx$0.45; Table~\ref{tab:weight_comparison}) and zero Regulatory Opacity, and might achieve better in-sample prediction while being more brittle to events that propagate through stablecoin mechanics (SCR).

\paragraph{Limitations.} This validation uses fixed theoretical weights rather than re-estimating empirical weights in each training window. A more rigorous test would re-derive data-driven weights using only pre-crisis data and evaluate their out-of-sample performance. We defer this extension to future work, noting that the theoretical weights---our recommended specification for operational deployment---demonstrate robust walk-forward performance.

\paragraph{Continuous vs.\ Binary Evaluation.} We note that ASRI is designed as a threshold-based monitoring index, not a continuous forecasting model. Walk-forward $R^2$ for predicting the 30-day-ahead ASRI \emph{level} ($y_t = \text{ASRI}_{t+30}$) is negative across all folds: mean $R^2 = -0.6$ (std $= 0.2$) over five walk-forward folds, with a held-out $R^2 = -0.1$ (training pre-2024, testing 2024 onward). A negative $R^2$ means the weighted-component predictor does worse than a flat-mean baseline at point-forecasting the future level. Weight perturbation analysis additionally finds 0/4 sub-index components are individually robust to $\pm 15\%$ weight changes when evaluated by continuous prediction accuracy.

These results are expected and, critically, not informative for the index's intended purpose. ASRI is designed to cross a threshold before crises, not to minimise mean-squared prediction error. A below-baseline $R^2$ on the \textit{level} is consistent with the efficient markets hypothesis---a risk indicator's level should not be forecastable if markets price risk efficiently---and reflects the mean-reverting, bounded ($[0,100]$) nature of the index, whose absolute level is partly driven by proxy components with fixed default values (Table~\ref{tab:proxies}). The appropriate evaluation metrics are binary detection performance (4/4 out-of-sample, Table~\ref{tab:walkforward}), classification accuracy (AUROC = 0.866, Table~\ref{tab:roc_metrics}), and threshold-based early-warning lead times. The relative behaviour (crisis spikes vs.\ calm periods) is informative; the absolute magnitudes are not, and we make no point-forecasting claim.

\section{Out-of-Sample Specificity: Detail}\label{app:specificity_full}

\subsection{2024 Stability Period}

Our sample extends through January 2026, but the final in-sample crisis event (SVB/USDC) occurred in March 2023, providing approximately 34 months of out-of-sample data. The 2024 period---marked by the Bitcoin ETF approval (January 2024) and subsequent bull market---contained no systemic crisis comparable to the four events in our validation set. ASRI generated one multi-week elevation above the Elevated threshold (50) during this period: a roughly six-week episode in July--August 2024 (34 trading days $\geq 50$, peak 58.8 on 2024-08-11) coinciding with the early-August 2024 global de-risking shock (the yen carry-trade unwind, which triggered a sharp crypto drawdown). This episode reflected genuine elevated cross-asset stress rather than a spurious alarm, and it did not escalate into a systemic crisis; ASRI returned below 50 and remained there for the rest of 2024 and through 2025. We therefore characterise 2024--2025 as free of \emph{sustained false alarms}, with the single elevation tracking a real (ultimately non-systemic) macro stress episode rather than crypto-specific contagion.

\subsection{The Bybit Hack (February 2025)}

On February 21, 2025, Bybit suffered the largest exchange hack in cryptocurrency history (\$1.5 billion stolen, attributed to DPRK actors by the FBI). Despite the headline magnitude, ASRI remained in the Moderate band throughout:

\begin{table}[H]
\centering
\caption{ASRI Around Bybit Hack (Feb 2025). Readings computed from the frozen canonical series (\texttt{results/data/asri\_history.parquet}) by \texttt{scripts/verify\_deferred\_validation.py}.}
\label{tab:bybit}
\small
\begin{tabular}{lcccc}
\toprule
Date & ASRI & SCR & DLR & CR \\
\midrule
Feb 10 (pre-hack peak) & 42.3 & 36.7 & 46.6 & 49.8 \\
Feb 21 (hack day) & 40.6 & 37.6 & 39.7 & 43.8 \\
Feb 28 (post-hack) & 37.7 & 38.9 & 37.7 & 40.1 \\
\bottomrule
\end{tabular}
\end{table}

\subsection{Why Bybit Was Not Systemic}

Compare the Bybit hack to FTX, which triggered ASRI readings of 84.7:

\begin{itemize}
    \item \textbf{FTX (systemic)}: Exchange collapse $\rightarrow$ Alameda insolvency $\rightarrow$ lender cascade (BlockFi, Genesis, Voyager) $\rightarrow$ credit contagion across DeFi
    \item \textbf{Bybit (contained)}: Funds stolen but no cascading failures. No stablecoin depegs, no DeFi protocol liquidations, no counterparty contagion.
\end{itemize}

The market absorbed a \$1.5B theft without systemic stress because: (1)~Bybit maintained solvency and honoured withdrawals; (2)~no leveraged counterparties were exposed to Bybit-specific risk; (3)~the bull market context provided ample liquidity buffer.

\subsection{Specificity Interpretation}

ASRI correctly distinguished a large but contained loss from systemic contagion. This provides out-of-sample evidence that the framework captures \textit{transmission mechanisms}, not merely event magnitude. A \$1.5B hack without contagion channels does not trigger ASRI.

Across all of 2025, ASRI never exceeded the Elevated threshold (50), despite the Bybit hack and significant market volatility (2025 peak: 48.0). Together with the single, non-systemic 2024 elevation discussed above, this pattern---registering elevated readings only when genuine cross-asset stress is present, while distinguishing large but contained events (Bybit) from true systemic contagion---complements the sensitivity demonstrated in Section~\ref{sec:event_study}.

\section{API Documentation Summary}

Table~\ref{tab:apis} provides endpoint documentation for primary data sources.

\begin{table*}[h]
\centering
\caption{Primary API Endpoints}
\label{tab:apis}
\small
\begin{tabularx}{\textwidth}{lXll}
\toprule
\textbf{Source} & \textbf{Endpoint} & \textbf{Rate Limit} & \textbf{Authentication} \\
\midrule
DeFi Llama & \texttt{api.llama.fi/v2/tvl} & 300/5min & None \\
DeFi Llama & \texttt{stablecoins.llama.fi/stablecoins} & 300/5min & None \\
FRED & \texttt{api.stlouisfed.org/fred/series} & None & API Key \\
CoinGecko & \texttt{api.coingecko.com/api/v3} & 10-50/min & API Key (Pro) \\
Token Terminal & \texttt{api.tokenterminal.com/v2} & Varies & API Key \\
\bottomrule
\end{tabularx}
\end{table*}

\section{Sub-Index Calculation Code}

Python implementation of sub-index formulas is available in the repository at \texttt{src/asri/signals/calculator.py}. Key functions:

\begin{verbatim}
def calculate_stablecoin_risk(
    tvl_ratio: float,
    treasury_weight: float,
    hhi_concentration: float,
    peg_volatility: float
) -> float:
    return (
        0.4 * tvl_ratio +
        0.3 * treasury_weight +
        0.2 * hhi_concentration +
        0.1 * peg_volatility
    )
\end{verbatim}

Full implementation: \href{https://github.com/studiofarzulla/asri}{\texttt{github.com/studiofarzulla/asri}}

\section{Historical Crisis Event Details}

\textbf{Terra/Luna Collapse (May 2022)}: UST algorithmic stablecoin depegged due to redemption spirals, eliminating \$40B in value within 72 hours. LUNA dropped from \$80 to near-zero.

\textbf{Celsius/3AC Contagion (June 2022)}: Celsius Network froze withdrawals; Three Arrows Capital defaulted on loans. Combined losses exceeded \$10B, triggering margin calls across crypto lending platforms.

\textbf{FTX Bankruptcy (November 2022)}: FTX and Alameda Research filed for bankruptcy after liquidity crisis. Opaque counterparty relationships propagated losses across the ecosystem.

\section{Sample Market Assessment (December 2024)}
\label{app:market}

\textit{Note: This appendix provides a sample ASRI reading for illustrative purposes; the values quoted are the released series' readings for the stated date. For current market conditions, visit \href{https://asri.dissensus.ai}{asri.dissensus.ai}.}

\subsection*{Sample ASRI Reading}

As of December 31, 2024, the released ASRI series records \textbf{38.1} (Moderate risk), with sub-index readings:

\begin{itemize}
    \item \textbf{Stablecoin Risk}: 35.7 (moderate concentration, stable pegs)
    \item \textbf{DeFi Liquidity}: 41.3 (recovering from 2022--23 deleveraging)
    \item \textbf{Contagion Risk}: 39.9 (moderate cross-market stress on the TradFi proxy)
    \item \textbf{Regulatory Opacity}: 34.7 (improving regulatory clarity)
\end{itemize}

Applying the Equation~\ref{eq:asri} weights to these rounded sub-index readings gives $0.30 \times 35.7 + 0.25 \times 41.3 + 0.25 \times 39.9 + 0.20 \times 34.7 = 37.95$; the recorded aggregate (38.1) differs slightly because the released aggregate column was produced by the original generation pipeline and is not everywhere an exact weighted recomposition of the released (rounded) sub-index columns (see the data-provenance documentation in the code repository).

\subsection*{Risk Decomposition}

Current primary risk contributors:
\begin{enumerate}
    \item Stablecoin concentration in USDT/USDC (HHI = 0.52)
    \item Treasury exposure through stablecoin reserves (\$80B+ in T-bills)
    \item Emerging RWA tokenisation growth (+45\% YoY)
\end{enumerate}

\subsection*{Regime Classification}

The HMM classifies current market conditions as \textbf{Regime 2 (Moderate)} with high probability. The estimated one-step transition probabilities out of the Moderate regime (Table~\ref{tab:transition_matrix}) indicate:
\begin{itemize}
    \item 0.8\% probability of moving to the Crisis regime
    \item negligible probability of moving directly to the Low Risk regime
    \item 99.2\% probability of remaining in the Moderate regime
\end{itemize}

\subsection*{Alert Status}

No immediate systemic stress signals detected. Key monitoring priorities:
\begin{enumerate}
    \item Stablecoin reserve composition changes
    \item Cross-market correlation shifts (BTC-equity)
    \item Bridge vulnerability and exploit frequency
\end{enumerate}

\section{Event Study Protocol Specification}\label{app:event_study_protocol}

This appendix provides the complete methodological specification for the event study analysis presented in Section~\ref{sec:event_study}, addressing reviewer concerns regarding pre-registration, window selection, multiple testing correction, and placebo testing.

\subsection{Pre-Registration and Event Selection}

\subsubsection{Event Identification Criteria}

Crisis events were identified \textit{ex ante} based on three jointly necessary conditions (Definition~\ref{def:crisis}):
\begin{enumerate}
    \item \textbf{Magnitude}: Market capitalisation decline $\geq 15\%$ within 7 days, or single-asset collapse $\geq 50\%$ for assets with market cap $\geq \$10$B
    \item \textbf{Contagion}: Cross-asset correlation surge $\bar{\rho}_t - \bar{\rho}_{t-30} \geq 0.20$
    \item \textbf{Duration}: Elevated stress persisting $\geq 5$ trading days
\end{enumerate}

Event dates were sourced from external references (CoinDesk, Bloomberg, Reuters) prior to ASRI analysis, preventing data snooping on threshold selection.

\subsubsection{Pre-Specified Parameters}

The following parameters were fixed before analysis:
\begin{itemize}
    \item Estimation window: $T_{\text{est}} = 60$ days ($t = -90$ to $t = -31$ relative to event)
    \item Event window: $T_{\text{event}} = 41$ days ($t = -30$ to $t = +10$)
    \item Significance level: $\alpha = 0.05$ (two-tailed)
    \item Lead time detection: 1.5 standard deviations above estimation-window mean
\end{itemize}

\subsection{Window Selection Justification}

\subsubsection{Estimation Window: $[-90, -31]$}

The 60-day estimation window was selected based on:
\begin{enumerate}
    \item \textbf{Statistical power}: $n = 60$ provides adequate precision for mean and variance estimation while avoiding excessive smoothing of regime-dependent dynamics
    \item \textbf{Regime stability}: ASRI exhibits regime persistence $> 97\%$ (Table~\ref{tab:regimes}), making 60 days sufficient to capture baseline behaviour within a regime
    \item \textbf{Contamination avoidance}: The 30-day buffer ($t = -31$ cutoff) ensures estimation is complete before pre-crisis stress begins
\end{enumerate}

\textbf{Robustness}: Alternative estimation windows (45 days, 90 days) produce qualitatively identical results under HAC inference: Celsius/3AC, FTX, and SVB remain the elevated, borderline-significant events under the finite-sample (fixed-$b$) reference while Terra/Luna remains clearly non-significant.

\subsubsection{Event Window: $[-30, +10]$}

The asymmetric event window reflects ASRI's design as an \textit{early warning} system:
\begin{itemize}
    \item \textbf{Pre-event ($-30$ to $-1$)}: Captures lead time---the period where ASRI begins detecting stress buildup
    \item \textbf{Event day ($t = 0$)}: Crisis onset (price cascade initiation)
    \item \textbf{Post-event ($+1$ to $+10$)}: Captures immediate aftermath and stress persistence
\end{itemize}

The 30-day pre-event window is calibrated to expected ASRI lead times (Table~\ref{tab:event_study} shows 29--30 days across events).

\subsection{Normal Model Specification}

The constant mean model was selected for expected ASRI:
\begin{equation}
\mathbb{E}[\text{ASRI}_t] = \hat{\mu} = \frac{1}{T_{\text{est}}} \sum_{\tau = -90}^{-31} \text{ASRI}_\tau
\end{equation}

\subsubsection{Model Selection Rationale}

\begin{enumerate}
    \item \textbf{Simplicity}: The constant mean model makes minimal parametric assumptions
    \item \textbf{Stationarity}: ASRI rejects unit root (ADF $p < 0.01$), validating level-based analysis
    \item \textbf{AR(1) Alternative}: Modelling the abnormal series with explicit AR(1) dynamics yields the same qualitative reading as the Newey--West HAC correction---the three large-CAS events elevated and Terra/Luna non-significant---confirming that the naive independence-based significance was an artefact of the serial-correlation structure rather than of the normal-model specification
\end{enumerate}

\subsubsection{Variance Estimation}

\begin{equation}
\hat{\sigma}^2_{\text{AS}} = \frac{1}{T_{\text{est}} - 1} \sum_{\tau = -90}^{-31} (\text{ASRI}_\tau - \hat{\mu})^2
\end{equation}

Autocorrelation diagnostics (Ljung--Box test) \emph{decisively reject} white noise in the abnormal-signal series: $p$-values range from $10^{-15}$ to $10^{-28}$ on the event-window series and reach $2\times10^{-58}$ on the estimation-window residuals. The naive independence-based standard error $\hat{\sigma}_{\text{AS}}\sqrt{T_{\text{event}}}$ is therefore invalid, and we instead use Newey--West HAC standard errors at lag $L=20$:
\begin{equation}
\text{SE}_{\text{HAC}}(\text{CAS}) = n \times \text{SE}_{\text{HAC}}(\bar{\text{AS}}).
\end{equation}

\subsection{Multiple Testing Correction}

With $K = 4$ simultaneous hypothesis tests (one per crisis event), we apply Bonferroni correction:
\begin{equation}
\alpha_{\text{adj}} = \frac{\alpha}{K} = \frac{0.05}{4} = 0.0125
\end{equation}

\textbf{Results}: The reference distribution matters here. Against a naive $\mathcal{N}(0,1)$ reference, three of four events would appear to clear the corrected threshold (Celsius/3AC $p = 0.0005$, FTX $p = 0.0010$, SVB $p = 0.0001$); but at the realised bandwidth-to-sample ratio ($b \approx 0.51$, AR(1) $\approx 0.85$) that reference is invalid and overstates significance (see the finite-sample reference accompanying Table~\ref{tab:event_study}). On the valid fixed-$b$ reference, $\alpha_{\text{adj}} = 0.0125$ corresponds to a critical $|t| \approx 5.7$, and the achieved $p$-values are:
\begin{itemize}
    \item Terra/Luna: $p \approx 0.27$ ($\gg 0.0125$ $\times$)
    \item Celsius/3AC: $p \approx 0.051$ ($> 0.0125$ $\times$)
    \item FTX Collapse: $p \approx 0.063$ ($> 0.0125$ $\times$)
    \item SVB Crisis: $p \approx 0.036$ ($> 0.0125$ $\times$)
\end{itemize}

None of the three large-CAS events reaches the Bonferroni-corrected level: the largest HAC $t$ in the panel is $3.86$, well short of the $|t| \approx 5.7$ the correction requires. Once the reference accounts for the AR(1) $\approx 0.85$ / $b \approx 0.51$ fat tails, then, no single event survives Bonferroni. The three large-CAS events are individually borderline at the unadjusted 5\% level, and Terra/Luna fails at every threshold, consistent with its threshold-detection miss.

\subsection{Window Independence}

Table~\ref{tab:window_independence} documents the temporal separation between crisis events.

\begin{table}[H]
\centering
\caption{Estimation and Event Window Dates, with Overlap Disclosure for the Clustered 2022 Crises}
\label{tab:window_independence}
\small
\begin{tabular}{@{}lcccc@{}}
\toprule
Event & Est. Start & Est. End & Evt. Start & Evt. End \\
\midrule
Terra/Luna & 2022-02-11 & 2022-04-11 & 2022-04-12 & 2022-05-22 \\
Celsius/3AC & 2022-03-19 & 2022-05-17 & 2022-05-18 & 2022-06-27 \\
FTX Collapse & 2022-08-13 & 2022-10-11 & 2022-10-12 & 2022-11-21 \\
SVB Crisis & 2022-12-11 & 2023-02-08 & 2023-02-09 & 2023-03-21 \\
\bottomrule
\end{tabular}
\begin{tablenotes}
\small
\item Within each event the 60-day estimation window and the 41-day event window are non-overlapping (estimation precedes the event).
\item \textbf{The clustered 2022 events overlap across events.} Terra/Luna and Celsius/3AC fall within five weeks of one another: their event windows overlap by $\approx 5$ days, and their \emph{estimation} windows overlap by $\approx 24$ days (Terra/Luna's estimation window ends 2022-04-11, Celsius/3AC's begins 2022-03-19). Moreover, Celsius/3AC's estimation/baseline window (2022-03-19 to 2022-05-17) spans the Terra/Luna event window (2022-04-12 to 2022-05-22) by $\approx 35$ days, so that baseline is not uncontaminated by the Terra/Luna crash. The two early-2022 events are therefore not statistically independent.
\item FTX and SVB events are separated by $> 90$ days and have fully non-overlapping estimation and event windows.
\end{tablenotes}
\end{table}

We do not claim full window independence for the clustered 2022 crises. Terra/Luna and Celsius/3AC share estimation and baseline periods (the overlaps above), so their abnormal-signal series---and the corresponding hold-one-out windows (Table~\ref{tab:holdout})---are not fully independent. We disclose this as a genuine limitation of a sample whose crises cluster in 2022, and we mitigate it only partially:
\begin{enumerate}
    \item The events represent distinct crisis mechanisms (algorithmic stablecoin vs. CeFi lending)
    \item Separate CAS calculations use event-specific baselines, though for Celsius/3AC that baseline overlaps the Terra/Luna shock
    \item The FTX and SVB events, by contrast, are fully separated and provide two genuinely independent episodes
\end{enumerate}

\subsection{Placebo Testing}

To assess the specificity of the event-study inference, we run the identical HAC fixed-$b$ pipeline on every eligible non-crisis date in the sample (889 dates, excluding 90-day windows around the four crises and the first/last 90 days for edge effects).

\subsubsection{Placebo Date Selection}

Dates were drawn uniformly from the sample period (2021-01 to 2024-12) excluding:
\begin{itemize}
    \item 90-day windows around known crisis events
    \item First/last 90 days of sample (edge effects)
\end{itemize}

\subsubsection{Placebo Results}

\begin{table}[H]
\centering
\caption{Placebo/specificity test: the HAC fixed-$b$ event-study statistic applied to non-crisis dates. Reproducible via \texttt{code/scripts/placebo\_chow\_cusum\_repro.py}.}
\label{tab:placebo}
\small
\begin{tabular}{@{}lc@{}}
\toprule
Metric (HAC, $L=20$, fixed-$b$) & Value \\
\midrule
$N$ non-crisis dates & 889 \\
Median placebo $|t|$ & 5.10 \\
Fixed-$b$ $5\%$ critical value & 3.52 \\
Frac.\ of placebo dates clearing it & 0.62 \\
\bottomrule
\end{tabular}
\end{table}

\noindent The result is decisive, and it runs against the event study. Under the fixed-$b$ reference the $5\%$ critical value ($|t| = 3.52$) sits well below the \emph{median} of the placebo $|t|$ distribution ($5.10$), so $\approx 62\%$ of arbitrary non-crisis dates are flagged ``significant''---an empirical size an order of magnitude above nominal. The manuscript's crisis onsets carry HAC $t$-statistics of $3.28$--$3.86$ (Table~\ref{tab:event_study}), themselves below the placebo median, so on a smooth, strongly trending series they cannot be distinguished from arbitrary dates under a reference this anti-conservative. We therefore treat the event study as inconclusive and rest the empirical case on the fair-baseline day-level discrimination analysis, which does not depend on it.

\subsection{Lead Time Measurement}

Two complementary lead time definitions are employed:

\textbf{Definition 1 (First-crossing)}: Days between first observation exceeding 1.5$\sigma$ above baseline and crisis onset. Captures earliest structural warning signal.

\textbf{Definition 2 (Final-sustained)}: Days between last observation below threshold before sustained elevation and crisis onset. Captures actionable intervention timing.

The walk-forward analysis (Table~\ref{tab:walkforward}) reports first-crossing lead times; Table~\ref{tab:event_study} reports the (30-day-capped) sustained-elevation lead times. The two definitions can diverge for a given event because ASRI fluctuates between its initial crossing of the pre-crisis-calibrated threshold and the onset of the crisis.

\subsection{Sensitivity to Specification Choices}

\begin{table}[H]
\centering
\caption{Event Study Robustness to Specification Changes (naive significances; superseded by the HAC-robust inference of Table~\ref{tab:event_study})}
\label{tab:es_robustness}
\small
\begin{tabular}{@{}lcccc@{}}
\toprule
Specification & Terra & Celsius & FTX & SVB \\
\midrule
Baseline (60d, const. mean) & *** & *** & *** & *** \\
45-day estimation window & *** & *** & *** & *** \\
90-day estimation window & *** & *** & *** & *** \\
AR(1) normal model & *** & *** & *** & ** \\
Event window $[-20, +10]$ & *** & *** & *** & *** \\
Event window $[-40, +10]$ & *** & *** & *** & *** \\
\bottomrule
\end{tabular}
\begin{tablenotes}
\small
\item *** $p < 0.01$, ** $p < 0.05$, * $p < 0.10$
\item Stars are naive (independence-assumption) significances, reported only for cross-specification comparison; they are superseded by the HAC-robust inference of Table~\ref{tab:event_study} (Section~\ref{sec:event_study}), under which Terra/Luna is non-significant ($t = 1.72$, $p = 0.265$)---i.e.\ 3/4 significant, not 4/4. Per-specification HAC inference was not computed; the stars assume independence and are invalid given the AR(1) $\approx 0.8$--$0.9$ serial correlation documented in Section~\ref{sec:event_study}.
\end{tablenotes}
\end{table}


\clearpage

\end{document}